\documentclass[final,1p,times,authoryear]{elsarticle}

\usepackage[utf8]{inputenc} 
\usepackage[T1]{fontenc}

\usepackage{epsfig}
\usepackage{amssymb}
\usepackage{amsmath}
\usepackage{graphicx}
\usepackage{multirow}
\usepackage[labelformat=simple]{subcaption}

\usepackage{placeins}
\usepackage{algorithm2e}
\SetKwInput{KwModel}{Model}
\makeatletter

\usepackage[dvipsnames]{xcolor}
\usepackage{soul}
\definecolor{myblue}{HTML}{2E59A7} 
\definecolor{myred}{HTML}{D23918} 
\definecolor{myblack}{HTML}{151D29} 
\definecolor{myyellow}{HTML}{E5A84B} 
\definecolor{mygreen}{HTML}{057748} 
\definecolor{mybrown}{HTML}{9F6027} 
\definecolor{mypurple}{HTML}{674196} 
\definecolor{mypink}{HTML}{FF0097} 
\usepackage[normalem]{ulem}

\newcommand{\cgreen}[1]{{\color{mygreen} {#1}}}

\newcommand{\cred}[1]{{\color{red} {#1}}}

\usepackage[pdfencoding=auto, psdextra, colorlinks = true]{hyperref}
\hypersetup{
  colorlinks=true,
  citecolor=violet,
  linkcolor=red,
  urlcolor=green,
  }
\usepackage{cleveref}

\crefname{figure}{fig.}{figs.} 
\Crefname{figure}{Fig.}{Figs.} 
\crefname{table}{tab.}{tabs.} 
\Crefname{table}{Tab.}{Tabs.} 

\usepackage{amsthm}

\journal{International Journal of Heat and Fluid Flow}

\begin{document}

\begin{frontmatter}




\title{Separation control applied to the turbulent flow around a NACA4412 wing section}


\author[inst1]{Yuning Wang}
\author[inst1,inst2]{Fermin Mallor}
\author[inst3]{Carlos Guardiola}
\author[inst1]{Raffaello Mariani}
\author[inst1]{Ricardo Vinuesa}

\affiliation[inst1]{organization={Department of Engineering Mechanics},
            addressline={KTH Royal Institute of Technology}, 
            postcode={SE-100 44},
            city={Stockholm},
            country={Sweden}}

\affiliation[inst2]{
              organization = {PredictiveIQ},
              addressline = {FL 33067, Parkland},
              country = {USA}}

\affiliation[inst3]{
              organization = {Departamento de M{\'a}quinas y Motores T{\'e}rmicos},
              addressline = {Universitat Polit{\`e}cnica de Val{\`e}ncia, Camino de Vera, s.n., 46022 Valencia},
              country = {Spain}}

\begin{abstract}
    We carried out high-resolution large-eddy simulations (LESs) to investigate the effects of several separation-control approaches on a NACA4412 wing section with spanwise width of $L_z = 0.6$ at an angle of attack of $AoA=11^{\circ}$ {at} a Reynolds number of $Re_c = 200,000$ based on chord length $c$ and free-stream velocity $U_{\infty}$.
    Two control strategies were considered: (1) steady uniform blowing and/or suction on the suction and/or pressure sides, and (2) periodic control on the suction side. 
    A wide range of control configurations were evaluated in terms of {aerodynamic efficiency (i.e., lift-to-drag ratio) and separation delay.} 
    Uniform blowing and/or suction effectively delayed flow separation, leading to a lift increase of up to $11\%$, but 
    yielded only marginal improvements in aerodynamic efficiency. 
    In contrast, periodic control neither enhanced separation delay nor improved efficiency.
    A detailed analysis of the interaction between uniform blowing and/or suction 
    and turbulent boundary layers (TBLs) over the wing was performed, including assessments of (1) integral boundary-layer quantities, (2) turbulence statistics, and (3) power-spectral densities. Significant modifications in Reynolds stresses and spectral characteristics were observed.
    To the authors' best knowledge, this is the first numerical study utilizing high-resolution LESs to provide comprehensive assessments on separation control.
\end{abstract}


\begin{highlights}
  \item High-fidelity simulations of a {separated} wing section with a wide spanwise width.
  \item Steady and oscillatory uniform blowing/suction {control strategies} are analyzed.
  \item A wide range of control {parameters} are evaluated for separation control.
  \item Combining steady uniform blowing and suction achieves optimal performance.
  \item Flow separation is eliminated, with a lift increase of up to $11\%$.
  \item Blowing and suction show evident impacts on turbulent boundary layers over the wing.
\end{highlights}


\end{frontmatter}

{
  \section*{Nomenclature}
  \vspace{-0.3cm}
  {
    \renewcommand\arraystretch{1.2}
  \noindent
  \textbf{Latin symbols} \\
  \begin{tabular}{ll}
  $b$               & Spanwise width \\
  $c$               & Chord length \\
  $C_d$             & Drag coefficient \\
  $C_l$             & Lift coefficient \\
  $c_f$             & Skin-friction coefficient \\
  $c_p$             & Pressure coefficient \\
  $C_{\mu}$         & Momentum coefficient \\
  $f$               & Frequency \\
  $H_{12}$          & Shape factor \\
  $k$               & Wave number  \\
  $L$               & Length scale \\
  $l^*$             & Viscous length scale \\
  $L/D$             & Lift-to-drag ratio \\
  $P$               & Mean pressure \\ 
  $P_e$             & Boundary-layer-edge pressure \\ 
  $p$               & Fluctuating pressure \\ 
  $q$               & Free-stream dynamic pressure \\ 
  $Re_c$            & Chord-length-based Reynolds number \\
  $Re_\tau$         & Friction Reynolds number \\
  $Re_\theta$       & Momentum-thickness-based Reynolds number \\
  $t$               & Flow-over time unit \\
  $U_t, V_n, W$     & Mean component of wall-tangential, wall-normal and spanwise velocity \\
  $\overline{u^2_t}, \overline{v^2_n}, \overline{w^2}$   & Fluctuation component of wall-tangential, wall-normal and spanwise velocity \\
  $\overline{u_tv_n}$   & Reynolds-shear stress \\
  $u$               & Fluctuating streamwise velocity \\
  $U_e$             & Edge velocity of boundary layer  \\
  $U_{\infty}$      & Free-stream velocity \\
  $u_{\tau}$        & Friction velocity  \\
  $x,y,z$           & Streamwise, vertical and spanwise distance \\
  $x_t,y_n$         & Wall-tangential and wall-normal distance \\
  \end{tabular}
  
  \vspace{0.3cm}
  \noindent
  \textbf{Greek symbols} \\
  \begin{tabular}{ll}
  $\beta$           & Clauser pressure-gradient parameter \\
  $\delta$          & Boundary-layer thickness \\
  $\delta^*$        & Displacement thickness \\
  $\delta_{99}$     & $99\%$ boundary-layer thickness \\
  $\epsilon$        & Dissipation rate \\
  $\ell_{\rm ctrl}$ & Streamwise extent of control region \\ 
  $\ell_{\rm sep}$  & Streamwise length of flow separation \\ 
  $\eta$            & Kolmogorov length scale \\
  $\Gamma$          & Local force \\
  $\lambda$         & Wave length \\
  $\rho$            & Density             \\
  $\theta$          & Momentum thickness  \\
  $\nu$             & Kinematic viscosity \\
  $\psi$            & Control intensity  \\
  $\tau_w$          & Wall-shear stress \\
  \end{tabular}
  
  \vspace{0.3cm}
  \noindent
  \textbf{Abbreviations} \\
  \begin{tabular}{ll}
  AFC               & Active flow control \\
  AMR               & Adaptive mesh refinement \\
  APG               & Adverse pressure gradient \\
  BDF               & Backward differentiation formula \\
  DNS               & Direct numerical simulation \\
  DRL               & Deep reinforcement learning \\
  FFT               & Fast Fourier transform \\
  FPG               & Favorable pressure gradient \\
  GL                & Gauss--Legendre \\
  GLL               & Gauss--Lobatto--Legendre \\
  LE               & Leading edge \\
  LES               & Large-eddy simulation \\
  PS               & Pressure side \\
  PSD               & Power-spectral density \\
  RANS              & Reynolds-averaged Navier--Stokes \\
  SEM               & Spectral-element method \\
  SGS               & Subgrid scale \\
  SS               & Suction side \\
  TBL               & Turbulent boundary layer \\
  TE               & Trailing edge \\ 
  U.B               & Uniform blowing \\
  U.S               & Uniform suction \\
  \end{tabular}
  
  \vspace{0.3cm}
  \noindent
  \textbf{Subscripts/Superscripts/Operators} \\
  \begin{tabular}{ll}
  $_{\rm ctrl}$     & Controlled flow      \\
  $_\infty$         & Free-stream quantity \\
  $_d$              & Drag  \\ 
  $_n$              & Wall-normal direction \\ 
  $_f$              & Friction  \\ 
  $_l$              & Lift  \\ 
  $_p$              & Pressure  \\ 
  $_t$              & Wall-tangential direction \\
  $_w$              & Wall quantity \\
  $_x \  _y \  _z$  & Streamwise, vertical and spanwise direction \\
  $\overline{(\cdot)}$ & Time-averaged quantity \\
  $^+$              & Inner-scaled quantity \\ 
  \end{tabular}
  }
}

\section{Introductions}
\label{sec:intro}

Flow separation refers to the detachment of a fluid from a solid surface~\citep{maskell_separation_1955}, 
which occurs due to a severe adverse pressure gradient (APG)~\citep{simpson_APGsep_1989}. 
This phenomenon leads to a significant reduction in near-wall momentum and flow velocity, 
often resulting in decreased aerodynamic efficiency~\citep{abdolahipour_review_2024}. 
In aviation, flow separation is particularly critical during take-off, where low-speed conditions 
require maximizing lift by increasing the angle of attack ($AoA$) of the wing. However, as the $AoA$ 
increases, the APG intensifies, often triggering flow separation before the maximum lift is achieved. 
This disrupts lift generation and degrades aerodynamic performance. Consequently, separation control 
is essential for optimizing aerodynamic efficiency, improving emission control, and enhancing economic 
viability~\citep{fukagata_reviewCTRL_2024}.

{The key to delaying flow separation lies in augmenting near-wall momentum~\citep{greenblatt_control_2000}.} 
Over the past decades, researchers have explored various separation control strategies. These methods 
fall into two main categories: passive and active flow control (AFC), based on whether they require 
an external energy input~\citep{gad_modern_1996}.
{Passive control techniques do not require external energy but instead redistribute momentum 
to influence flow behavior.} A common example is the vortex generator, 
which creates streamwise vortices that energize the boundary layer, enabling it to resist APGs~\citep{lin_reviewVortexGen_2002}. 
While effective in delaying flow separation~\citep{neves_vortexGen_2024}, vortex generators can increase drag 
in scenarios where separation is absent~\citep{amitay_aerodynamic_2001}.

In contrast, AFC methods utilize external energy sources to modify momentum exchange. Compared to passive methods, 
AFC offers greater efficiency and adaptability since it can be activated or deactivated as needed~\citep{brunton_closed-loop_2015}. 
{A variety of active control strategies have been investigated, including vibrating flaps~\citep{nishri_flap_1998} 
and acoustic excitation~\citep{huang_Acousticseparation_1987}.}
Recently, deep reinforcement learning (DRL) has emerged as an innovative AFC technique, demonstrating promising results 
in separation control~\citep{suarez_highRe_2024,suarez_transition_2024,font_sepBubbleDRL_2024}. DRL leverages the capability 
of neural networks to model complex nonlinear interactions~\citep{vinuesa_flow_2022}.

Among AFC techniques, steady uniform suction and blowing have been widely utilized, with a long history 
of application~\citep{prandtl_uniform_1904}.
{Steady suction removes low-momentum fluid from the boundary layer while deflecting high-momentum 
fluid toward the wall, effectively stabilizing the boundary layer~\citep{schlichting_BLTheory_2016}.}
This technique is typically employed on the suction side of airfoils to mitigate separation. Numerous 
experimental and numerical studies confirm its effectiveness. For instance, \citet{hunter_flight_1954} 
used leading-edge suction on a thin airfoil, successfully eliminating separation. Similar enhancements 
in maximum lift coefficients have been demonstrated for gliders~\citep{cornish_prevention_1953} and motor 
planes~\citep{raspet_delay_1956}. Parametric studies~\citep{goodarzi_SuctionInvestigation_2012} and large-eddy simulations 
(LES)~\citep{zhi_lowResuction_2018} further explored suction control performance on various airfoil shapes.
Steady uniform blowing increases near-wall momentum by enhancing wall-normal convection. 
{While primarily employed for circulation control~\citep{chen_BlowingSeparation_2012}, 
steady blowing has also been used for drag reduction~\citep{vinuesa_skin_2017}.}
Combinations of blowing and suction have also been extensively studied. {For example, simulations 
of NACA0012 airfoils~\citep{yousefi_BlwSct3D_2015} demonstrated improvements in aerodynamic efficiency and separation delay through optimized control configurations.}

Beyond steady control methods, periodic-control techniques use oscillatory blowing or suction to introduce 
additional momentum at specific frequencies, modifying mixing-layer development and delaying separation~\citep{greenblatt_control_2000}. 
Experimental studies on NACA0015 airfoils~\citep{seifert_delay_1996} showed that oscillatory blowing outperformed steady 
blowing in enhancing maximum lift. {Wind-tunnel tests~\citep{tang_PeriodicUse_2014} and numerical investigations~\citep{abdolahipour_effects_2023} 
further highlight the effectiveness of periodic control.}

Despite these advancements, several limitations remain. Experimental studies primarily focus on 
aerodynamic performance metrics (e.g., lift and drag) but lack high-fidelity flow measurements. 
Moreover, most numerical studies employ Reynolds-averaged Navier--Stokes (RANS) models with two-dimensional 
domains, potentially overestimating control effectiveness by neglecting three-dimensional effects.
To address these gaps, the present study conducts high-resolution LES of a NACA4412 wing section at 
$AoA = 11^{\circ}$ and $Re_c = 200{,}000$ to evaluate steady and oscillatory blowing/suction configurations. 
This study focuses on a single airfoil type and angle of attack to facilitate direct comparisons with 
existing studies~\citep{atzori_aerodynamic_2020,mallor_bayesian_2024} while maintaining computational feasibility. 
For higher angles of attack and Reynolds numbers, we refer to the comprehensive database documented in~\citet{mallor_IJHFF_2024}. 
To the best of the authors' knowledge, this is the first numerical study employing high-resolution LES 
to provide comprehensive assessments of separation control.

The paper is organized as follows. In \S~\ref{sec:method}, we introduce the numerical setup and control configurations. In \S~\ref{sec:results}, we present findings on (1) control impacts on aerodynamic characteristics, (2) streamwise development of the boundary layer, (3) inner- and outer-scaled wall-normal profiles of turbulence statistics, and (4) spectral analysis. Finally, in \S~\ref{sec:conclusion}, we summarize the results and provide further discussion. 

\section{Methodology}\label{sec:method}
\subsection{High-resolution large-eddy simulation (LES)}\label{sec:sim}
We perform high-resolution LES of flow around a NACA4412 wing section at an angle of attack ($AoA$) of $11^{\circ}$ and a Reynolds number of $Re_c = 200{,}000$ using the incompressible Navier--Stokes solver Nek5000~\citep{nek5000}. 
The solver employs the spectral-element method (SEM)~\citep{patera_spectral_1984}, which combines high accuracy with computational efficiency.

The computational domain consists of hexahedral elements, where velocity and pressure are represented using Lagrange interpolants of polynomial order $N=7$. 
Following the $\mathbb{P}_N \mathbb{P}_{N-2}$ formulation~\citep{maday_spectral_1989}, velocity is evaluated at $N^3$ points based on the Gauss--Lobatto--Legendre (GLL) quadrature rule, while pressure is defined on a staggered grid of 
$(N-2)^3$ points using the Gauss--Legendre (GL) quadrature. 
Time integration is performed using an explicit third-order extrapolation (EXT3) scheme for nonlinear terms and an implicit third-order backward differentiation formula (BDF3) for viscous terms. 
To minimize aliasing errors, overintegration is applied by oversampling nonlinear terms by a factor of $3/2$ in each direction.

{An implicit subgrid-scale (SGS) model, based on a time-dependent relaxation-term filter~\citep{negi_unsteady_2018}, accounts for the dissipation of unresolved scales, which is able to resolve $\approx 90\%$ of the total dissipation~\citep{eitel_LESSGS_2014}.
This model operates through a volume force applied to a subset of spectral-element modes and has been validated in wing simulations by~\citet{vinuesa_turbulent_2018}. We refer to~\citet{negi_unsteady_2018} for the details of implementation.
}
Boundary-layer tripping is implemented via volume forcing~\citep{hosseini_DNSwing_2016}, applied at the streamwise location $x/c = 0.1$ on both suction and pressure sides. 
The tripping parameters follow~\citet{hosseini_DNSwing_2016} and are adjusted to the present setup using local laminar {boundary-layer} scales (i.e., the edge velocity $U_e$ and boundary-layer thickness $\delta^{*}$). 
This formulation ensures consistency across different computing environments and compilers~\citep{mallor_IJHFF_2024}. Further details on the Nek5000 implementation are provided by~\citet{massaro_NekFrame_2024}.
{Note that the tripping force, imposed at $x/c = 0.1$, ensures flow reattachment shortly downstream of the leading edge. As a result, the presence of a leading-edge separation bubble does not affect the flow development further downstream.}

\begin{figure}[ht]
  \centering
  \includegraphics[width=0.6\textwidth]{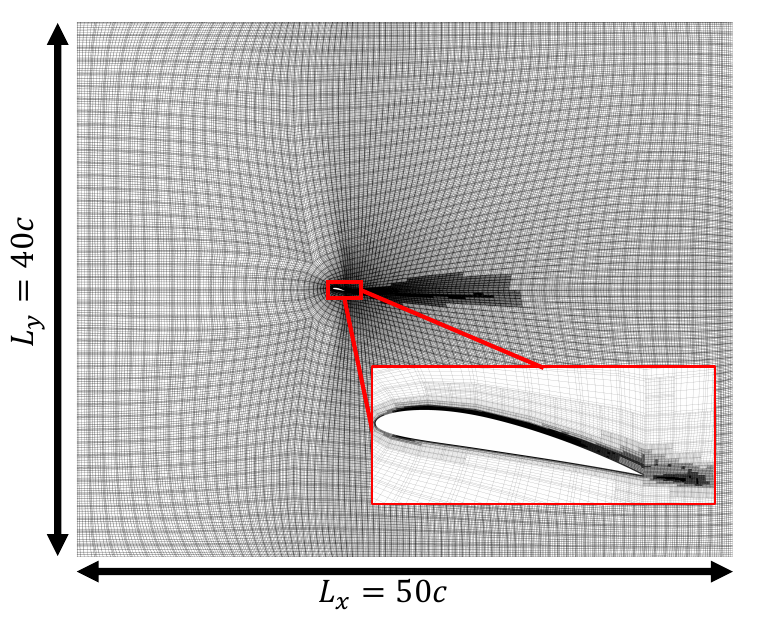}
  \caption{
    {Two-dimensional plane of the spectral-element mesh used in the computational domain. 
  The inset illustrates the mesh refinement near the airfoil surface. 
  The domain is three-dimensional with a spanwise width of $L_z = 0.6c$, containing 
  approximately $3.9 \times 10^{8}$ grid points after applying adaptive mesh 
  refinement (AMR).}
  }
  \label{fig:mesh_illust}
\end{figure}

The computational domain (see~\cref{fig:mesh_illust}) is a rectangular box with streamwise and vertical dimensions of $L_x = 50$ and $L_y = 40c$~\citep{vinuesa_minimumAR_2015} with a spanwise length of $L_z = 0.6c$. 
The leading (LE) and trailing edges (TE) of the airfoil are positioned at $-0.5c$ and $0.5c$ from the domain center, respectively. 
The airfoil is rotated to achieve $AoA = 11^{\circ}$. 
Boundary conditions are defined as follows: the front, upper, and lower boundaries use Dirichlet conditions, with a constant inflow velocity $U_{\infty}$ applied at the left boundary. 
The top and bottom boundaries enforce a normal outflow condition with tangential velocity $U_{\infty}$. 
For the right (outflow) boundary, a modified Neumann condition~\citep{dong_robustoutlet_2014} 
is used to prevent uncontrolled influx of kinetic energy.

The simulation employs adaptive mesh refinement (AMR) and non-conformal meshing to optimize accuracy while maintaining computational efficiency. 
Unlike prior studies that relied on conformal meshes~\citep{hosseini_DNSwing_2016,vinuesa_turbulent_2018}, which often led to excessive resolution and degraded pressure-solver performance, 
the present setup allows for a larger spanwise extent of $L_z = 0.6c$. 
This spanwise width is sufficient to capture the dominant turbulent structures in the boundary layer~\citep{vinuesa_turbulent_2018}.

The choice of $L_z$ is motivated by the need to resolve separation physics, as smaller domains can artificially constrain separation dynamics. 
\citet{mallor_IJHFF_2024} suggest that a spanwise width of at least $0.4c$ is needed to prevent artificial domain-induced constraints on turbulence dynamics. 
The present setup satisfies this criterion and exceeds the spanwise extent used in previous high-fidelity wing simulations at high angles of attack \citep{sato_large_2017,asada_large_2018,tamaki_wall_2023}. 
These choices are further supported by recent experimental findings~\citep{mallor_experimental_2025}.

The $h$-type AMR strategy in Nek5000~\citep{offermans_amr_2019} ensures that the mesh resolution adheres to high-resolution LES standards, closely approaching direct numerical simulation (DNS) accuracy~\citep{negi_unsteady_2018}. 
The computational domain consists of approximately $3.9 \times 10^{8}$ grid points after AMR is applied to the non-conformal mesh.
Near-wall spatial resolutions are defined in viscous units as follows: $\Delta x^+_t < 18.0$, $\Delta y^+_n < (0.64, 11.0)$, and $\Delta z^+ < 9.0$ in the tangential, normal, and spanwise directions, respectively. 
The viscous length scale is defined as $l^* = \nu / u_{\tau}$, where $u_{\tau} = \sqrt{\tau_{w} / \rho}$ is the friction velocity, and $\tau_{w} = \rho \nu \left( {\rm d} U_t / {\rm d} y_n \right)_{ y_n = 0 }$ is the mean wall-shear stress. 
{In the wake, to ensure the streamwise resolution is fine enough to resolve the convecting structures~\citep{spalart_comments_1997}, the spatial resolution is designed to satisfy $\Delta x / \eta < 9$, where $\eta = (\nu^3/\epsilon)^{1/4}$ is the Kolmogorov length scale and $\epsilon$ is the local isotropic dissipation rate. }

The AMR process follows these steps: first, isotropic elements are clustered in an oct-tree structure using the {\it p4est} library~\citep{burstedde_p4est_2011}, with interpolation operators~\citep{kruse_parallel_1997} ensuring continuity at non-conformal interfaces. 
An $a$-posteriori spectral error indicator~\citep{mavriplis_posteriori_1990} identifies elements requiring refinement by measuring truncation and quadrature errors. 
Refinement is applied iteratively until convergence is reached and the mesh meets the resolution criteria. 
Note that the refinement is homogeneous in the spanwise direction.
For details on AMR implementation in Nek5000 and validation against conformal wing simulations, see~\citet{tanarro_AMR_2020}.

\subsection{Control Configurations}\label{sec:config}

\Cref{tab:ctrl_configs} summarizes the control configurations considered in this study, which employ steady uniform blowing and suction to delay flow separation. 
Note that we explored periodic control over the suction side and other configurations of uniform suction (see~\ref{sec:appendix-clcd}) to enhance reattachment and aerodynamic efficiency. 
However, {the periodic} configurations {generally did not outperform} the cases listed in~\cref{tab:ctrl_configs} in terms of separation delay or aerodynamic efficiency. Thus, the analysis focuses on Cases A--E.

The primary configuration combines uniform suction (U.S.) on the suction side (SS) with uniform blowing (U.B.) on the pressure side (PS). 
The control intensity $\psi$ varies as $\psi = 0.25\%U_{\infty}$, $0.50\%U_{\infty}$, and $1.0\%U_{\infty}$ corresponding to Case A, Case B, and Case C, respectively. 
Case A replicates the optimized configuration from \citet{mallor_bayesian_2024}, which was {obtained} using Bayesian optimization in a 2D RANS simulation of the NACA4412 airfoil at $AoA = 11^{\circ}$ and a higher Reynolds number ($Re_c = 1{,}000{,}000$). 
This setup demonstrated significant improvements in aerodynamic efficiency and separation delay. 
However, due to RANS limitations, the control intensity was restricted to $\psi = 0.25\%U_{\infty}$ to prevent non-physical solutions. 
The high-resolution LES in {the present} study allows for higher control intensities while maintaining physical consistency.
Additionally, configurations with only uniform suction on the suction side (Case D) and only uniform blowing on the pressure side (Case E) are considered to assess the individual effects of each control method. 
These cases provide insights into the efficiency of different strategies in improving aerodynamic performance and delaying separation.

\begin{table}[ht]
  \centering
  \def~{\hphantom{0}}
  \begin{tabular}{ccccc}
    Case notation  & Control method  & Control area & Input intensity ($\psi$)   &   Color code           \\[3pt]
    {\rm Ref}   & Uncontrolled  & $ - $  &    $ - $         &  \colorbox{myblack}{\makebox[0.1em][c]{\rule{0pt}{0.1em}}}                     \\
    {\rm Case A}  & U.S. at SS \& U.B. at PS & $0.25 \leq x/c \leq 0.86$  & $0.25\% U_{\infty}$  &  \colorbox{myblue}{\makebox[0.1em][c]{\rule{0pt}{0.1em}}}             \\
    {\rm Case B} & U.S. at SS \& U.B. at PS & $0.25 \leq x/c \leq 0.86$ & $0.50\% U_{\infty}$  &  \colorbox{myyellow}{\makebox[0.1em][c]{\rule{0pt}{0.1em}}}             \\
    {\rm Case C}  & U.S. at SS \& U.B. at PS & $0.25 \leq x/c \leq 0.86$  & $1.00\% U_{\infty}$  &  \colorbox{myred}{\makebox[0.1em][c]{\rule{0pt}{0.1em}}}             \\
    {\rm Case D}  & U.S. at SS & $0.25 \leq x/c \leq 0.86$  & $0.25\% U_{\infty}$  &  \colorbox{mygreen}{\makebox[0.1em][c]{\rule{0pt}{0.1em}}}             \\
    {\rm Case E}  & U.B. at PS & $0.25 \leq x/c \leq 0.86$  & $0.25\% U_{\infty}$  &  \colorbox{mypurple}{\makebox[0.1em][c]{\rule{0pt}{0.1em}}}             \\
  \end{tabular}
  \caption{Control configurations considered in the present study on a NACA4412 wing section at $AoA = 11^{\circ}$ and ${Re}_c = 200{,}000$. The colored squares denote the color code for each case.}
  \label{tab:ctrl_configs}
\end{table}

\begin{figure}[ht]
  \centering
  \includegraphics[width=0.5\textwidth]{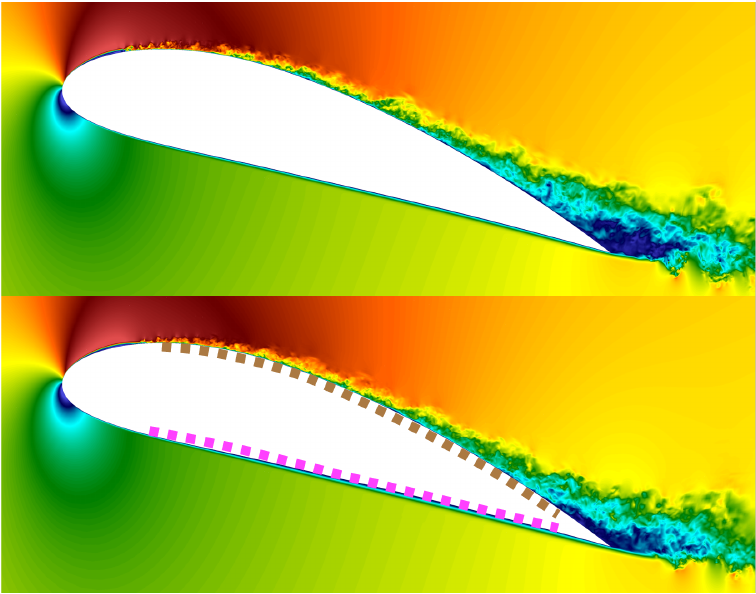}
  \caption{Snapshots of the horizontal velocity component at an arbitrary time step for (top) the uncontrolled case and (bottom) Case A. The velocity values range from $-0.2U_{\infty}$ {(dark blue)} to $1.8U_{\infty}$ {(dark red)}. Differences in {boundary-layer} thickness on both sides and the extension of attached flow downstream over the suction side are evident. The control regions for uniform suction on the SS. and uniform blowing on the PS. are indicated by dashed brown and magenta lines, respectively.}
  \label{fig:ctrl_illust}
\end{figure}

The control region {spans from }$x/c = 0.25$ to $0.86$ (see \cref{fig:ctrl_illust}), which aligns with previous studies~\citep{vinuesa_skin_2017,atzori_aerodynamic_2020,atzori_uniform_2021,wang_opposition_2024}, 
enabling a direct comparison of control effects. 
Blowing and suction are imposed as Dirichlet boundary conditions at the wall, where the horizontal and vertical velocity components are adjusted to ensure that the wall-normal velocity ($V_n$) matches the prescribed control intensity~\citep{wang_opposition_2024}. {Note that the boundary conditions for both the wall-tangential and spanwise velocity components are of Dirichlet type, and are set to zero in accordance with the no-slip condition.}

For each case, simulations were run for at least $2$ flow-over times (i.e., the time required for a fluid particle to traverse a distance of $c$ at velocity $U_{\infty}$) to ensure statistical convergence. 
Note that {for the} controlled cases{, the} simulations were initialized from a fully-developed turbulent field of the uncontrolled case. 
\citet{vinuesa_convergence_2016} suggested that statistical convergence depends on both time averaging and the 
spanwise width. 
\citet{vinuesa_turbulent_2018} reported that $\simeq 10$ flow-over times were required for a wing section with $L_z = 0.1c$ at $Re_c = 200{,}000$. 
Due to the larger spanwise width ($L_z = 0.6c$) in our setup, the achieved {convergence here} is equivalent to $\approx 12$ flow-over times for a domain with $L_z = 0.1c$.
However, for cases presented in \ref{sec:appendix-clcd}, a shorter averaging time of $\simeq 1.6$ flow-over 
times was used, which is sufficient for evaluating aerodynamic characteristics. 
The computational cost for simulating $2$ flow-over times is approximately 
$1.0$ million CPU hours on a Cray-XC40 system.

\section{Results}\label{sec:results}
In this section, we first evaluate the aerodynamic characteristics of the wing section under various control configurations. 
Next, we analyze the integral quantities of streamwise boundary-layer development, followed by an assessment of wall-normal profiles of turbulence statistics, including mean and fluctuating velocity components. 
Finally, we examine the time-series data using spectral analysis in terms of space and time.

\subsection{Aerodynamic characteristics}\label{sec:ctrl_eff}
We begin by assessing the control effect on aerodynamic characteristics, specifically focusing on separation delay and aerodynamic efficiency. 
The separation delay is quantified by the streamwise length of the separated flow 
($\ell_{\rm sep}$), defined as:
\begin{equation}
  \ell_{\rm sep} = (x_{\rm TE} - x_{\rm sep}) / c,
  \label{eq:lsep}
\end{equation}
\noindent where $x_{\rm TE}$ is the trailing-edge location, $x_{\rm sep}$ is the 
separation onset point where the skin-friction coefficient $c_f$ becomes 
negative, and $c$ is the chord length.

Another key metric for separation control is the momentum coefficient $C_{\mu}$, which governs the control effect through momentum addition~\citep{poisson_momentum_1948}, given by:
\begin{equation}
  C_{\mu} = \frac{\rho \psi^2 \ell_{\rm ctrl}}{\frac{1}{2}\rho U^2_{\infty}},
  \label{eq:Cmu}
\end{equation}
\noindent where $\ell_{\rm ctrl}$ is the streamwise extent of the control region. 
Note that for configurations that combine suction on the suction side and blowing on the pressure side, $C_{\mu}$ is doubled.

The aerodynamic efficiency is evaluated using the lift-to-drag ratio ($L/D$), where the lift and drag coefficients are defined as:
\begin{equation}
  C_l = \frac{f_l}{bq}, \quad C_d = \frac{f_d}{bq},
\end{equation}
\noindent where $b$ is the spanwise width, $f_l/b$ and $f_d/b$ are the lift and drag forces per unit length, and $q = \frac{1}{2} \rho U^2_{\infty}$ is the free-stream dynamic pressure. 
The drag coefficient is further decomposed into skin-friction drag ($C_{d,f}$) and pressure drag ($C_{d,p}$) such that $C_d = C_{d,f} + C_{d,p}$.

\begin{table}[ht]
  \begin{center}
  \resizebox*{\textwidth}{!}{
    \def~{\hphantom{0}}
  \begin{tabular}{ccccccccc}
    Name   & $\ell_{\rm sep}$ & $C_{\mu} [\times 10^{-5}]$ & $C_l$ & $C_{d,p}$ & $C_{d,f}$ & $C_d = C_{d,p} + C_{d,f}$ & $L/D$ \\ [3pt]
    Ref    & 0.14 & 0.0 & 1.314 & 0.0450  & 0.0078 & 0.0527  & 24.876 \\
    Case A & 0.02 (\cgreen{$-85\%$}) & $1.52$ & 1.355 (\cgreen{$+3.2\%$}) & 0.0473 (\cred{$+5.3\%$}) & 0.0083 (\cred{$+5.5\%$}) & 0.0556  (\cred{$+5.37\%$}) & 24.36 (\cred{$-2.06\%$})  \\
    Case B & 0.00 (\cgreen{$-100\%$}) & $6.10$ & 1.392 (\cgreen{$+5.9\%$}) & 0.0488 (\cred{$+8.55\%$}) & 0.0099 (\cred{$+26.46\%$}) & 0.0587 (\cred{$+11.21\%$}) & 23.67 (\cred{$-4.73\%$}) \\
    Case C & 0.00 (\cgreen{$-100\%$}) & $24.4$ & 1.463 (\cgreen{$+11.45\%$}) & 0.0519 (\cred{$+15.53\%$}) & 0.0145  (\cred{$+85.29\%$}) & 0.0665 (\cred{$+25.88\%$}) & 22.02 (\cred{$-11.47\%$})  \\
    Case D & 0.02 (\cgreen{$-85\%$}) & $0.76$ & 1.326 (\cgreen{$+0.95\%$}) & 0.0466 (\cred{$+3.57\%$}) & 0.0096  (\cred{$+15.75\%$}) & 0.0566 (\cred{$+5.38\%$}) & 23.829 (\cred{$-4.21\%$}) \\
    Case E & 0.05 (\cgreen{$-64\%$}) & $0.76$ & 1.312 (\cred{$-0.13\%$}) & 0.0457 (\cred{$+1.10\%$}) & 0.0069  (\cgreen{$-12.47\%$}) & 0.0523 (\cgreen{$-0.915\%$}) & 25.073 (\cgreen{$+0.79\%$}) \\
  \end{tabular}
  }
  \caption{Separation length ($\ell_{\rm sep}$), momentum coefficient ($C_{\mu}$), 
  lift coefficient ($C_l$), drag components ($C_{d,p}$ and $C_{d,f}$), total 
  drag ($C_d$), and aerodynamic efficiency ($L/D$) for the cases in~\cref{tab:ctrl_configs}. 
  Percent changes from the reference case are shown in parentheses.}
  \label{tab:ClCd}
  \end{center}
\end{table}

\begin{figure}[ht]
  \begin{center}
    \includegraphics[width=0.4\textwidth]{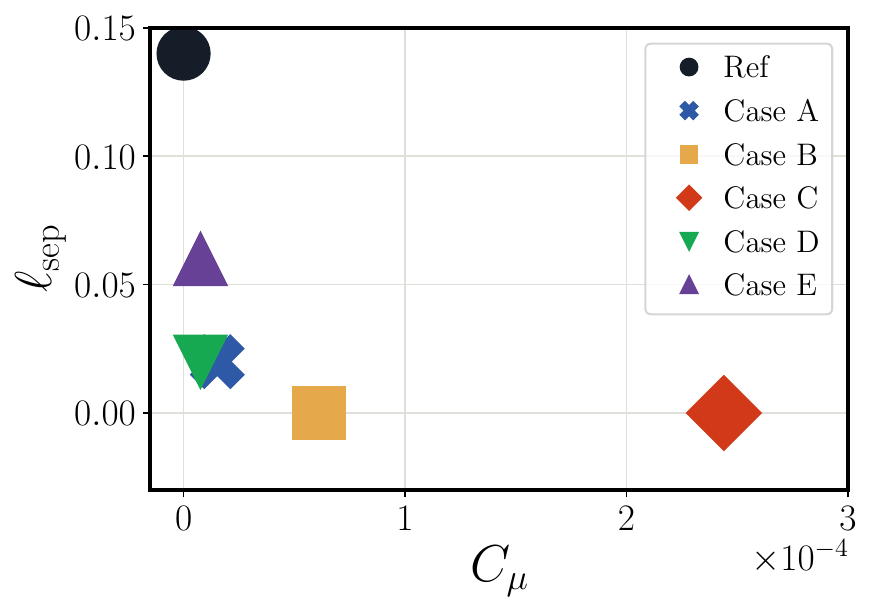}
    \caption{The length of separation region $\ell_{\rm sep}$ as a function of momentum coefficient $C_{\mu}$. The color code follows~\cref{tab:ctrl_configs}.}
    \label{fig:sep-loc}
  \end{center}
\end{figure}

\Cref{tab:ClCd} presents the control impact on aerodynamic performance. 
The uncontrolled flow separates at $x/c = 0.86$, yielding $\ell_{\rm sep} = 0.14$, which remains consistent across Reynolds numbers at $AoA = 11^{\circ}$~\citep{mallor_IJHFF_2024}. 
With control, separation is significantly delayed. 
{Case A reduces $\ell_{\rm sep}$ to $0.02$, whereas 
Cases B and C fully eliminate separation, achieving complete reattachment.}
This is directly linked to the increasing of $C_{\mu}$, whose relation with $\ell_{\rm sep}$ is clearly demonstrated by~\cref{fig:sep-loc}.
However, solely applying uniform suction over suction side (Case D) yield the same $\ell_{\rm sep}$ as combined control ({\rm Case A}) while the $C_{\mu}$ of Case D is half of that of {\rm Case A}. 
Moreover, solely uniform blowing over pressure side (Case E) yields a lower $\ell_{\rm sep} = 0.05$. 
The results indicate that suction over the suction side is dominant for delaying flow separation.

\begin{figure}[h!]
  \centering
  \begin{subfigure}{0.49\linewidth}
    \centering
    \includegraphics[width=\textwidth]{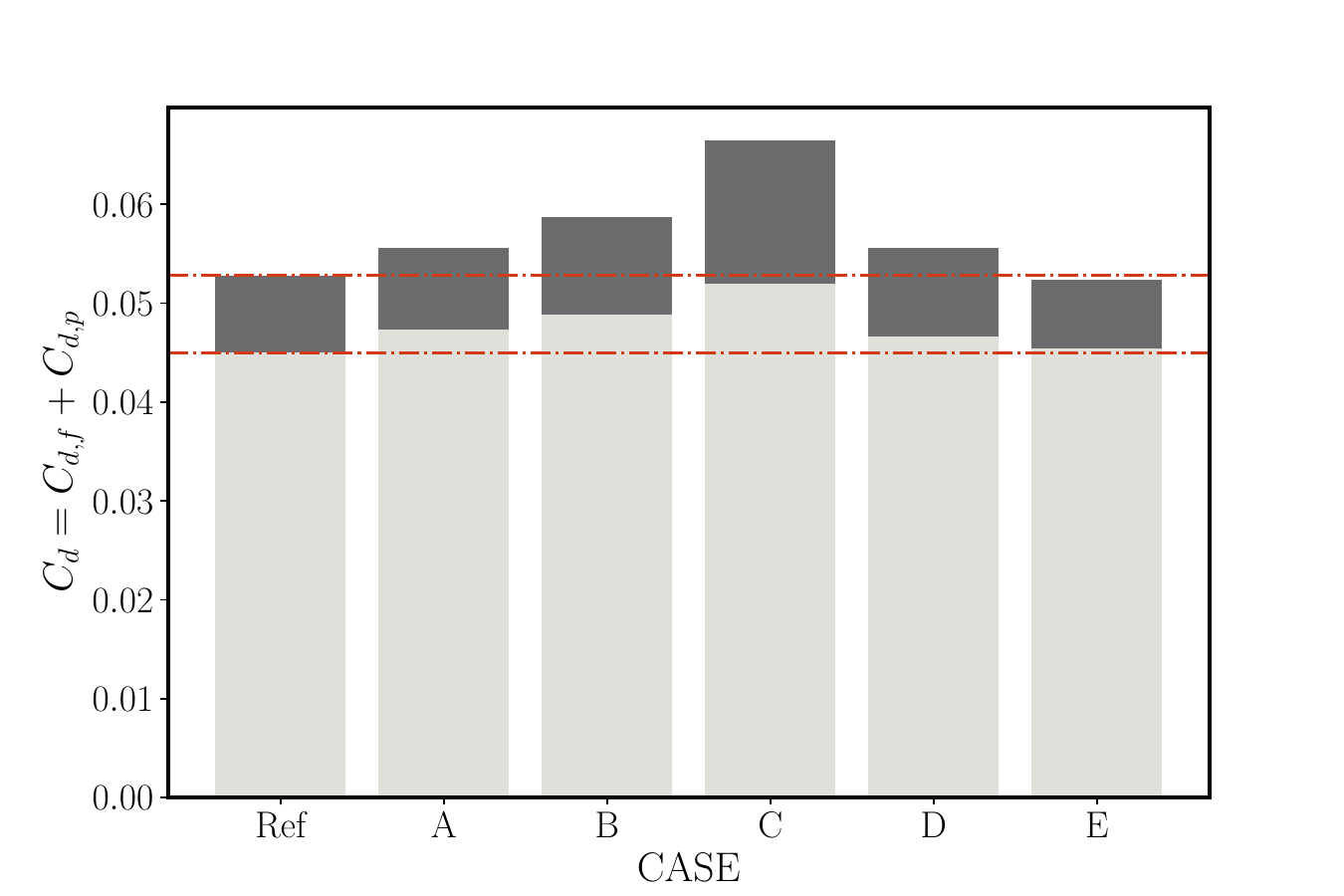}
    \subcaption{{Total drag and its $C_{d,f}$ and $C_{d,p}$ components.}}
    \label{fig:cl-bar}
  \end{subfigure}
  \hfill
  \begin{subfigure}{0.49\linewidth}
    \centering
    \includegraphics[width=\textwidth]{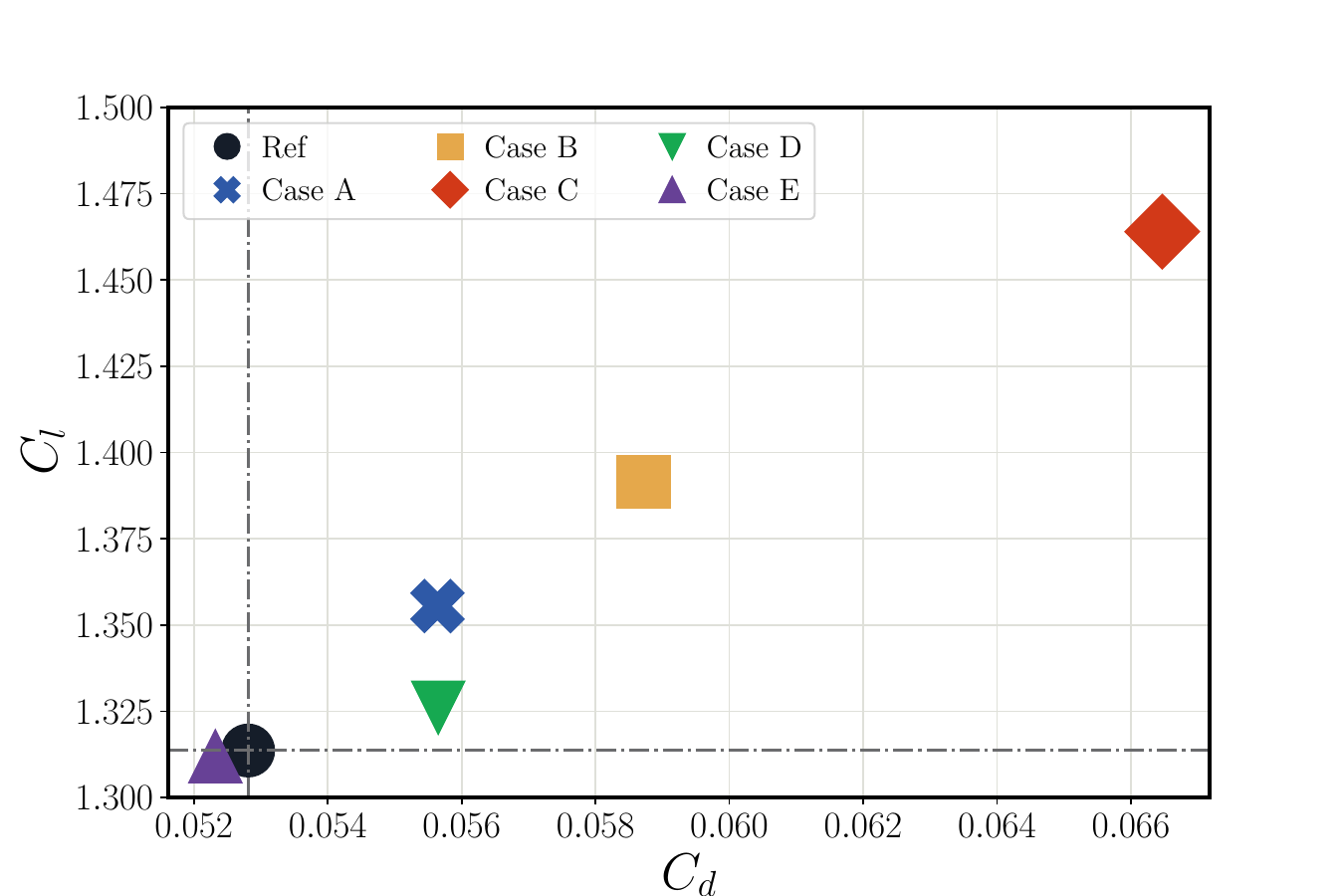}
    \subcaption{{$C_l$ as a function of $C_d$.}}
    \label{fig:cl-cd}
  \end{subfigure}
  \caption{(a) Total drag for all considered cases, where the $C_{d,f}$ and $C_{d,p}$ components are colored in light  and dark gray, respectively. Note that the uncontrolled $C_{d,f}$ and $C_{d,p}$ are denoted by red dash-dotted lines. (b) Relation between lift and drag, where the dash-dotted lines indicate the $C_l$ and $C_d$ of the uncontrolled case, respectively.}
  \label{fig:clcd}
  
\end{figure}

On the other hand, although the control consistently increases $C_l$ up to $11.45\%$, aerodynamic efficiency is more sensitive to changes in total drag. 
\Cref{fig:cl-bar} shows the the decomposition of the total drag into $C_{d,f}$ and $C_{d,p}$.
The control primarily increases skin-friction drag, with a smaller increase in pressure drag, which dominates at high angles of attack. 
As consequence, total drag increases significantly with higher control inputs, particularly in {\rm Case C}, where both drag components increase drastically.
\Cref{fig:cl-cd} illustrates the relationship between $C_l$ and $C_d$. 
The control configurations did not improve aerodynamic efficiency except {in} Case E.
In particular, at $\psi = 1.0\%U_{\infty}$ (Case C), the control drastically degrades the aerodynamic efficiency, reducing $L/D$ by $11.47\%$ although {\rm Case C} achieved the highest lift ($C_l = 1.46$), its $L/D$ decreased by $11.5\%$ due to a $25.8\%$ increase in total drag. 
This highlights the trade-off between lift enhancement and drag penalty at higher control intensities.
Note that the $L/D$ {values} of {Case A} is {in} contrast with those reported by~\citet{mallor_bayesian_2024}, where the same control configuration improved $L/D$ by $42\%$. 
However, their study was based on 2D RANS simulations at a higher Reynolds number ($Re_c = 1{,}000{,}000$), which may overestimate the control performance due to the absence of {spanwise and Reynolds number} effects{, as well as the effect of the turbulence model.}

Moreover, Case D reduces $L/D$ by $4.21\%$, with minimal lift improvement and a $1.1\%$ increase in $C_{d,p}$, while Case E is the only case that yields a marginal improvement in $L/D$ by reducing $C_{d,f}$ while minimally increasing $C_{d,p}$. 
These findings are distinct from those at $AoA = 5^{\circ}$ and the same Reynolds number.
Using $\psi = 0.2\%$ with same control area, uniform suction over the suction side improved $L/D$ by significantly increasing $C_l$ and reducing $C_{d,p}$, albeit with a slight increase in $C_{d,f}$, whereas uniform blowing over the pressure side improved $L/D$ by reducing both $C_{d,p}$ and $C_{d,f}$ outlined by~\citet{atzori_aerodynamic_2020}.
These results suggest that control strategies effective for attached flows may not be suitable for separation control, as flow properties differ significantly at high angles of attack.

{
Furthermore, we assess the distribution of lift and drag forces around the wing section. The integrated lift and drag forces can be expressed as:
\begin{equation}
 \begin{split}
   f_l &= \int_{\Xi} \Gamma_l \, {\rm d} \xi, \\
   f_d &= \int_{\Xi} \Gamma_d \, {\rm d} \xi,
 \end{split}
 \label{eq:lift-drag-integral}
\end{equation}
\noindent where $\Gamma_l$ and $\Gamma_d$ are the local lift and drag forces, respectively, and $\Xi$ is the perimeter of the airfoil with $\xi$ as the curvilinear coordinate along it~\citep{atzori_aerodynamic_2020}. The local forces can be further decomposed into contributions from friction and pressure forces as follows:

\begin{equation}
 \begin{split}
   \Gamma_l &= \Gamma_{l,f} + \Gamma_{l,p}, \\
   \Gamma_d &= \Gamma_{d,f} + \Gamma_{d,p},
 \end{split}
 \label{eq:lift-drag-decompose}
\end{equation}
\noindent where the subscripts $f$ and $p$ refer to friction and pressure force components, respectively. \Cref{fig:Force_distribute}~(a) and~(b) show the distributions of local lift and drag forces, along with the respective contributions of friction and pressure.

As shown in \cref{fig:Force_distribute-L}, the pressure component $\Gamma_{l,p}$ dominates the lift generation, while the friction component $\Gamma_{l,f}$ is negligible, on the order of $\sim \mathcal{O}(-6)$. Regarding drag (\cref{fig:Force_distribute-D}), the pressure component $\Gamma_{d,p}$ is also dominant, with $\Gamma_{d,f}$ smaller by an order of magnitude—consistent with the relationship between $C_{d,p}$ and $C_{d,f}$.

Notably, as the only case that improves aerodynamic efficiency, \textbf{Case D} shows a clear reduction in $\Gamma_d$, particularly by decreasing $\Gamma_{d,p}$ in the region $x/c = 0.1$ to $0.3$. This behavior is not observed in the profiles of the other control cases.

\begin{figure}[ht]
  \centering
  \begin{subfigure}[b]{0.49\linewidth}
    \centering
    \includegraphics[width=\textwidth]{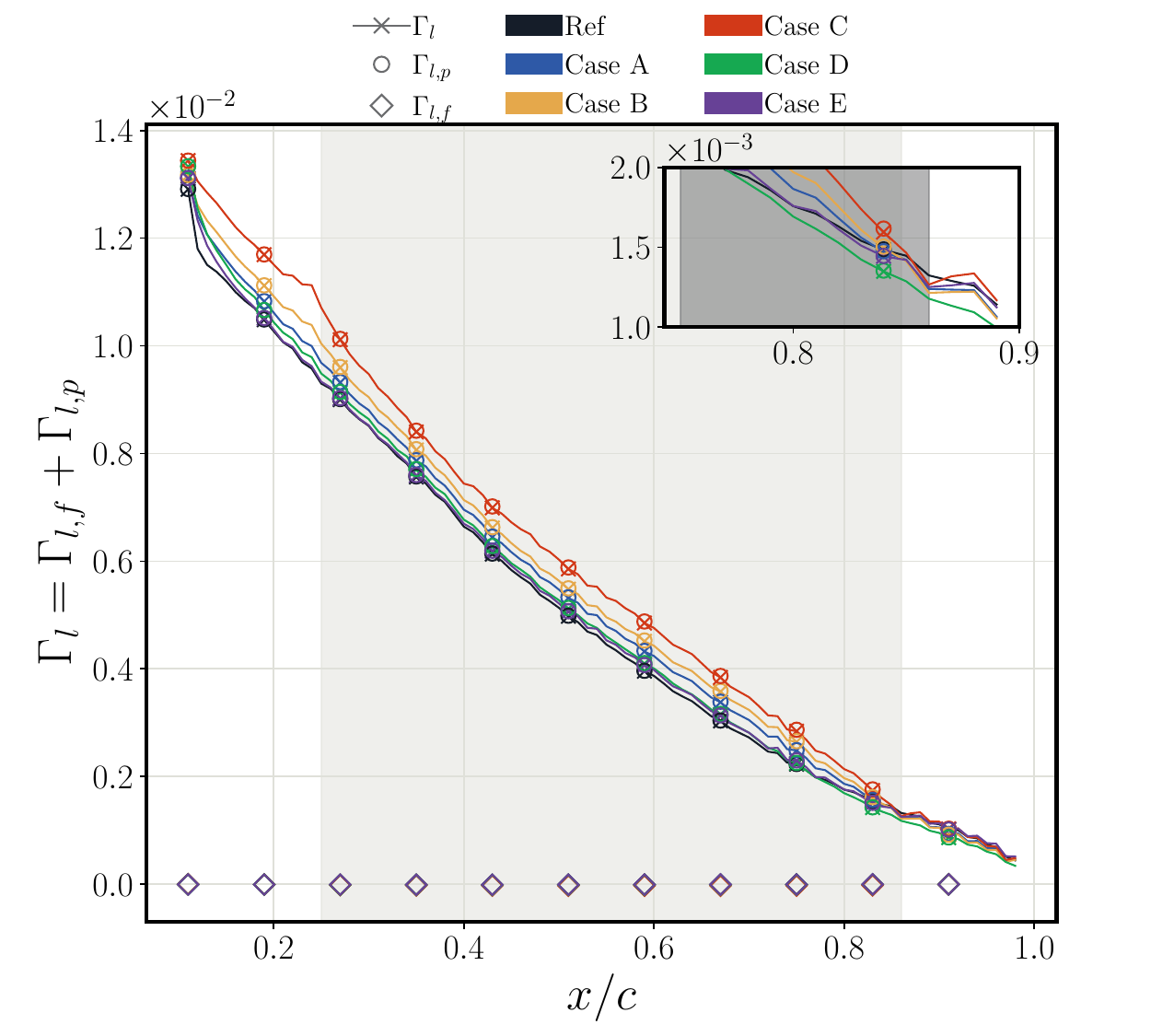}
    \subcaption{Lift-force distributions}
    \label{fig:Force_distribute-L}
  \end{subfigure}
  \hfill
  \begin{subfigure}[b]{0.49\linewidth}
    \centering
    \includegraphics[width=\textwidth]{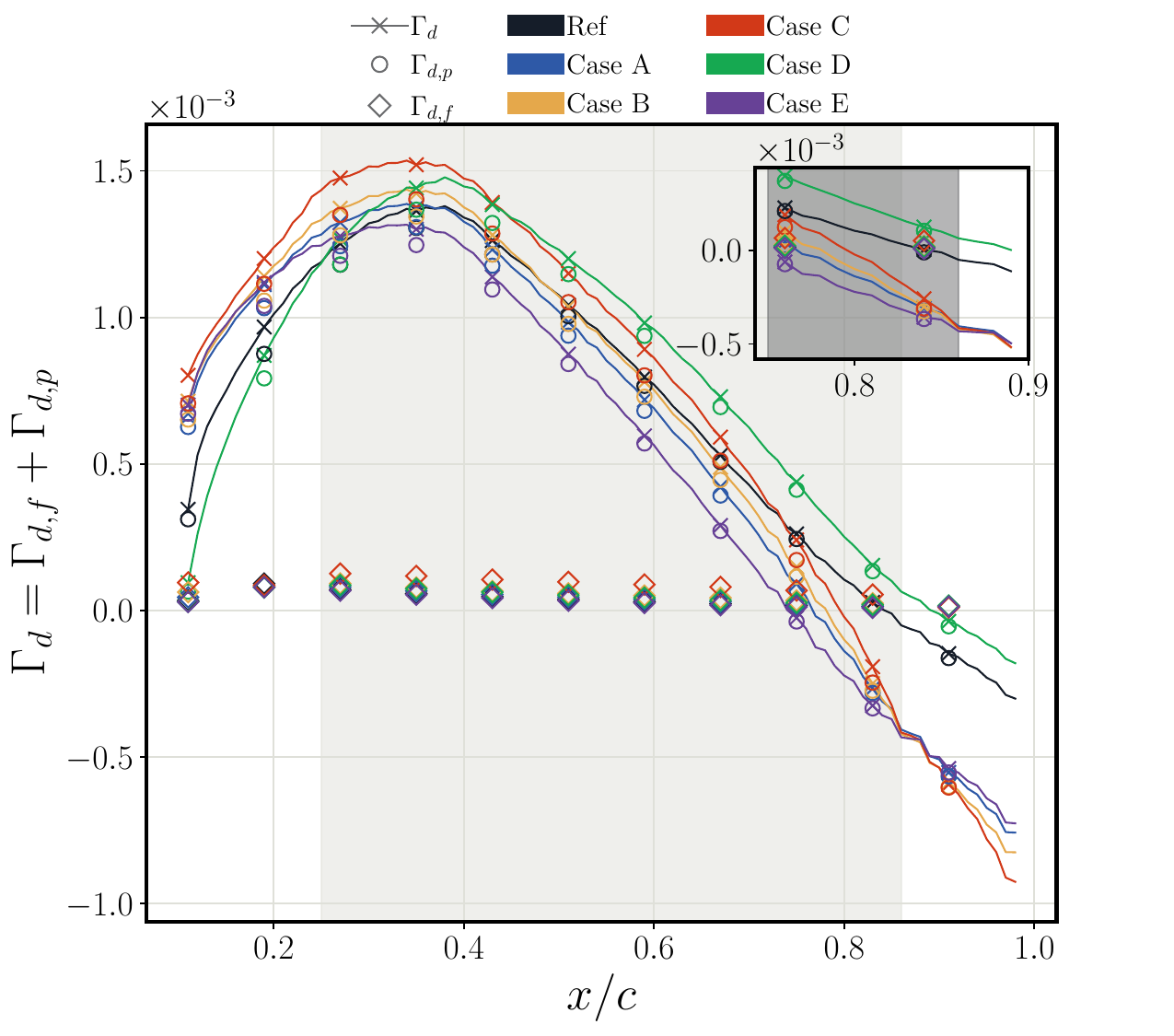}
    \subcaption{Darg-force distributions}
    \label{fig:Force_distribute-D}
  \end{subfigure}
  \caption{{(a) Lift ($\Gamma_l$) and (b) drag ($\Gamma_d$) force and its friction-force and pressure-force components around the airfoil as a function of $x/c$. The solid line with cross marker denotes the local force while the cricle and dimonand markers denote its pressure- and fricition-force components, respectively. 
  Note that the control region is indicated in light and deep gray for panels and inspections, respectively.   
  The color code follows~\cref{tab:ctrl_configs}.}}
  \label{fig:Force_distribute}
\end{figure}
}


\subsection{Streamwise development of the boundary layer}\label{sec:bl_develop}

The distinct modifications in friction and pressure drag achieved by different control configurations motivate an investigation into the interactions between control and turbulent boundary layers (TBLs) over the wing surfaces. 
In this section, we assess the effects of control on integral quantities of TBLs. 
The $99\%$ boundary-layer thickness ($\delta_{99}$) and the mean velocity at the boundary-layer edge ($U_e$) are determined using the method proposed by \citet{vinuesa_determining_2016}.

\begin{figure}[h!]
  \centering
  \begin{subfigure}{0.48\linewidth}
    \centering
    \includegraphics[width=\textwidth]{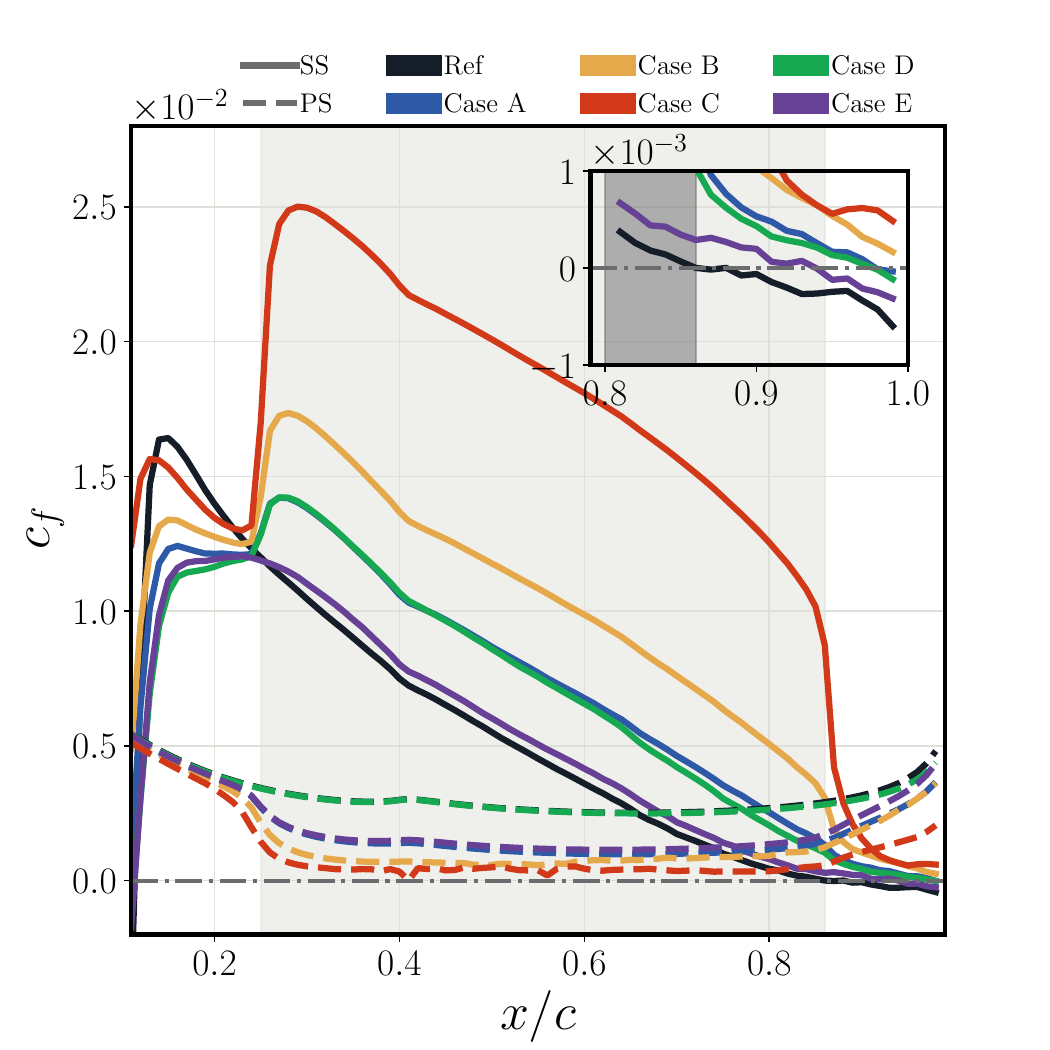}
    \subcaption{
      {
      Skin-friction coefficient ($c_f$) distributions.
      }
      }
    \label{fig:cf}
  \end{subfigure}
  \hfill
  \begin{subfigure}{0.48\linewidth}
    \centering
    \includegraphics[width=\textwidth]{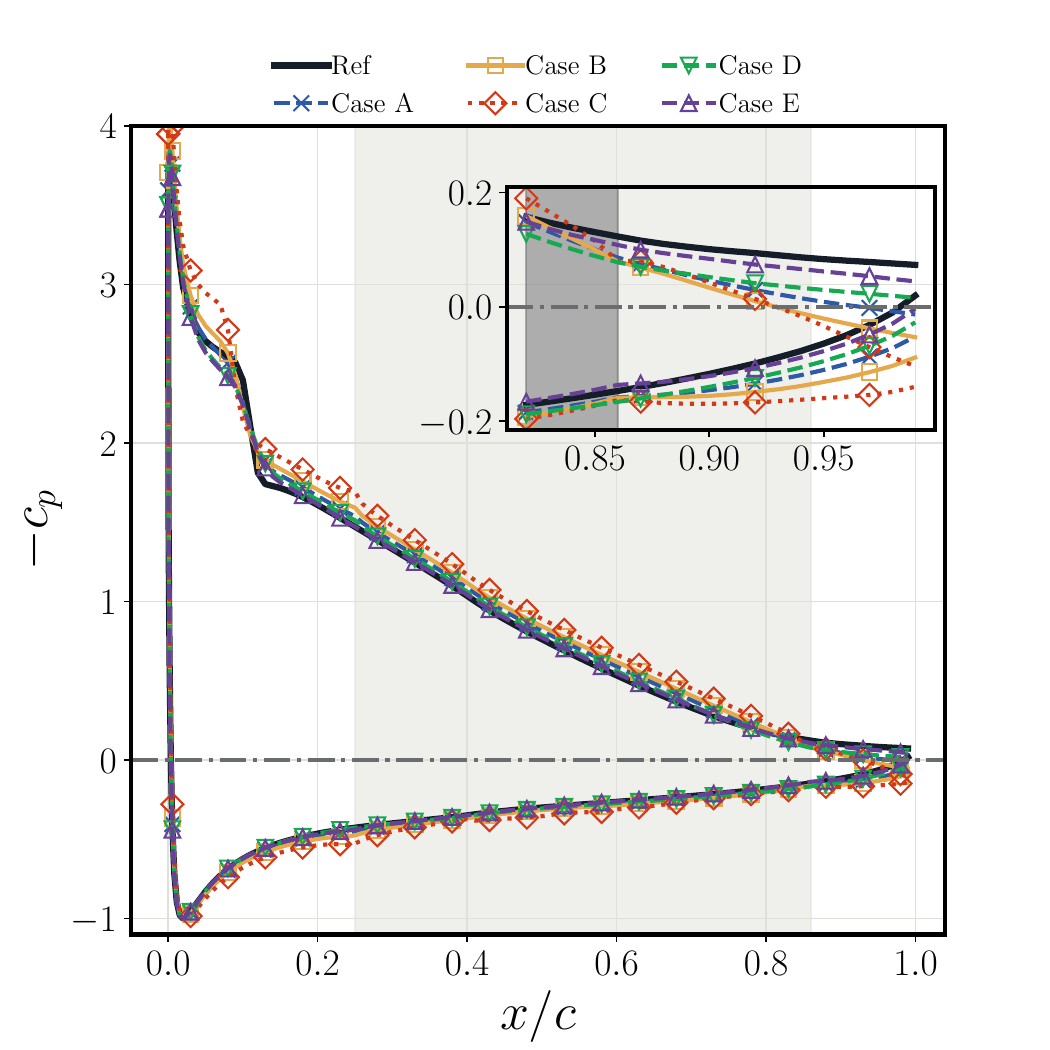}
    \subcaption{
    {  
    Pressure coefficient ($c_p$) distributions.
    }
    }
    \label{fig:cp}
  \end{subfigure}
  
  \caption{
    {
    (a) Skin-friction coefficient ($c_f$) distributions on the suction and pressure side of NACA4412 at $AoA = 11^{\circ}$ and $Re_c = 200,000$, where the solid and dashed lines {denote} the distributions on the suction and pressure side, respectively. The gray dash-dotted line denotes the position where $c_f = 0$. 
  (b) Pressure coefficient ($c_p$) distributions on the suction and pressure side.
  Note that the control region is indicated in light and deep gray for panels and inspections, respectively.   
  The color code follows~\cref{tab:ctrl_configs}.
    }
  }
  \label{fig:cfcp}
\end{figure}

\Cref{fig:cf} depicts the streamwise evolution of the skin-friction coefficient ($c_f = \tau_{w} / (\frac{1}{2} \rho U^2_{e})$, where $\tau_{w}$ is the wall-shear stress) on the suction and pressure sides of the airfoil.
The impact of control on separation delay is evident. 
Uniform suction on the suction side increases $c_f$ by enhancing near-wall momentum, with effects proportional to the input intensity ($\psi$). 
On the contrary, uniform blowing reduces $c_f$ on the pressure side by removing high-momentum flow away from the wall. 
These observations align with the findings of~\citet{kametani_effect_2015}.

Additionally, control influences flow regions upstream of the control area. 
On the suction side, a significant reduction in $c_f$ occurs upstream of the control region, proportional to $\psi$. This effect was not observed at $AoA = 5^{\circ}$~\citep{atzori_aerodynamic_2020}, suggesting distinct 
interactions between control and stronger adverse pressure gradients (APGs) in separated flows. 
Notably, individual applications of suction or blowing yield similar effects on $c_f$ as their combined use.

{\Cref{fig:cp} shows the pressure coefficient ($c_p = (P_w-P_0)/(\frac{1}{2} \rho U^2_{\infty})$, where $P_w$ and $P_0$ are the mean wall pressure and static pressure, respectively) on the suction and pressure sides.}
Control enhances $c_p$ on the pressure side, increasing its value from the stagnation point ($c_p = 1$) to the TE. 
On the suction side, $c_p$ exhibits a downward shift upstream of the control region, followed by an upward shift downstream, mimicking the effects of increased $AoA$~\citep{mallor_bayesian_2024}. 
Note that the Reynolds-number and pressure-gradient effects are disentangled for this wing profile~\citep{pinkerton_NACA4412_1937}, suggesting that the control effects on $c_p$ are consistent across different Reynolds numbers.

\begin{figure}[h!]
    \centering
    \begin{subfigure}[b]{0.42\linewidth}
      \centering
      \includegraphics[width=\textwidth]{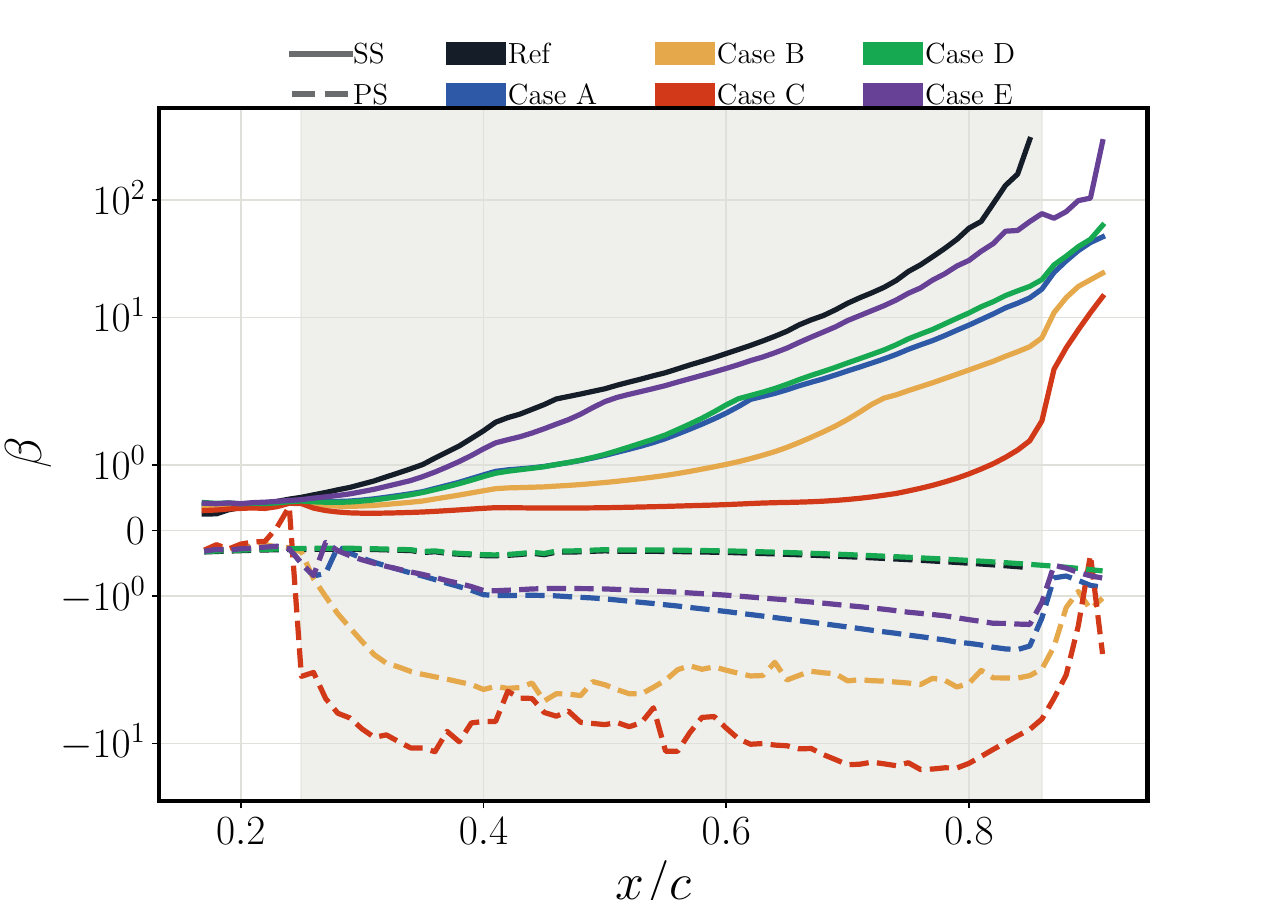}
      \subcaption{{Clauser pressure-gradient parameter ($\beta$).}}
      \label{fig:beta}
    \end{subfigure}
    \hspace{0.1em}
    \begin{subfigure}[b]{0.42\linewidth}
      \centering
      \includegraphics[width=\textwidth]{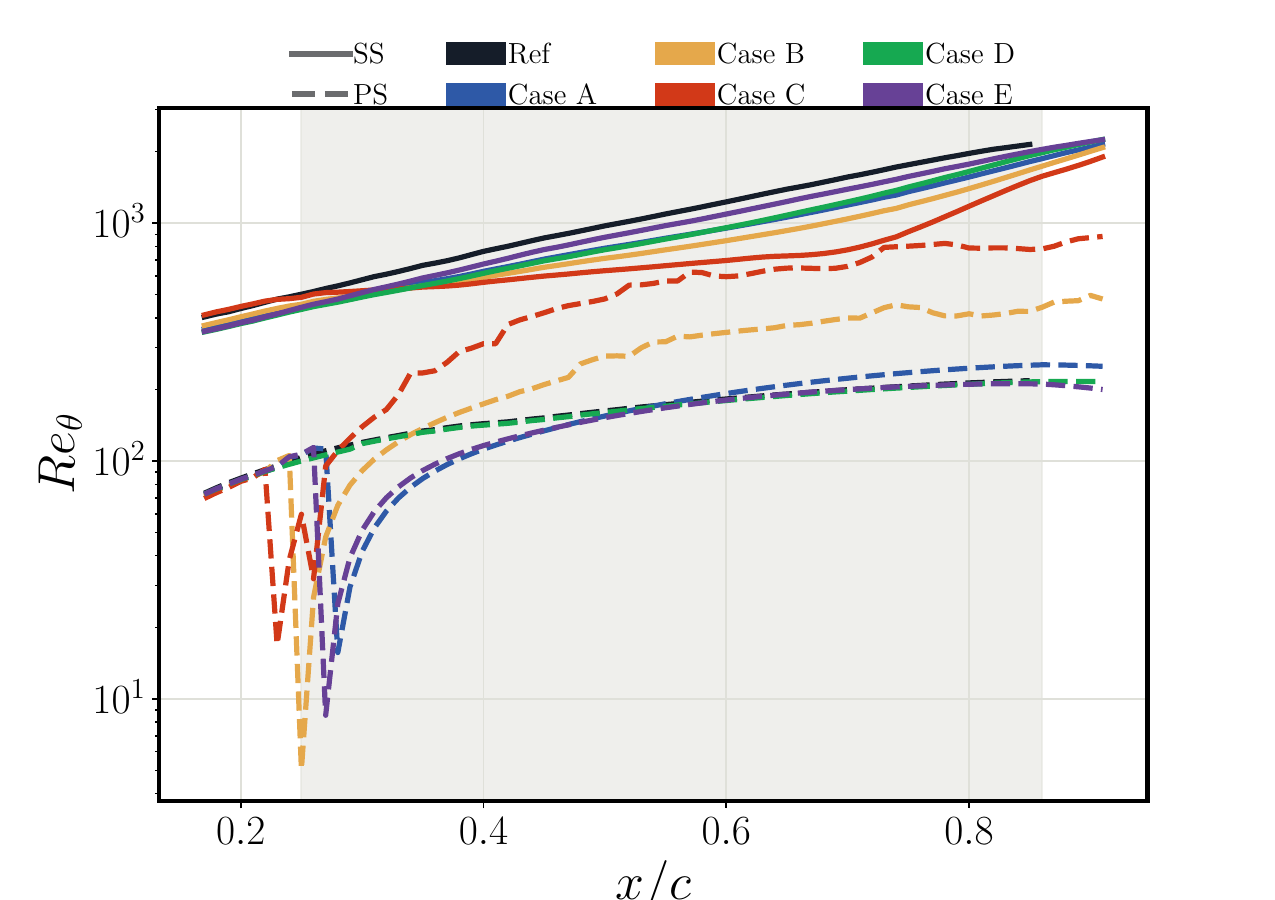}
      \subcaption{
      {Momentum-thickness-based Reynolds number ($Re_{\theta}$).}}
      \label{fig:retheta}
    \end{subfigure}
    \hfill
    \begin{subfigure}[b]{0.42\linewidth}
      \centering
      \includegraphics[width=\textwidth]{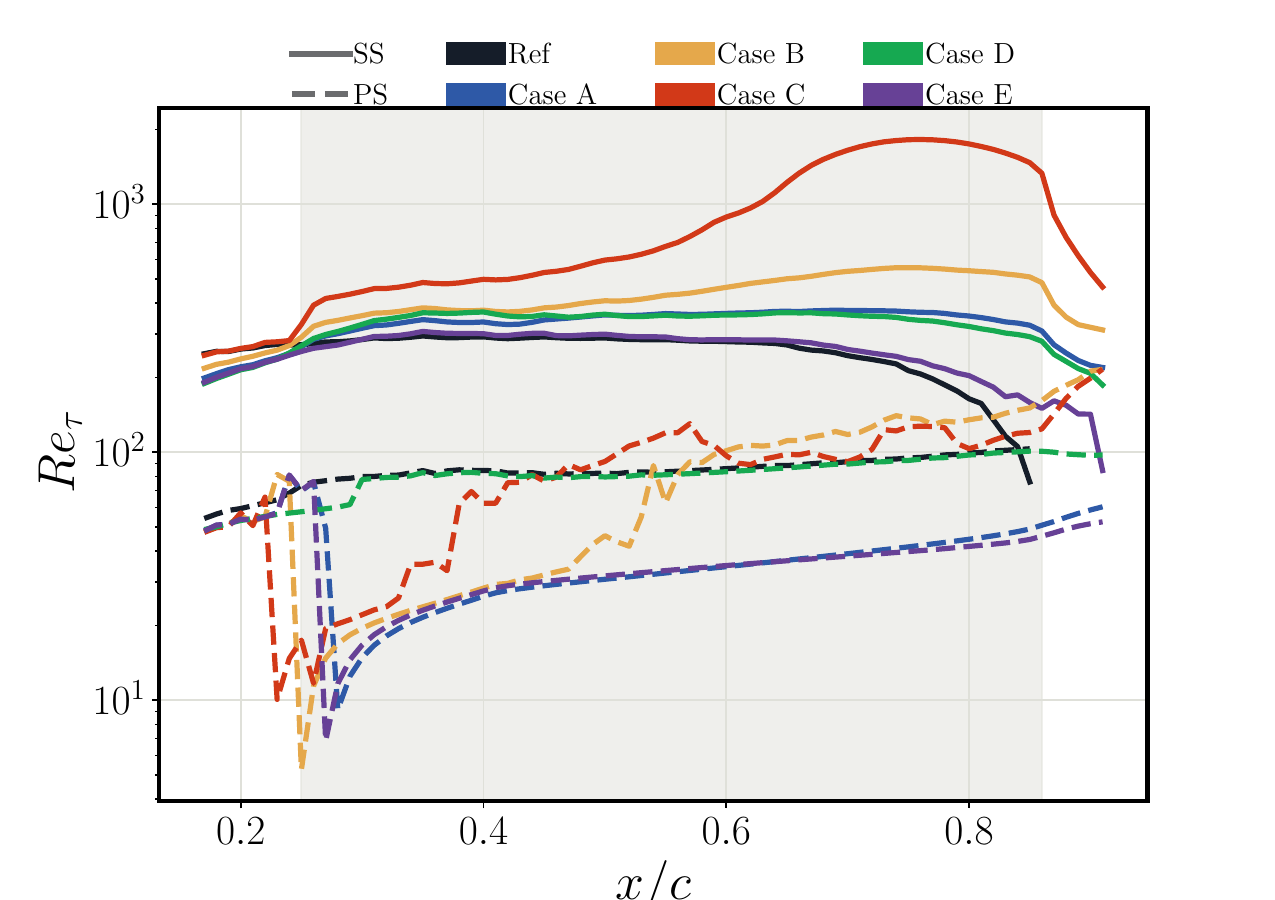}
      \subcaption{{Friction Reynolds number ($Re_{\tau}$).}}
      \label{fig:retau}
    \end{subfigure}
    \hspace{0.1em}
    \begin{subfigure}[b]{0.42\linewidth}
      \centering
      \includegraphics[width=\textwidth]{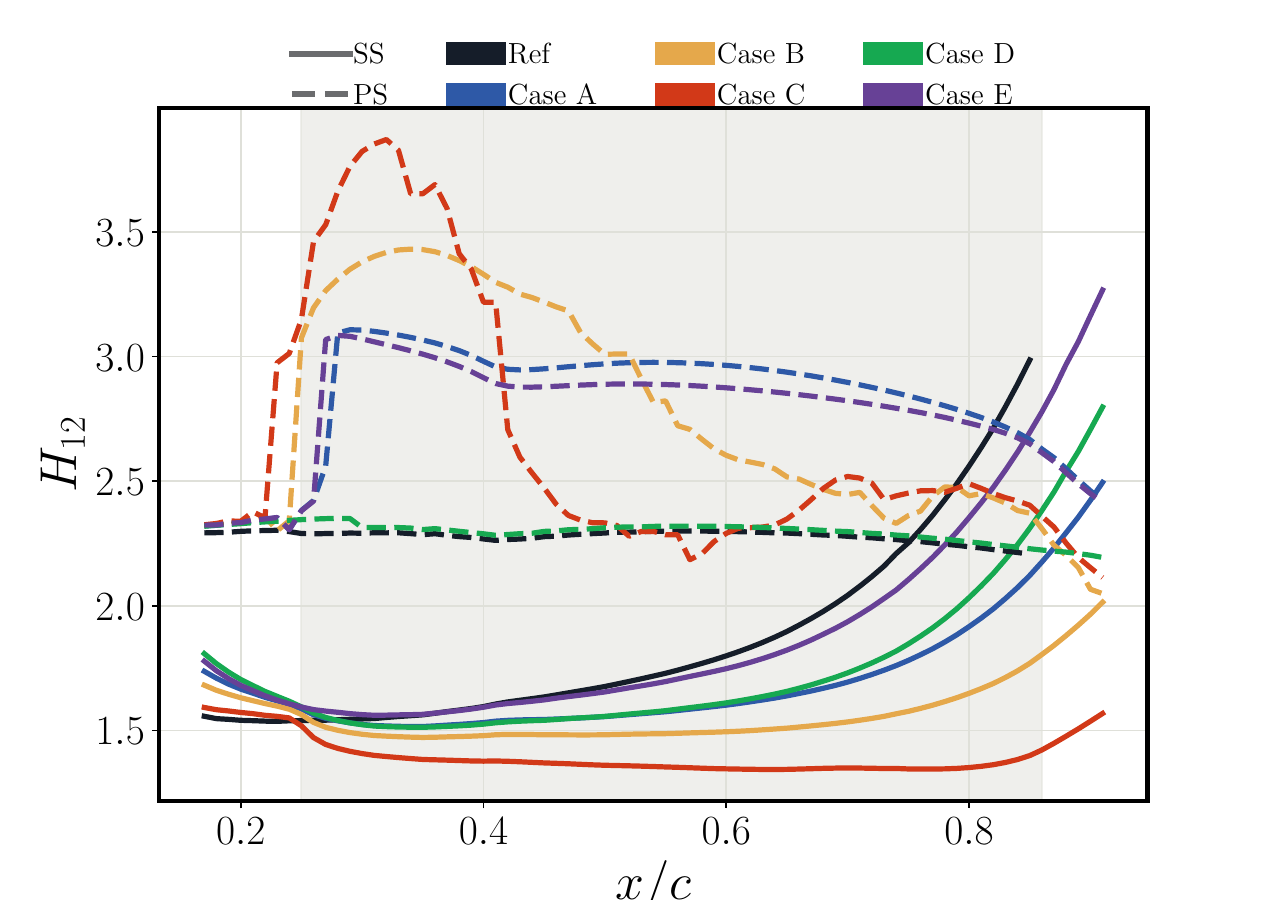}
      \subcaption{{Shape factor ($H_{12}$).}}
      \label{fig:H12}
    \end{subfigure}
    \hfill
    \begin{subfigure}[b]{0.42\linewidth}
      \centering
      \includegraphics[width=\textwidth]{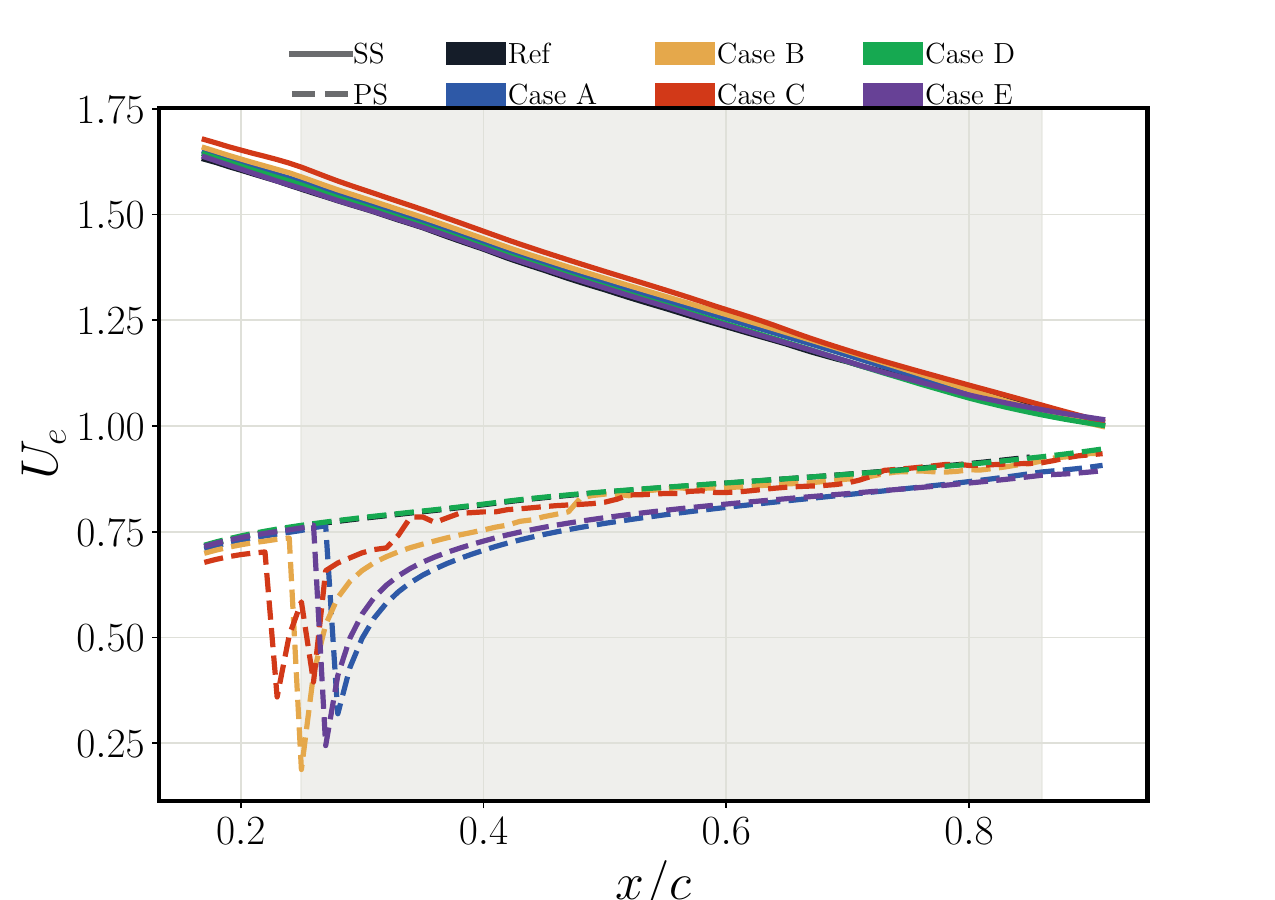}
      \subcaption{{Edge velocity ($U_{e}$).}}
      \label{fig:Ue}
    \end{subfigure}
    \hspace{0.1em}
    \begin{subfigure}[b]{0.42\linewidth}
      \centering
      \includegraphics[width=\textwidth]{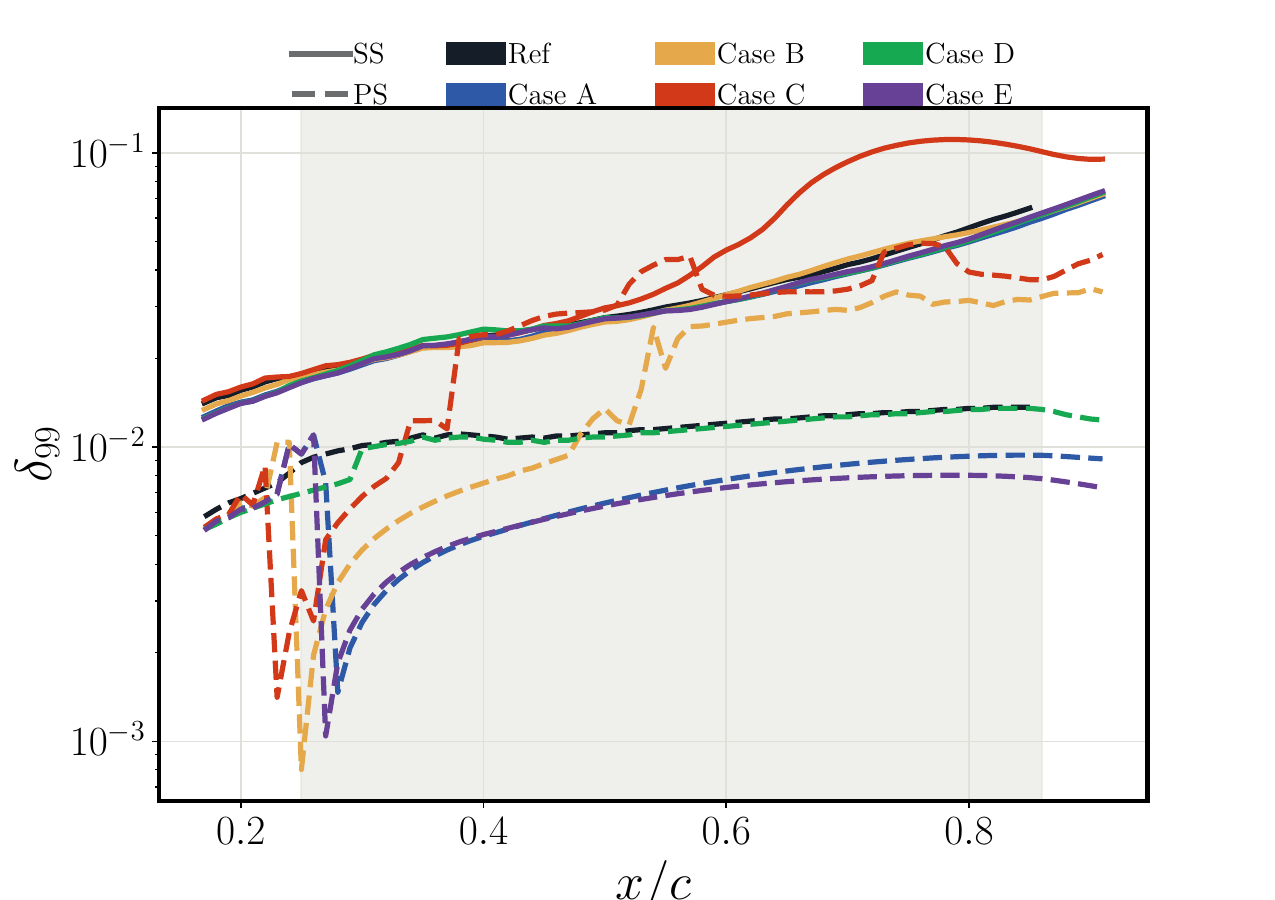}
      \subcaption{{Boundary-layer thickness ($\delta_{99}$).}}
      \label{fig:d99}
    \end{subfigure}

    \caption{
    {  
    (a) Clauser pressure-gradient parameter ($\beta$), (b) momentum-thickness-based Reynolds number ($Re_{\theta}$), (c) friction Reynolds number ($Re_{\tau}$), (d) shape factor ($H_{12}$), (e) Edge velocity ($U_e$) and (f) boundary-layer thickness ($\delta_{99}$) as a function of streamwise location $x/c$ on the suction side (solid lines) and pressure side (dashed lines) of NACA4412 at $AoA = 11^{\circ}$ and $Re_c = 200,000$. 
    Note that the control region is indicated {in} gray.  
    The color code follows~\cref{tab:ctrl_configs}.
    }
    }
    \label{fig:tbl_develop}
\end{figure}

Next, we further assess integral quantities of {TBLs} including the Clauser {pressure-gradient} parameter {$\beta = \delta^*/\tau_{w} {\rm d} P_e/ {\rm d} x_t$} 
(where $\delta^*$ is the displacement thickness and $P_e$ is the pressure at the boundary-layer edge), 
the {momentum-thickness-based} Reynolds number ${Re}_{\theta} = U_e \theta / \nu$ (where $\theta$ is the momentum thickness), 
the friction Reynolds number ${Re}_{\tau} = \delta_{99} u_{\tau} / \nu$ and 
the shape factor $H_{12} = \delta^* / \theta$. 
\Cref{fig:tbl_develop} shows their streamwise distributions on both suction and pressure side of the wing profiles. 
{Additionally, the profiles of local edge velocity ($U_e$) (\cref{fig:Ue}) and $99\%$ boundary-layer thickness ($\delta_{99}$) (\cref{fig:d99}) are provided for the sake of future reference, as these quantities are particularly sensitive to boundary-layer development and control effects.}

At $AoA = 11^{\circ}$, $\beta$ ranges from $-0.1$ to $300$, indicating highly non-equilibrium TBLs (\cref{fig:beta}). 
The uncontrolled case exhibits rapid increases on the suction side and gradual decreases on the pressure side, indicating the strong adverse-pressure-gradient (APG) and flavorable-pressure-gradient (FPG) conditions, respectively. 
Uniform suction over the suction side significantly attenuates $\beta$. 
In particular, the profiles of {Case C} {exhibit} a nearly constant $\beta \approx 0.64$ until $x/c = 0.8$, followed by a sharp increase, which indicates effective TBL stabilization. 
Uniform blowing on the pressure side increases $|\beta|$, intensifying the FPG conditions. 
Moreover, note that Cases B and C exhibit downstream effects on $\beta$, while the profiles of Case A and the reference case collapse downstream.

\Cref{fig:retheta} shows the streamwise {evolution} of $Re_{\theta}$. 
The uncontrolled case exhibits rapid growth on both sides, indicating strong boundary-layer development. 
Uniform suction over {the} suction side decelerates $Re_{\theta}$ growth by deflecting high-momentum flow toward the wall~\citep{kametani_effect_2015,atzori_uniform_2021}, while uniform blowing on the pressure side accelerates $Re_{\theta}$ growth by thickening the boundary layer.
Note that the spikes at the {beginning} of control area is due to the modification of the boundary-condition type.

\Cref{fig:retau} depicts the distributions of $Re_{\tau}$, directly linked to the variation of friction velocity ($u_{\tau}$). 
The uncontrolled case shows {a rapid decrease of $Re_{\tau}$} on the suction side due to the intensified APGs downstream.
Uniform suction mitigates this decline by increasing $u_{\tau}$, proportional to $\psi$, which is also demonstrated in~\cref{fig:cfcp} (a).
The effect of blowing over {the} pressure side depends on the input intensity: for Cases B and C, profiles start deviating 
above the uncontrolled case at $x/c = 0.45$ and $0.56$, respectively, due to 
increased $\delta_{99}$ and APG effects~\citep{harun_PGEffect_2013,pozuelo_adverse_2022}. 
In contrast, Case A consistently suppresses $Re_{\tau}$ within the control area.

\Cref{fig:H12} depicts the streamwise development of shape factor $H_{12}$. 
Higher $H_{12}$ values indicate more laminar-like flow, with $H_{12} \geq 2$ 
characteristic of a Blasius boundary layer. 
Suction significantly reduces $H_{12}$, preventing re-laminarization by adding 
near-wall momentum, whereas blowing increases $H_{12}$, promoting re-laminarization. 

Note that the profiles of Cases D and E closely match those of the reference case and Case A, respectively, indicating that suction over the suction side has minimal influence on the pressure side and vice versa. 
This observation aligns with {the conclusions} from \citet{atzori_aerodynamic_2020,fahland_drag_2023}.


\subsection{Wall-normal profiles of turbulence statistics}\label{sec:turb_stats}
The analysis of integral quantities has highlighted the interactions between control and TBLs in terms of streamwise development. 
In this section, we examine the inner- and outer-scaled profiles of turbulence statistics, focusing 
on mean velocity components and Reynolds stresses.

The assessment is conducted at $x/c = 0.75$, within the control area. 
This location is chosen for two reasons: (1) pressure-gradient effects are highly pronounced at this location, and (2) many previous studies~\citep{vinuesa_turbulent_2018,atzori_uniform_2021,fahland_drag_2023,mallor_IJHFF_2024,wang_opposition_2024} have documented profiles at this location, facilitating future comparisons.
\Cref{tab:tbl-val} summarizes the integral quantities obtained at $x/c = 0.75$ for all cases. 
The aforementioned control effects of boundary-layer developments are clearly quantified, leading to significant modifications in turbulence statistics.

\begin{table}[ht]
  \centering
  \def~{\hphantom{0}}
  \resizebox{\textwidth}{!}{
  \begin{tabular}{ccccccccccc}
  \textbf{Side} & \textbf{Case} & \textbf{$u_{\tau}$} &\textbf{$U_e$} &\textbf{$\delta_{99}$} & \textbf{$c_f$} & \textbf{$c_p$} & \textbf{$\beta$} & \textbf{$Re_{\theta}$} & \textbf{$Re_{\tau}$} & \textbf{$H_{12}$} \\ [3pt] 
  \multirow{6}{*}{SS} & Ref & 0.021 & 1.11 & 0.059 & 0.0009 & -0.23 & 27.97 & 1802.0 & 206.9 & 2.31 \\ 
                    & Case A & 0.041 & 1.11 & 0.045 & 0.0033 & -0.26 & 5.85 & 1395.3 & 366.9 & 1.80 \\ 
                    & Case B & 0.056 & 1.12 & 0.050 & 0.0061 & -0.28 & 2.58 & 1230.9 & 554.5 & 1.58 \\ 
                    & Case C & 0.083 & 1.13 & 0.120 & 0.0139 & -0.32 & 0.64 & 964.8  & 1825.5 & 1.35 \\ 
                    & Case D & 0.038 & 1.10 & 0.055 & 0.0029 & -0.23 & 7.21 & 1462.3  & 340.4 & 1.88 \\ 
                    & Case E & 0.026 & 1.10 & 0.056 & 0.0013 & -0.23 & 17.80 & 1611.3  & 232.9 & 2.15 \\ [2pt]
   \multirow{6}{*}{PS} & Ref    & 0.036 & 0.90 & 0.013 & 0.0026 & 0.20 & -0.46 & 208.1 & 95.3 & 2.26 \\ 
                        & Case A & 0.023 & 0.86 & 0.009 & 0.0011 & 0.21 & -1.62 & 237.8 & 42.1 & 2.83 \\ 
                        & Case B & 0.021 & 0.89 & 0.033 & 0.0009 & 0.22 & -3.15 & 442.5 & 136.3 & 2.38 \\ 
                        & Case C & 0.013 & 0.90 & 0.049 & 0.0003 & 0.22 & -16.63 & 806.5 & 127.5 & 2.46 \\ 
                        & Case D & 0.036 & 0.90 & 0.013 & 0.0026 & 0.21 & -0.42 & 205.4  & 93.8 & 2.28 \\ 
                        & Case E & 0.025 & 0.85 & 0.008 & 0.0013 & 0.19 & -1.27 & 207.2  & 40.1 & 2.78 \\ [2pt]
   
  \end{tabular}
  }
  \caption{{Integral quantities of streamwise development of turbulent boundary layer over suction (SS) and pressure side (PS) obtained at $x/c = 0.75$.}}
  \label{tab:tbl-val}
\end{table}

\begin{figure}[h!]
  \centering
  \begin{subfigure}{0.42\linewidth}
    \centering
    \includegraphics[width=\textwidth]{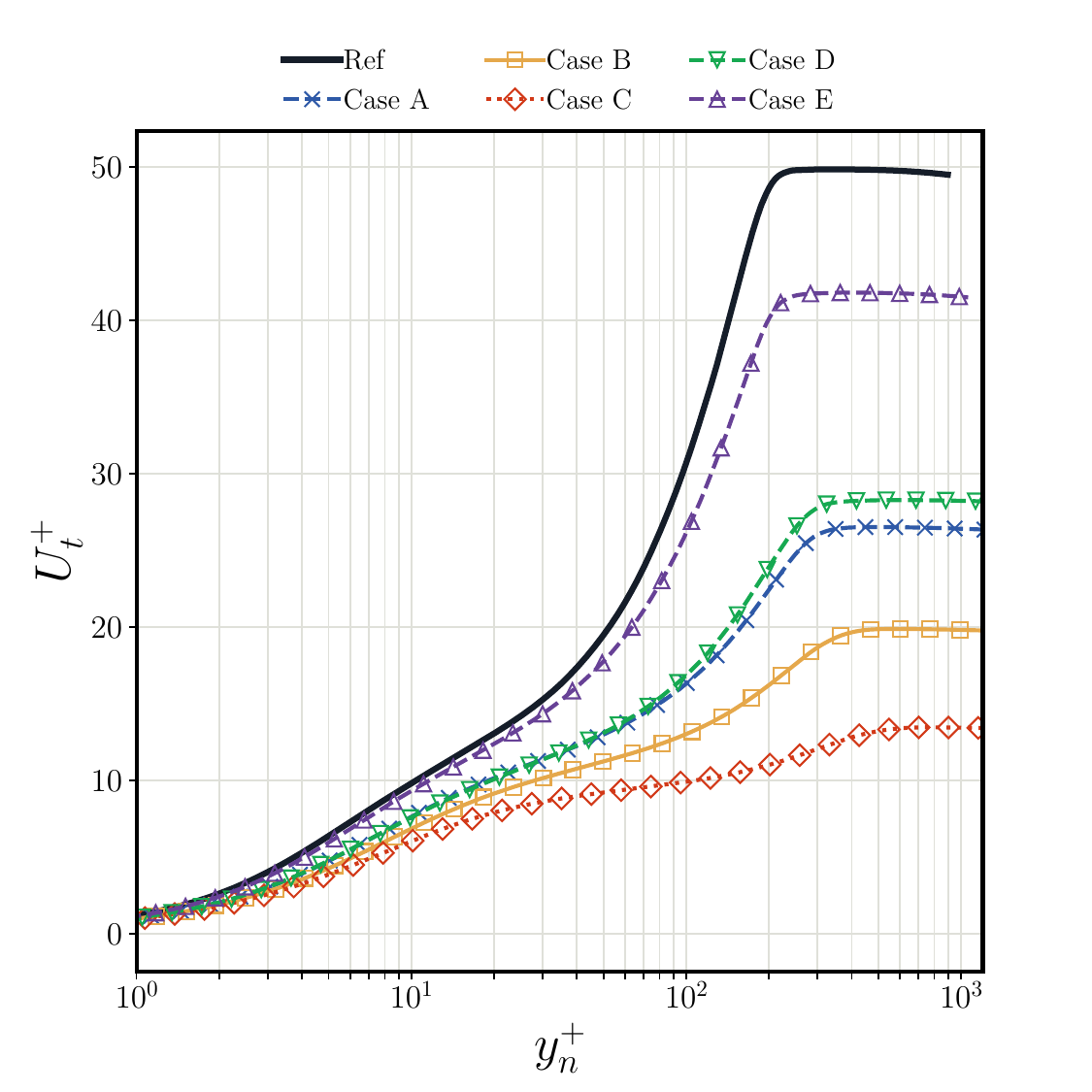}
    \subcaption{{Inner-scaled mean components of wall-tangential velocity on the SS.}}
    \label{fig:U-inner-SS}
  \end{subfigure}
  \hspace{0.5em}
  \begin{subfigure}{0.42\linewidth}
    \centering
    \includegraphics[width=\textwidth]{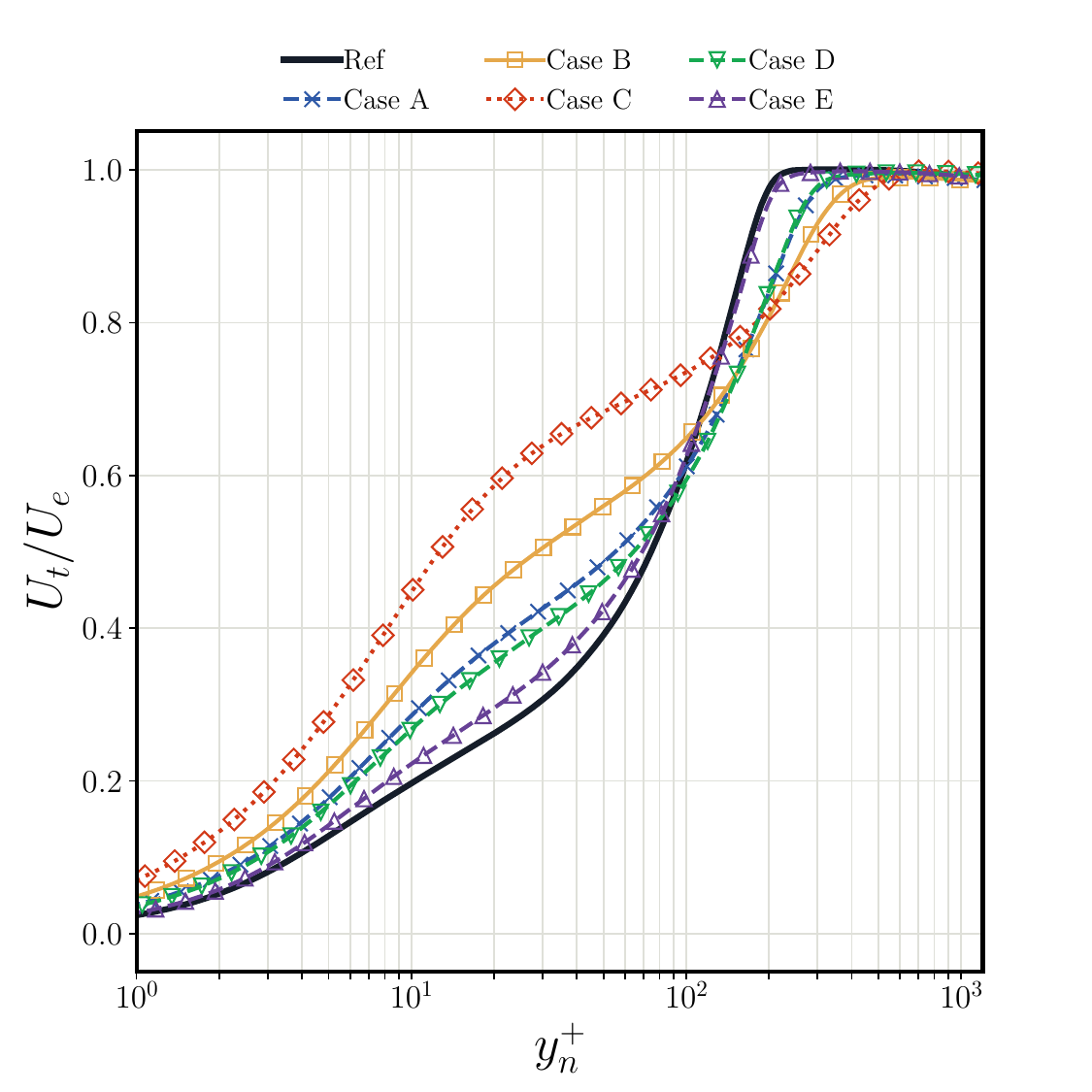}
    \subcaption{{Outer-scaled mean components of wall-tangential velocity on the SS.}}
    \label{fig:U-outer-SS}
  \end{subfigure}
  
  \vspace{0.5em}

  \begin{subfigure}{0.42\linewidth}
    \centering
    \includegraphics[width=\textwidth]{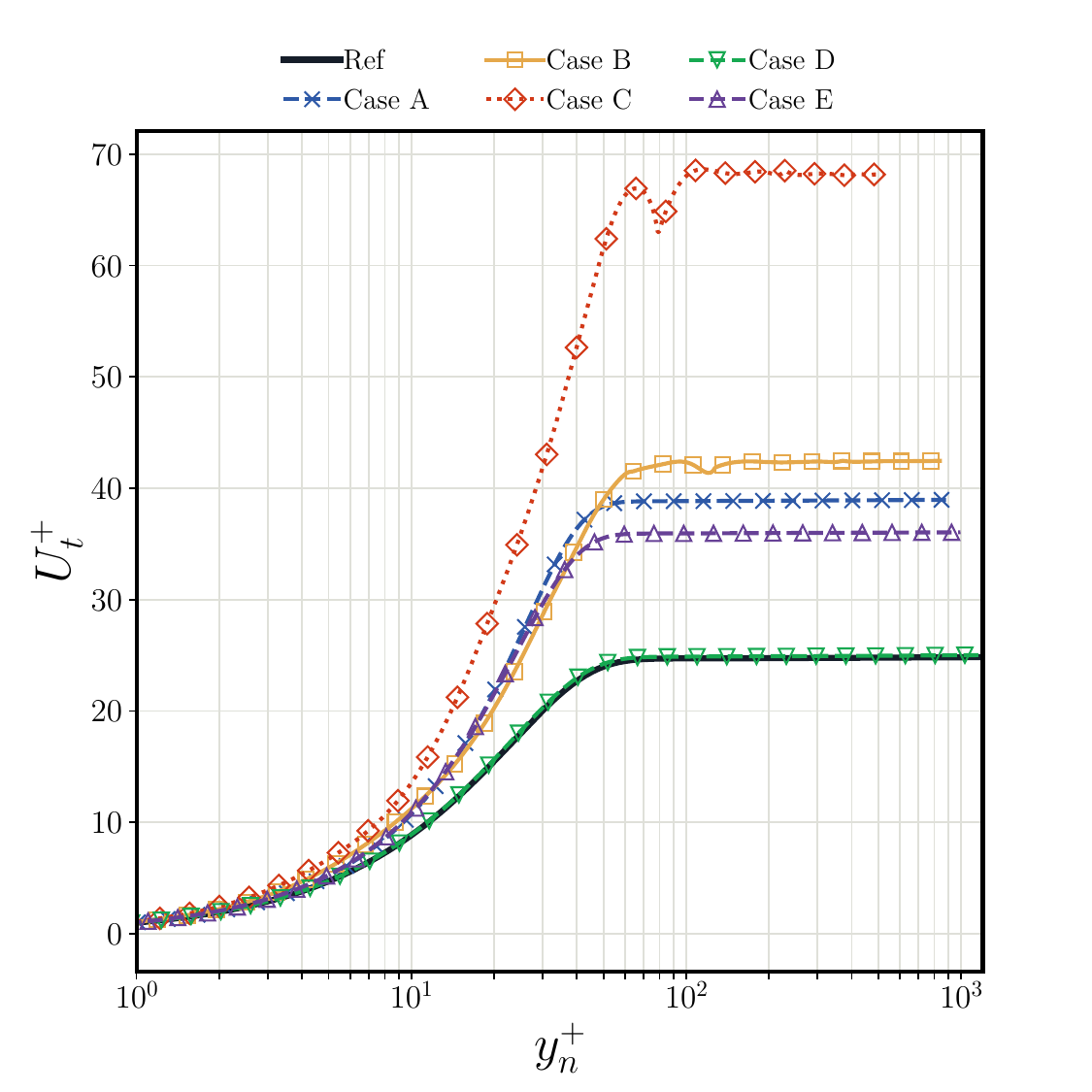}
    \subcaption{{Inner-scaled mean components of wall-tangential velocity on the PS.}}
    \label{fig:U-inner-PS}
  \end{subfigure}
  \hspace{0.5em}
  \begin{subfigure}{0.42\linewidth}
    \centering
    \includegraphics[width=\textwidth]{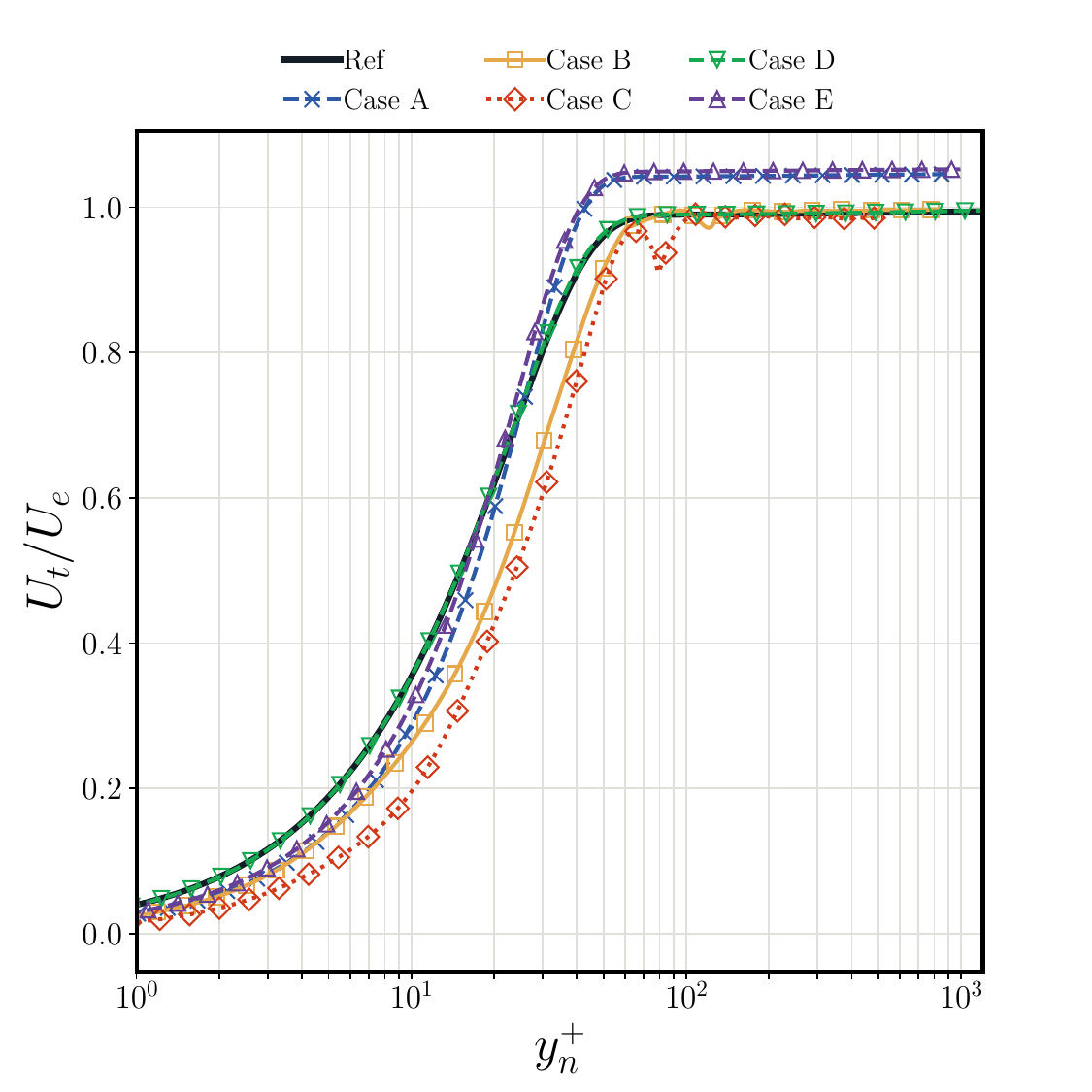}
    \subcaption{{Outer-scaled mean components of wall-tangential velocity on the PS.}}
    \label{fig:U-outer-PS}
  \end{subfigure}
  \caption{ 
  {(a, c) Inner- and (b, d) outer-scaled mean components of wall-tangential velocity ($U_t$) as a function of inner-scaled wall-normal distance ($y^+_n$) on the (a, b) suction and (c, d) pressure side at streamwise location of $x/c = 0.75$, respectively. The color code follows~\cref{tab:ctrl_configs}.}}
  \label{fig:U-vel}
\end{figure}

\Cref{fig:U-inner-SS,fig:U-inner-PS} depict the inner-scaled mean wall-tangential velocity on the suction side and pressure side at $x/c = 0.75$, respectively.
Note that the development of log-layer is less pronounced due to the low-Reynolds-number effects~\citep{vinuesa_turbulent_2018}.
Uniform suction significantly modifies the inner-scaled profile on the suction side, attenuating the wake 
region by approximately $50\%$, $40\%$, and $28\%$ for Cases A, B, and C, respectively.
The undershooting of log-layer becomes more evident with increasing control intensity ($\psi$). 
Within the buffer layer, the profiles shift downward toward the wall, with Cases A, B, and C aligning closely.

On the pressure side, uniform blowing has the opposite impact, intensifying $U^+_t$ in the overlap and wake regions. 
Regardless of the deviations in the wake region, the profiles of Case D on the suction side and E on the pressure side align well with that of Case A.
Note that Cases B and C exhibit a slight dip near the potential region, which is caused by excessive control intensity.

The control effects on $U_t$ are further revealed by evaluating the outer-scaled profiles, which are demonstrated in~\cref{fig:U-outer-SS,fig:U-outer-PS}.
The wake regions of all cases collapse, indicating that inner-scaled wake modifications are due to changes in $u_{\tau}$.
The results also suggest that the deviation between profiles of Case D/E and A on the suction/pressure side is caused by the variation of $u_{\tau}$.
The overlap region of controlled profiles overshoot the log-layer, being opposite to the inner-scaled profiles. 
Additionally, within the buffer layer, controlled profiles exhibit higher magnitudes, where the variations are proportional to $\psi$.

\begin{figure}[h!]
  \centering
  \begin{subfigure}{0.42\linewidth}
    \centering
    \includegraphics[width=\textwidth]{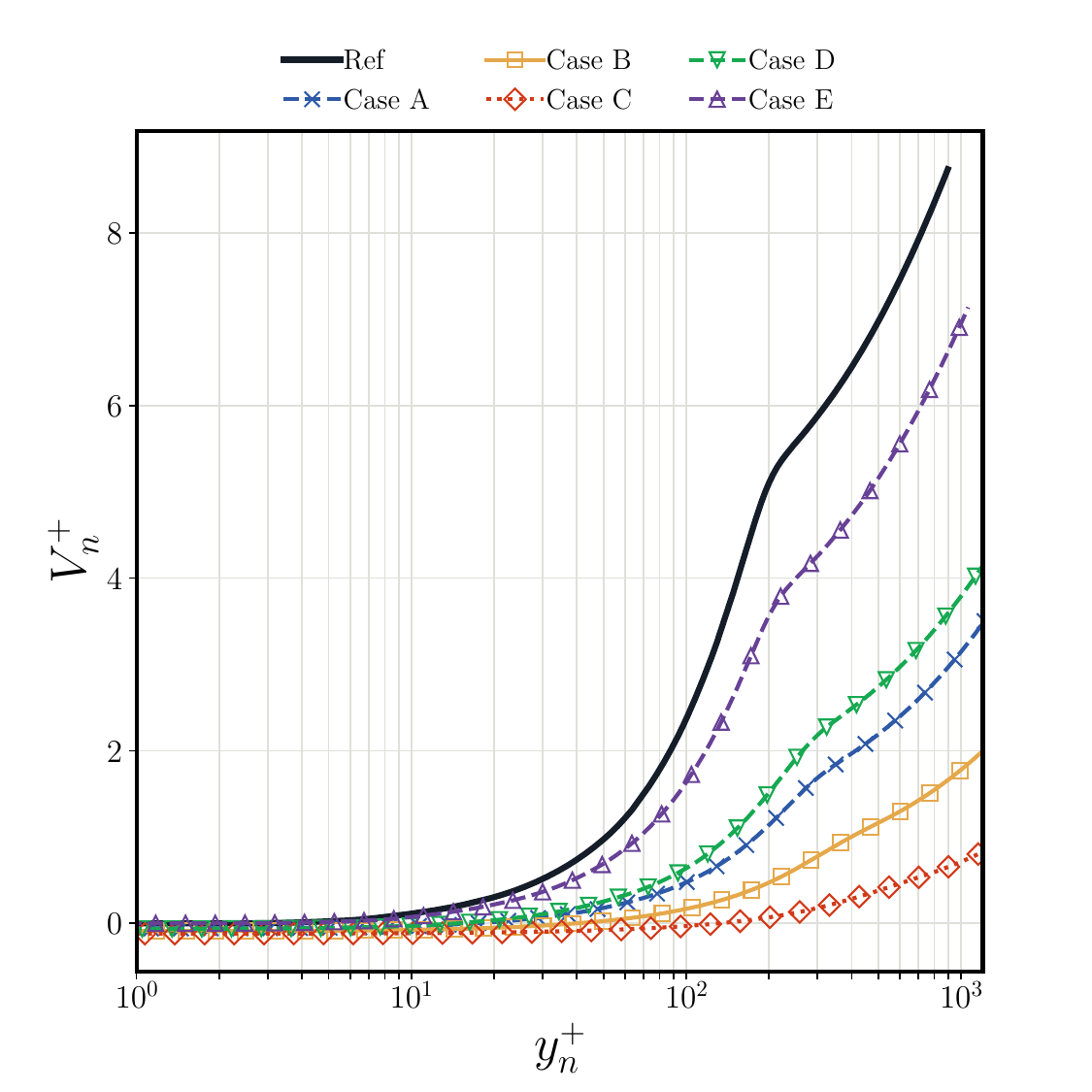}
    \subcaption{{Inner-scaled mean components of wall-normal velocity on the SS.}}
    \label{fig:V-inner-SS}
  \end{subfigure}
  \hspace{0.5em}
  \begin{subfigure}{0.42\linewidth}
    \centering
    \includegraphics[width=\textwidth]{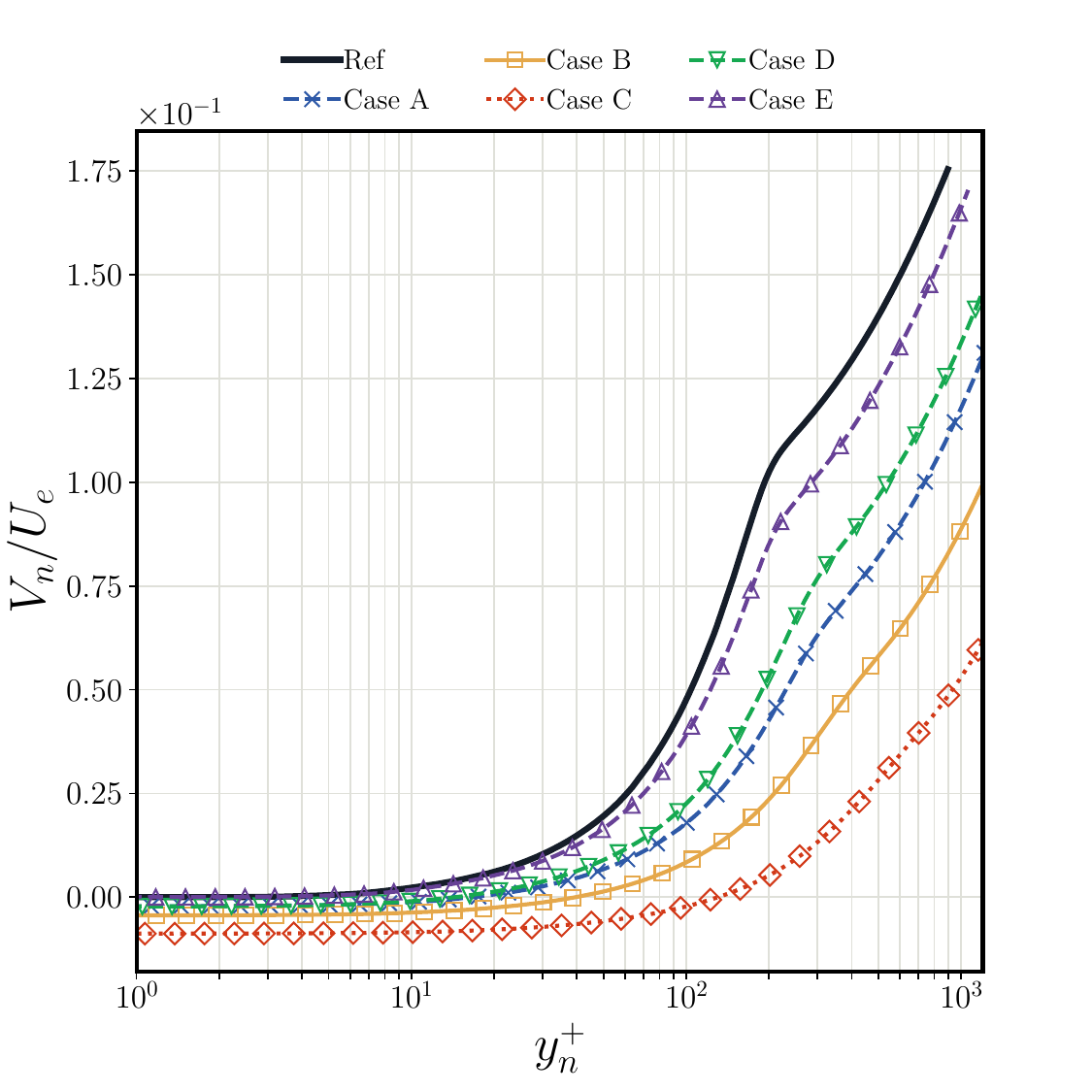}
    \subcaption{{Outer-scaled mean components of wall-normal velocity on the SS.}}
    \label{fig:V-outer-SS}    
  \end{subfigure}
  \vspace{0.5em}
  \begin{subfigure}{0.42\linewidth}
    \centering
    \includegraphics[width=\textwidth]{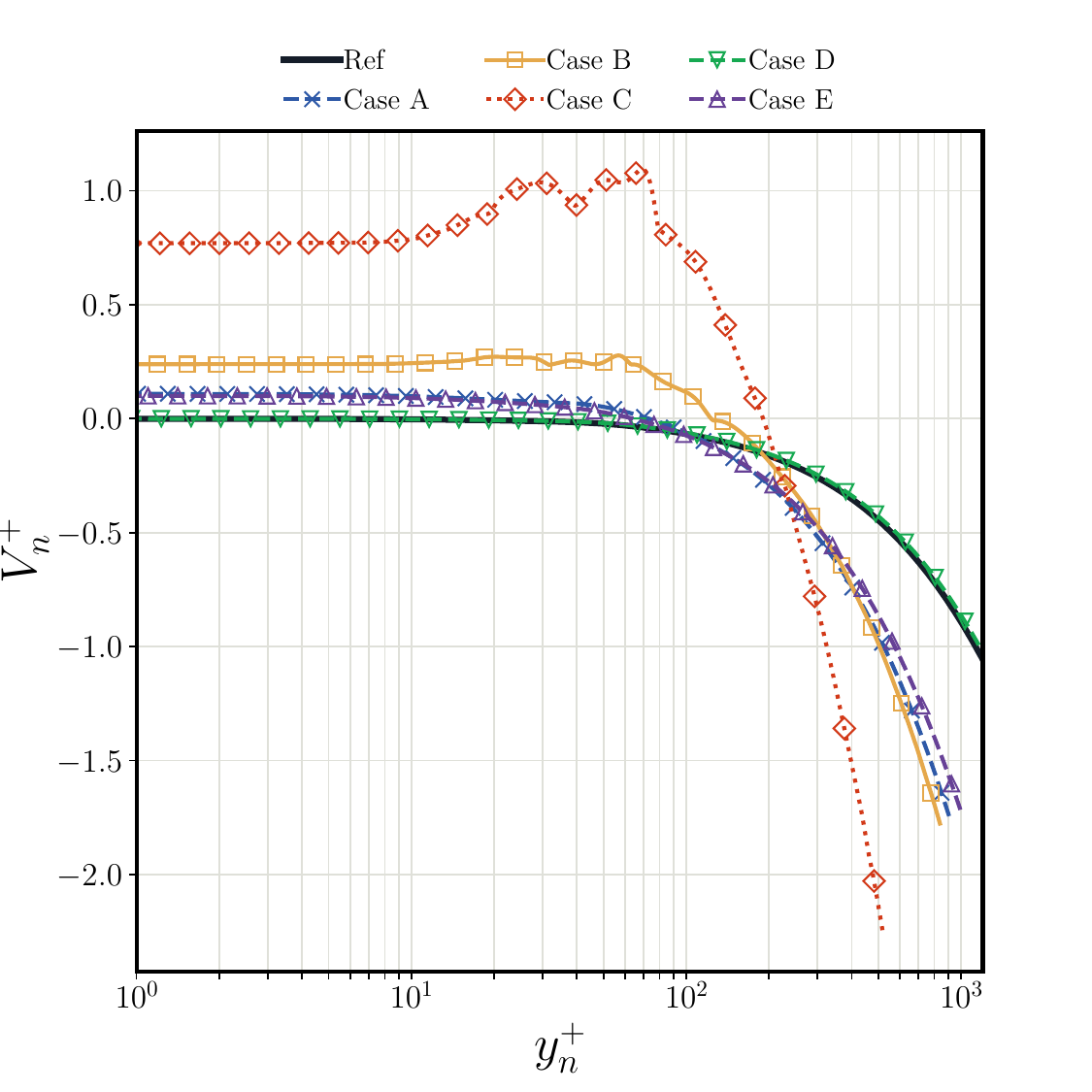}
    \subcaption{{Inner-scaled mean components of wall-normal velocity on the PS.}}
    \label{fig:V-inner-PS}
  \end{subfigure}
  \hspace{0.5em}
  \begin{subfigure}{0.42\linewidth}
    \centering
    \includegraphics[width=\textwidth]{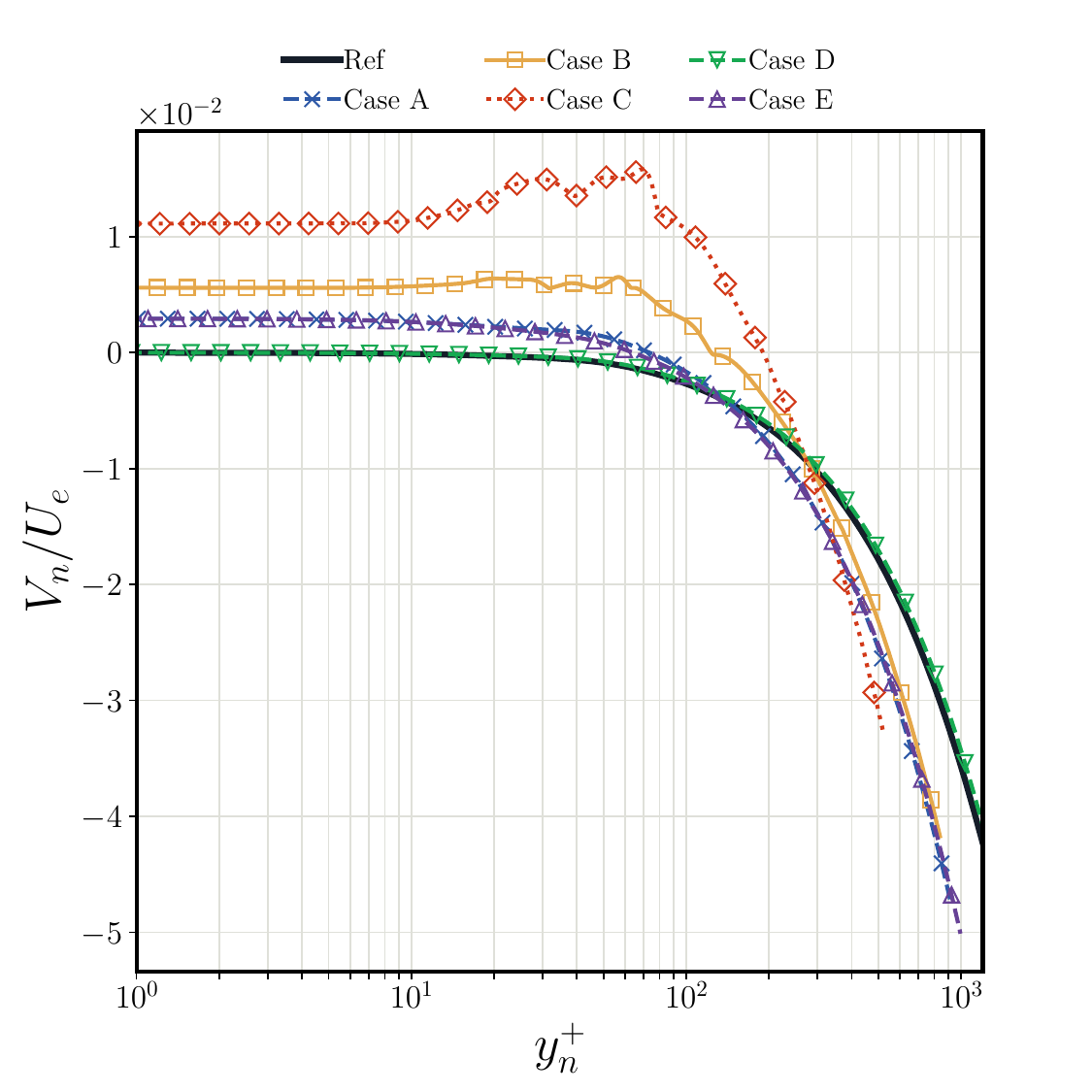}
    \subcaption{{Outer-scaled mean components of wall-normal velocity on the PS.}}
    \label{fig:V-outer-PS}
  \end{subfigure}
  \caption{{(a, c) Inner- and (b, d) outer-scaled mean components of wall-normal velocity ($V_n$) as a function of inner-scaled wall-normal distance ($y^+_n$) on the (a, b) suction and (c, d) pressure side at streamwise location of $x/c = 0.75$, respectively.
  The color code follows~\cref{tab:ctrl_configs}.}}
  \label{fig:V-vel}
\end{figure}

Adverse pressure gradient (APG) and FPG decelerate and accelerate TBL development, respectively, significantly impacting the wall-normal velocity ($V_n$).  
\Cref{fig:V-vel} depicts the inner- and outer-scaled $V_n$ profiles at $x/c = 0.75$. 
On the suction side, uniform suction significantly attenuates $V_n$, with the effect increasing with wall-normal distance and control intensity.
On the pressure side, uniform blowing intensifies $V_n$ by increasing the wall-normal convection, being similar to the effect of strong APGs. 
Interestingly, the profiles of Cases A and B align closely in the wake region, while Case C deviates due to distinct differences in $\beta$ at this location ($-1.62$, $-3.15$, and $-16.63$ for Cases A, B, and C, respectively). 
The profiles of Case D on the suction and Case E on pressure side deviate {with respect to} that of Case A, which is evident for inner- and outer-scaled profiles, being connected to the variation of $\beta$. 
Note that the sink observed in $U_t$ profiles on the pressure side is also present in $V_n$ profiles. 
Moreover, the magnitude of controlled $V_n/U_e$ at the wall matches the imposed boundary conditions described in \S~\ref{sec:config}, validating the implementation. 

\begin{figure}[h!]
  \centering
  \begin{subfigure}{0.42\linewidth}
    \centering
    \includegraphics[width=\textwidth]{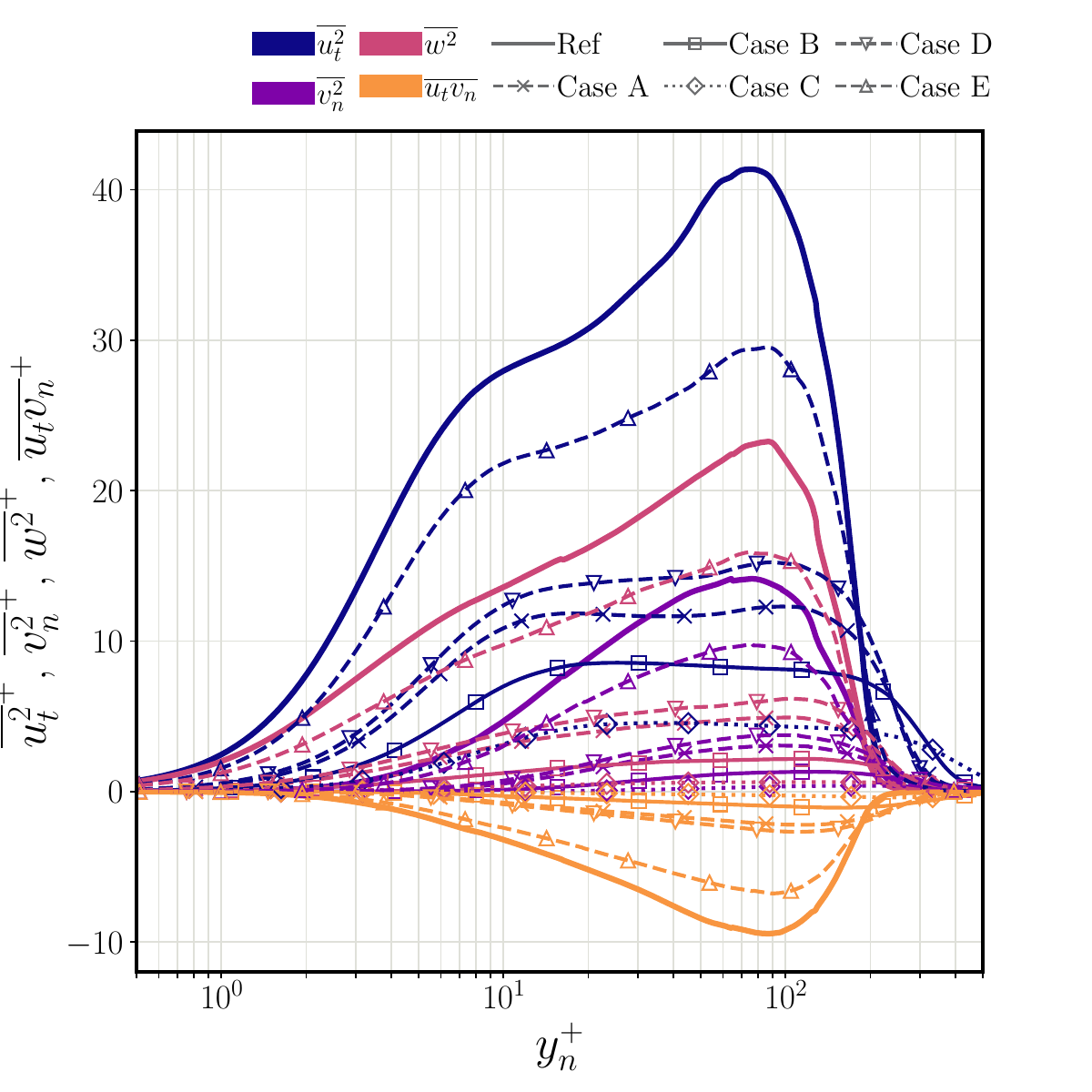}
    \subcaption{{Inner-scaled Reynolds-stress components on the SS.}}
    \label{fig:uiuj-inner-SS}
  \end{subfigure}
  \hspace{0.5em}
  \begin{subfigure}{0.42\linewidth}
    \centering
    \includegraphics[width=\textwidth]{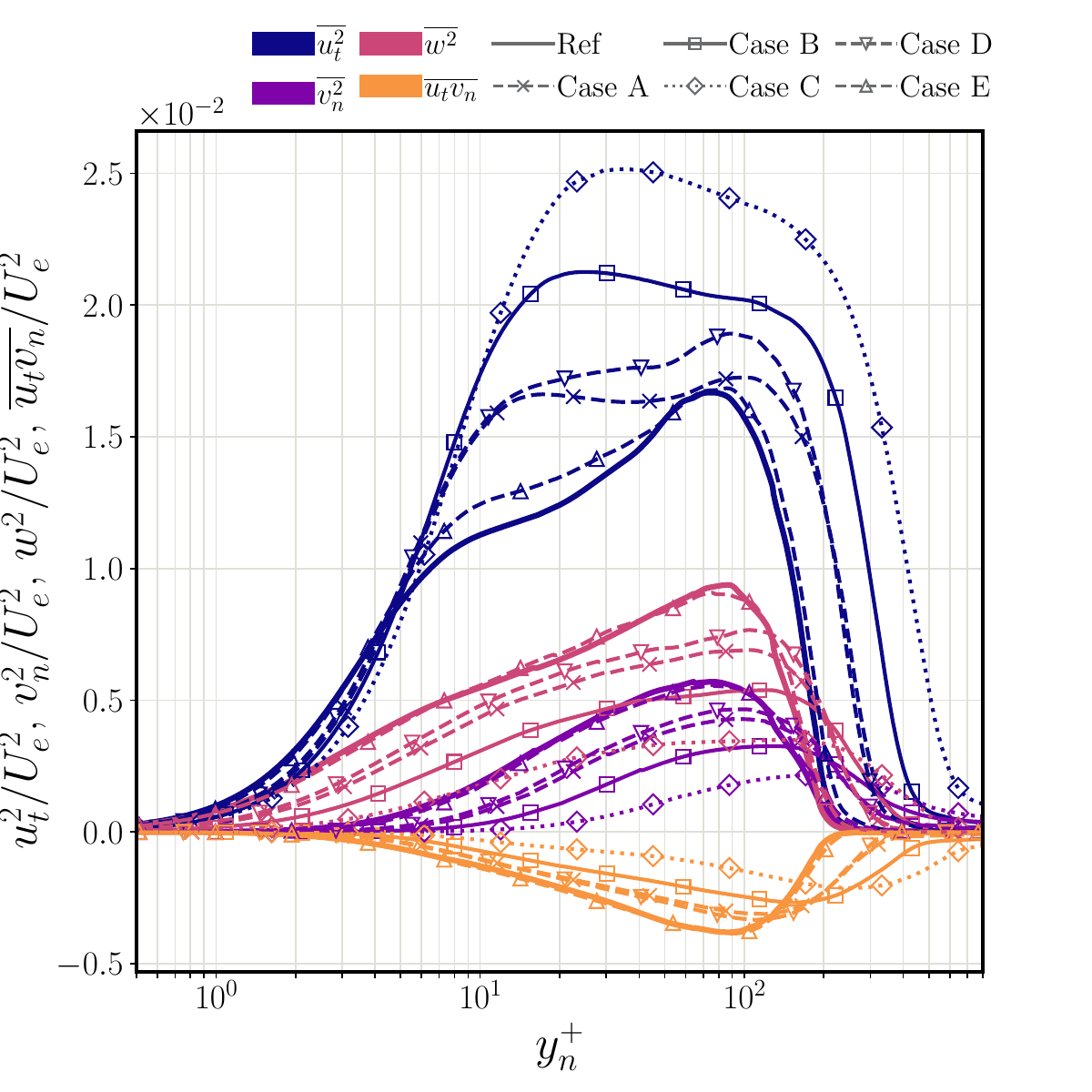}
    \subcaption{{Outer-scaled Reynolds-stress components on the SS.}}
    \label{fig:uiuj-outer-SS}
  \end{subfigure}
  \vspace{0.5em}
  \begin{subfigure}{0.42\linewidth}
    \centering
    \includegraphics[width=\textwidth]{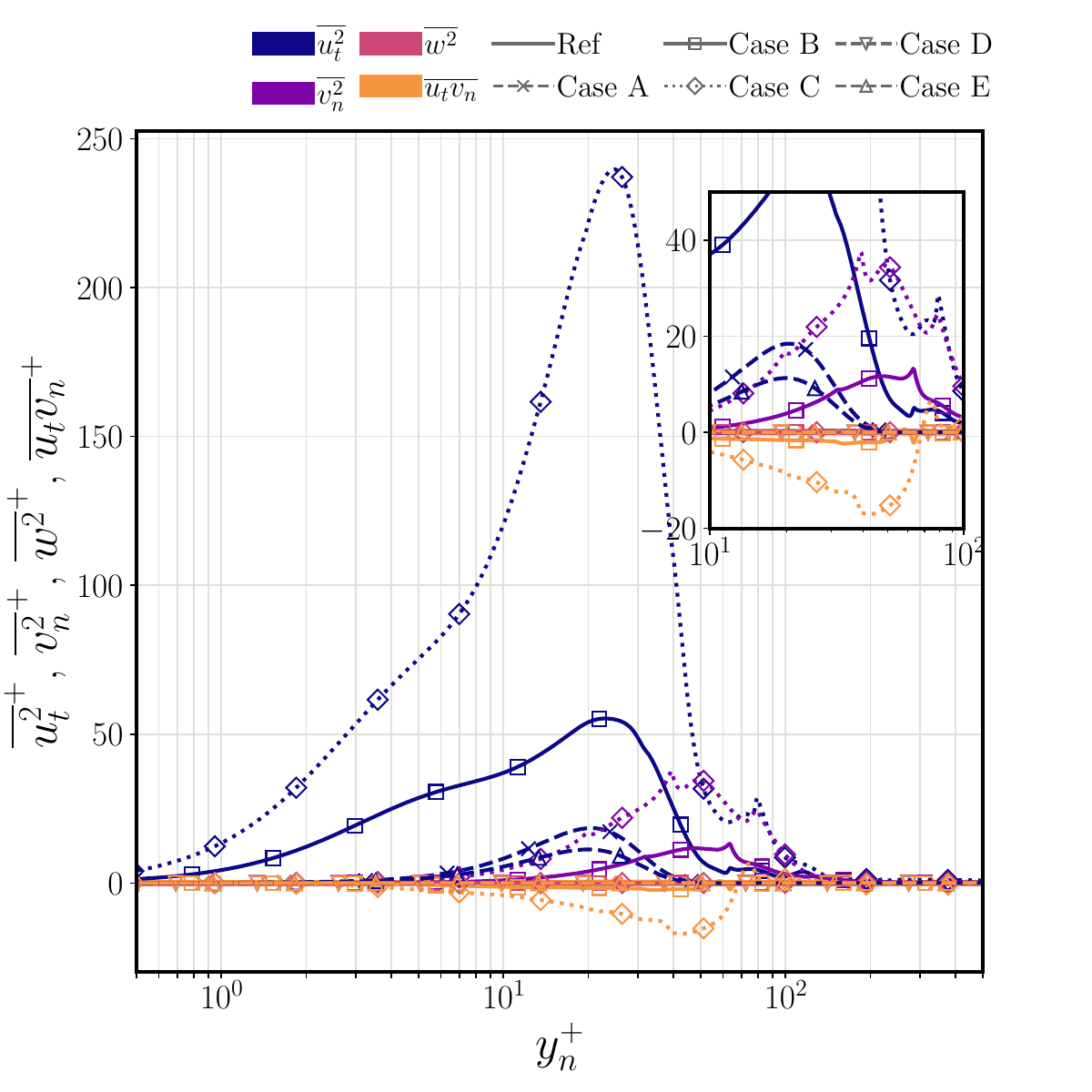}
    \subcaption{{Inner-scaled Reynolds-stress components on the PS.}}
    \label{fig:uiuj-inner-PS}
  \end{subfigure}
  \hspace{0.5em}
  \begin{subfigure}{0.42\linewidth}
    \centering
    \includegraphics[width=\textwidth]{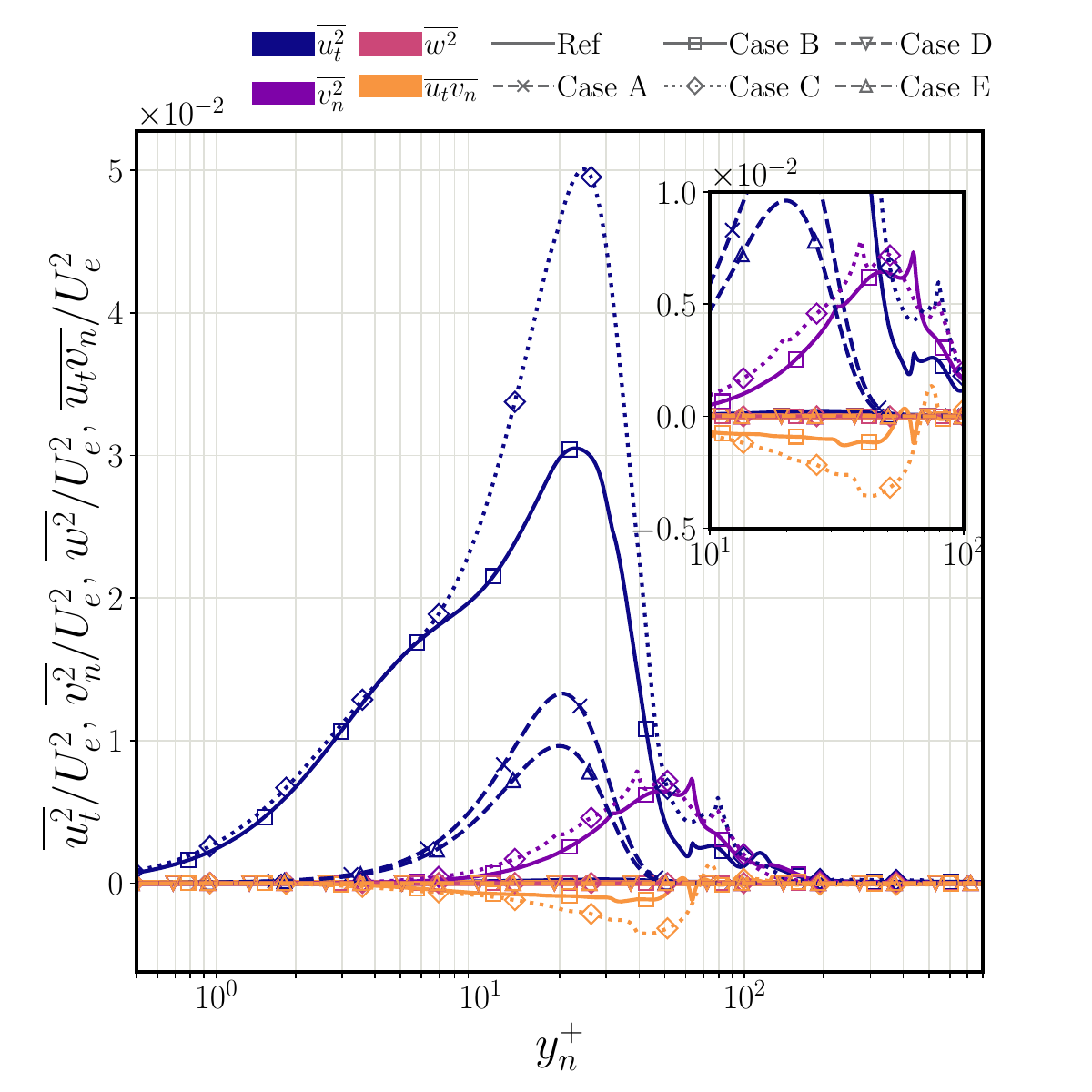}
    \subcaption{{Outer-scaled Reynolds-stress components on the PS.}}
    \label{fig:uiuj-outer-PS}
  \end{subfigure}
    \caption{{((a), (c)) Inner- and ((b), (d)) outer-scaled fluctuation components of wall-tangential ($\overline{u^2_t}$), wall-normal ($\overline{v^2_n}$) and spanwise ($\overline{w^2}$) velocity as well as Reynolds-shear stress ($\overline{u_tv_n}$) as a function of inner-scaled wall-normal distance ($y^+_n$) on the ((a), (b)) suction and ((c), (d)) pressure side at streamwise location of $x/c = 0.75$, respectively.}}
    \label{fig:uiuj-vel}
\end{figure}

Apart from the mean-velocity profiles, we assess the Reynolds stresses to investigate the momentum distribution along the wall-normal direction, which is critical to understand the control effects. 
\Cref{fig:uiuj-inner-SS,fig:uiuj-outer-SS} depict the inner- and outer-scaled Reynolds stress profiles at $x/c = 0.75$ on the suction side, respectively. 
The reference case ($\beta = 27.97$ for the reference case) exhibit an outer peak in the overlap region for the $\overline{u^2_t}$ profiles, being more prominent than the inner peak in the buffer layer, which is pronounced in both inner- and outer-scaled profiles.
Uniform suction significantly attenuates both peaks and intensifies the wake region. 
The modified $\overline{u^2_t}/U^2_e$ profiles exhibit intensified inner and outer peaks, indicating that the effects are primarily due to modifications in $u_{\tau}$, while the prominent wake region of modified $\overline{u^2_t}/U^2_e$ remains pronounced, being evident for inner- and outer-scaled profiles.  

It is worth noting that, at lower $AoA$ ($5^{\circ}$), \citet{atzori_aerodynamic_2020} reported that uniform suction enhances the inner peak while suppressing the outer peak by adopting a similar input intensity $\psi =0.2\%$ with the same control area. 
The major discrepancy is attributed to the larger $\beta$ in the present study ($\beta = 27.97$ vs. $\beta = 6.6$), which is nearly five times higher.  
This suggests that high $\beta$ significantly influences the control performance. 

On the pressure side, $\overline{u^2_t}$ profiles exhibit low magnitudes of $\sim \mathcal{O}(10^{-3})$ due to low Reynolds numbers and strong FPGs. 
Uniform blowing significantly intensifies $\overline{u^2_t}$ with substantially high magnitude, producing a peak in the overlap region, which is pronounced for inner- and outer-scaled profiles. 
Additionally, Cases B and C exhibit a similar buffer-layer response, being consistent with the observations in~\citet{atzori_aerodynamic_2020}.

Moreover, the outer-scaled profiles of {Case E} produces a less prominent outer peak than that of {Case A} on the pressure side, which is connected to the variation of $u_{\tau}$.
On the contrary, the {Case D} produces a more prominent outer peak than that of {Case A} on the suction side, while the inner peaks of both profiles are collapsed. 
At this location, Case D achieves a lower $\beta$ than Case A on the suction side.
Therefore, the results implies that control effectiveness depends on flow history and $\beta$ variations~\citep{atzori_uniform_2021,pozuelo_adverse_2022,wang_opposition_2024}. 

Regarding other Reynolds stresses, namely the fluctuation components of wall-tangential ($\overline{v^2_n}$) and spanwise ($\overline{w^2}$) velocity as well as the Reynolds-shear stress ($\overline{u_tv_n}$), the modifications are simpler to describe than that of $\overline{u^2_t}$. 
On the suction side, due to the increased wall-normal convection caused by strong APGs, the wall-normal position of the outer {peaks} align with that of $\overline{u^2_t}$ in the overlap region. 
The effects of uniform suction vary with the wall-normal distance. 
In particular, below the wall-normal plane of the outer peak, uniform suction remarkably attenuates the $\overline{v^2_n}$, $\overline{w^2}$ and $\overline{u_tv_n}$, which is observed for both inner- and outer-scaled profiles. 
Above this wall-normal position, the uniform suction intensifies the profiles of Reynolds stresses, which can be more clearly observed on the outer-scaled profiles in~\cref{fig:uiuj-outer-SS}. 

On the other hand, uniform blowing on the pressure side significantly intensifies the $\overline{v^2_n}$ and $\overline{u_t v_n}$ {values within} the inner region, while the modification on $\overline{w^2}$ is very {negligible}.
Note that the intensified $\overline{v^2_n}$ exhibits an outer peak at a wall-normal position that is farther from the wall compared to the one of $\overline{u^2_t}$.

{
Additionally, to evaluate the control mechanism in terms of momentum exchange, we assess the visualization of vortical structures identified using the $\lambda_2$ method~\citep{jeong_la2_identification_1995}. \Cref{fig:la2} shows isosurfaces of these structures, colored by streamwise velocity. The region of interest spans from $x/c = 0.5$ to $1.5$, encompassing the separated flow region and the near-wake area downstream of the TE.

The control effect on momentum exchange becomes evident when examining the spatial development of high- and low-speed regions along the wing surface. Under uniform suction applied to the SS, both the off-wall high-speed regions and near-wall low-speed regions are attenuated as suction intensity increases. This behavior is attributed to the mechanism by which steady suction removes low-momentum fluid and redirects high-momentum fluid toward the wall. Consequently, the regions with the lowest streamwise velocity (i.e., shown in dark blue) are significantly reduced, which corresponds to delayed separation. On the other hand, uniform blowing on the PS generates near-wall vortical structures by enhancing wall-normal momentum transfer. This displaces near-wall flow to the outer region and becomes more pronounced with increasing control intensity, consistent with the observation shown in~\cref{fig:uiuj-outer-PS}. 

Interestingly, Cases B and C also exhibit a remarkable impact on the vortex shedding in the near wake, with a dramatic modification of flow structures. This is likely due to the combined effects of the control type and intensity, as such changes are not observed when applying a lower intensity ($\psi = 0.25\%$) or using either uniform suction or blowing alone.
  
\begin{figure}[ht]
  \centering
  \begin{subfigure}{0.48\linewidth}
    \centering
    \includegraphics[width=\textwidth]{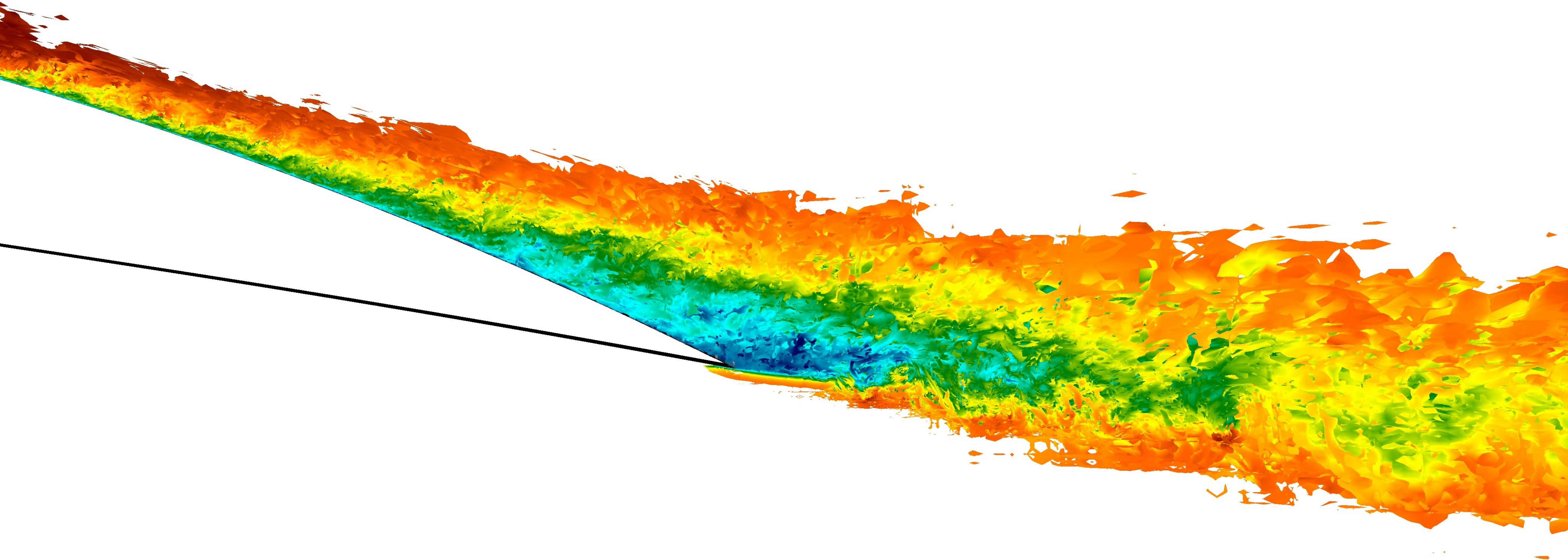}
    \subcaption{Uncontrolled case}
  \end{subfigure}
  \hfill
  \begin{subfigure}{0.48\linewidth}
    \centering
    \includegraphics[width=\textwidth]{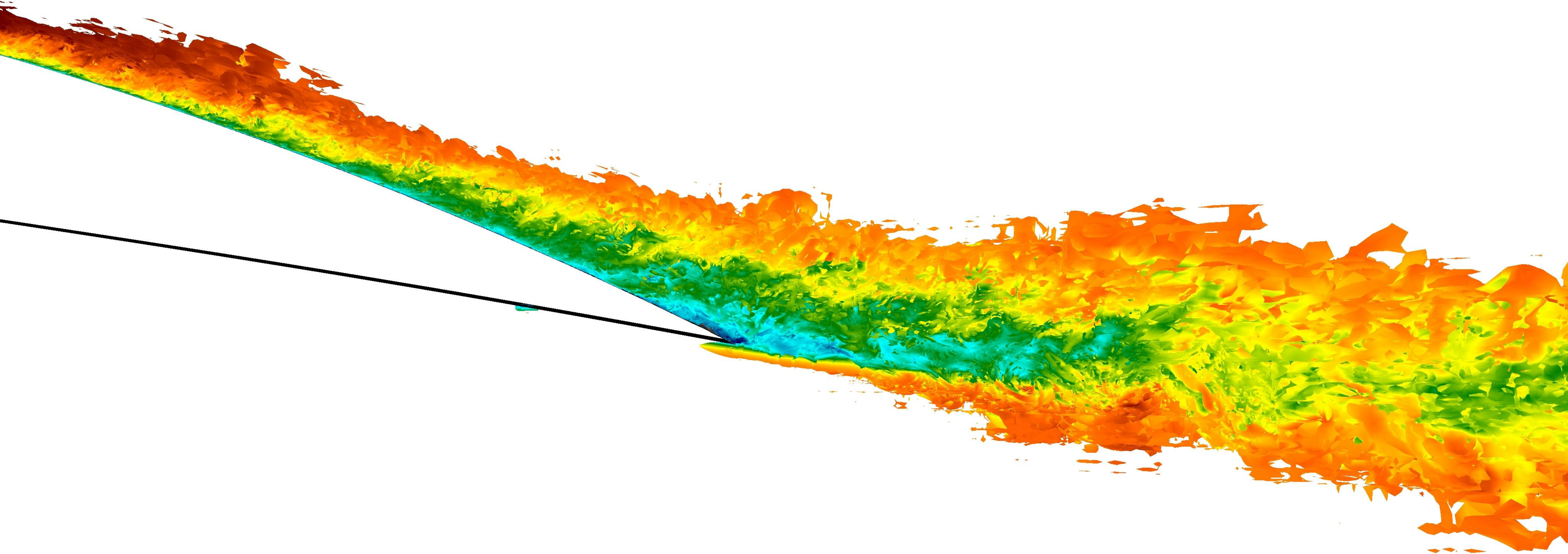}
    \subcaption{Case A}
  \end{subfigure}
  \hfill
  \begin{subfigure}{0.48\linewidth}
    \centering
    \includegraphics[width=\textwidth]{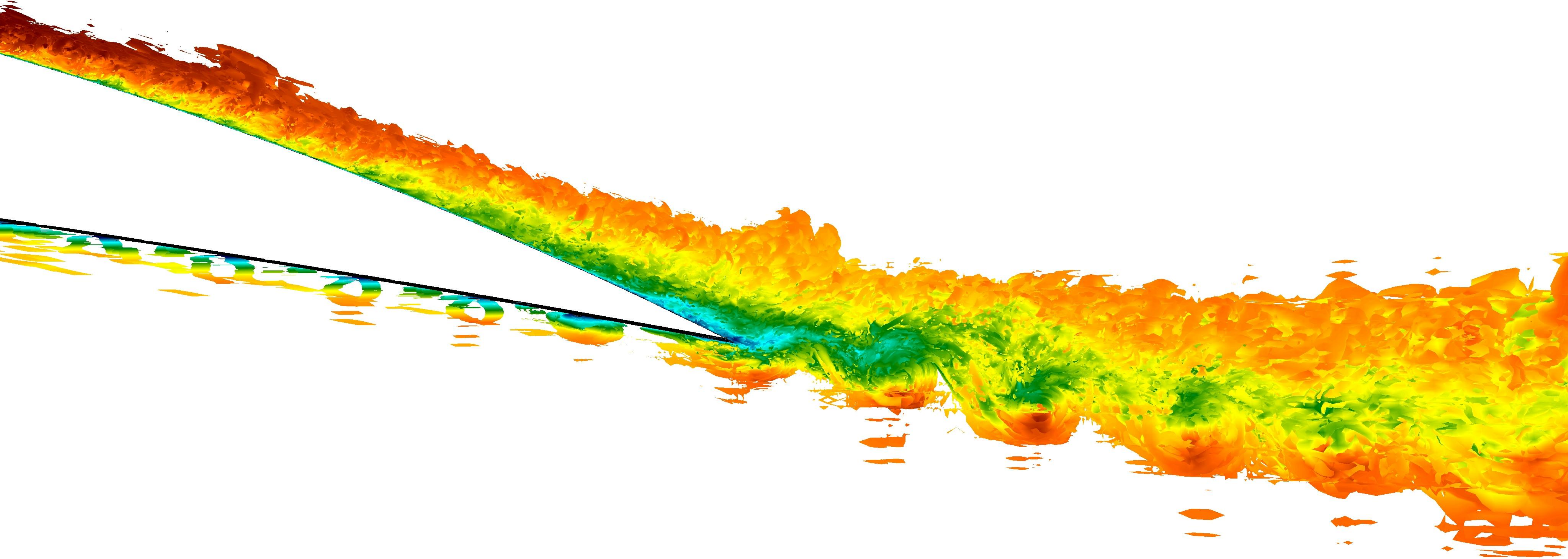}
    \subcaption{Case B}
  \end{subfigure}
  \hfill
  \begin{subfigure}{0.48\linewidth}
    \centering
    \includegraphics[width=\textwidth]{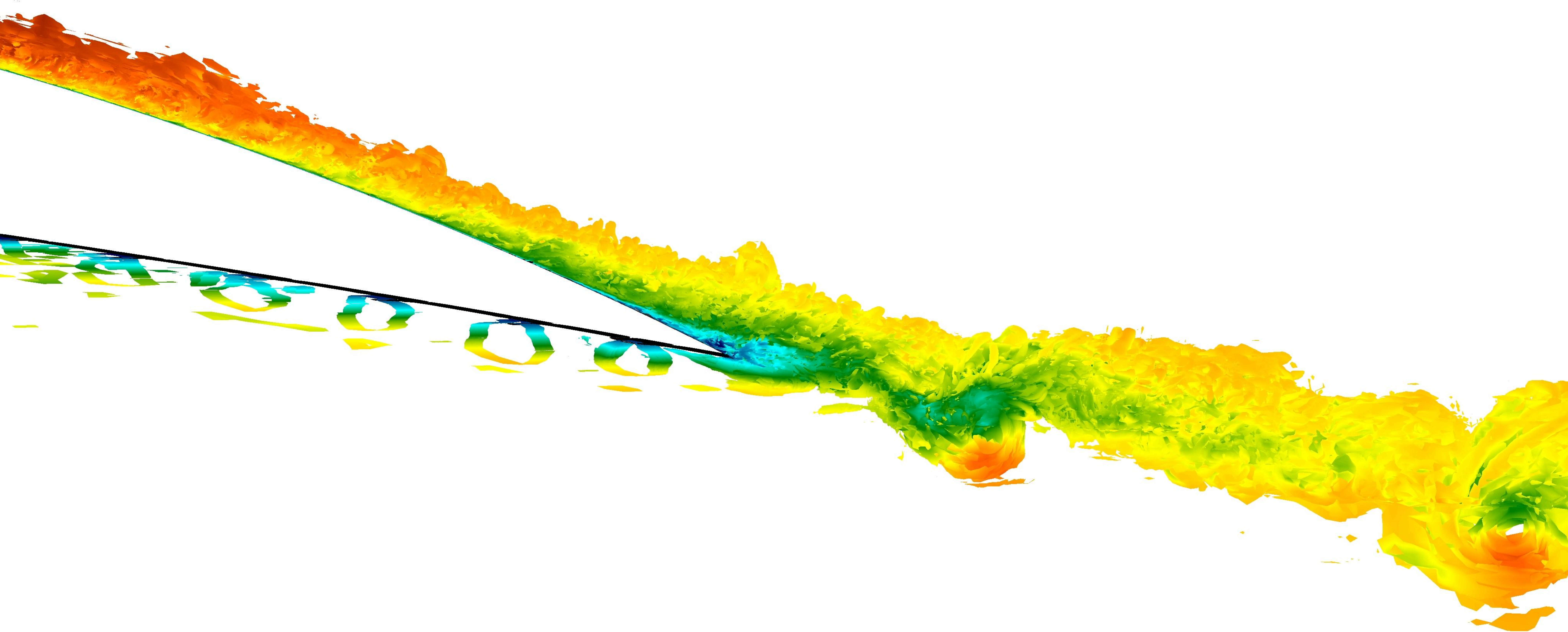}
    \subcaption{Case C}
  \end{subfigure}
  \hfill
  \begin{subfigure}{0.48\linewidth}
    \centering
    \includegraphics[width=\textwidth]{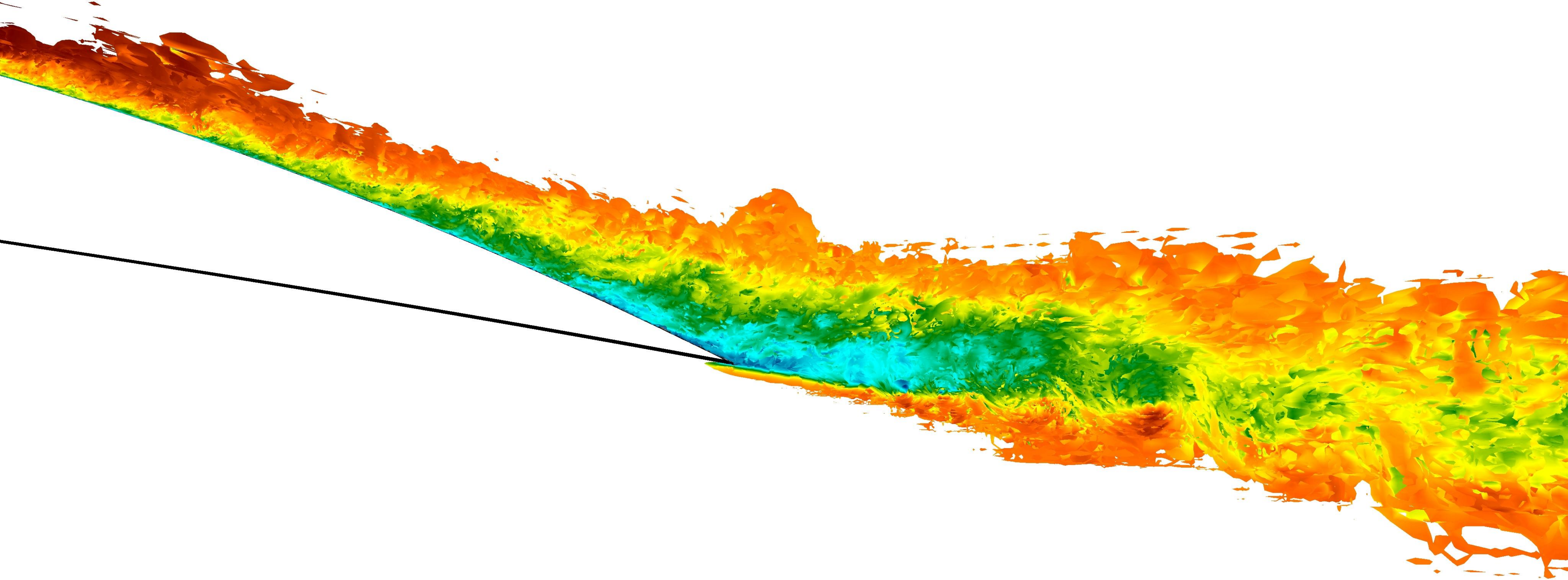}
    \subcaption{Case D}
  \end{subfigure}
  \hfill
  \begin{subfigure}{0.48\linewidth}
    \centering
    \includegraphics[width=\textwidth]{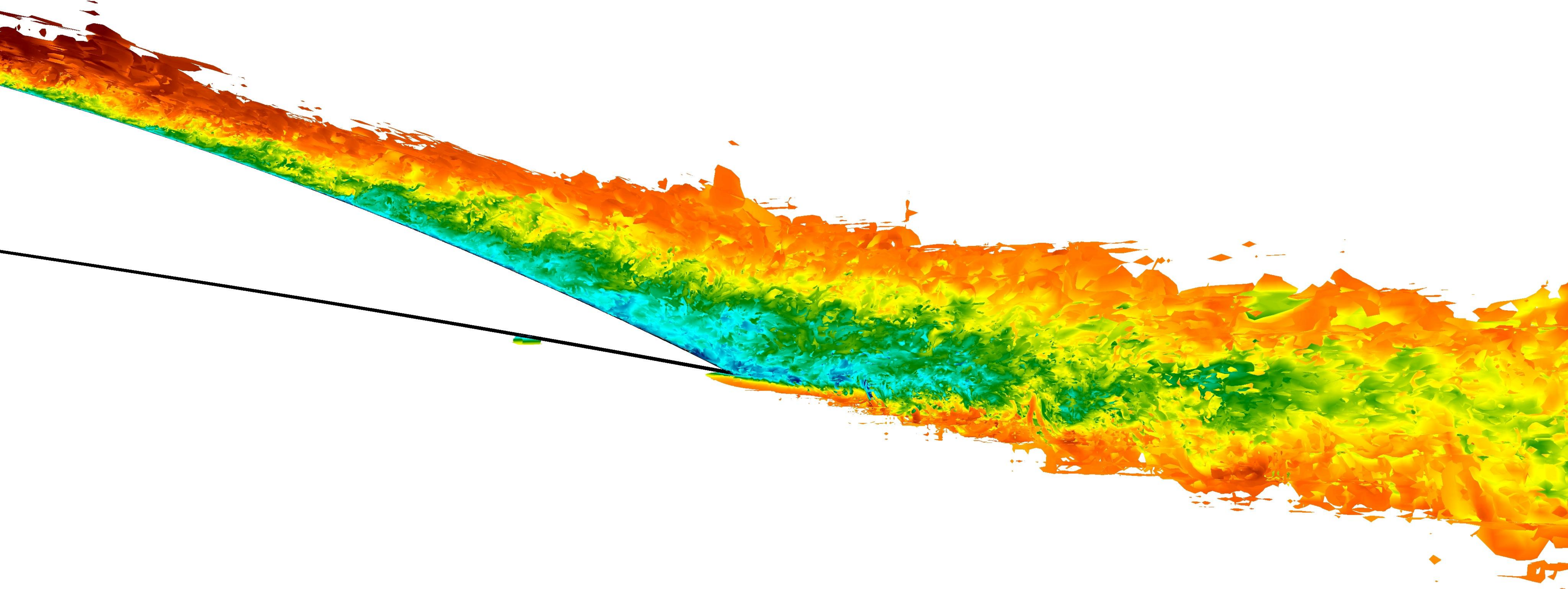}
    \subcaption{Case E}
  \end{subfigure}
  \hfill
  \caption{{Visualization of vortical structures identified by the $\lambda_2$ method~\citep{jeong_la2_identification_1995} between $x/c = 0.5$ and $1.5$, colored by the instantaneous streamwise velocity at an arbitrary time step. 
  An isosurface of $\lambda_2 = -10$ is shown in all the cases. The vortical motions are colored by their streamwise velocity value, ranging from $-0.4U_{\infty}$ (dark blue) to $1.6 U_{\infty}$ (dark red).}}
  \label{fig:la2}
\end{figure}
}

\subsection{Time-series analysis} \label{sec:psd}

In this section, we employ the spectral analysis to investigate the temporal and scale-dependent interactions between control and TBLs. 
To this end, time series of the velocity components and pressure were collected for multiple wall-normal profiles across various streamwise locations, extending to $x/c = 1.74$ in the wake. 
The dataset spans a total of $2$ flow-over times{after excluding initial transients}, with a sampling rate of $2\times10^{-4}$ flow-over times.

\begin{figure}[ht]
  \begin{center}
    \includegraphics[width=0.6\textwidth]{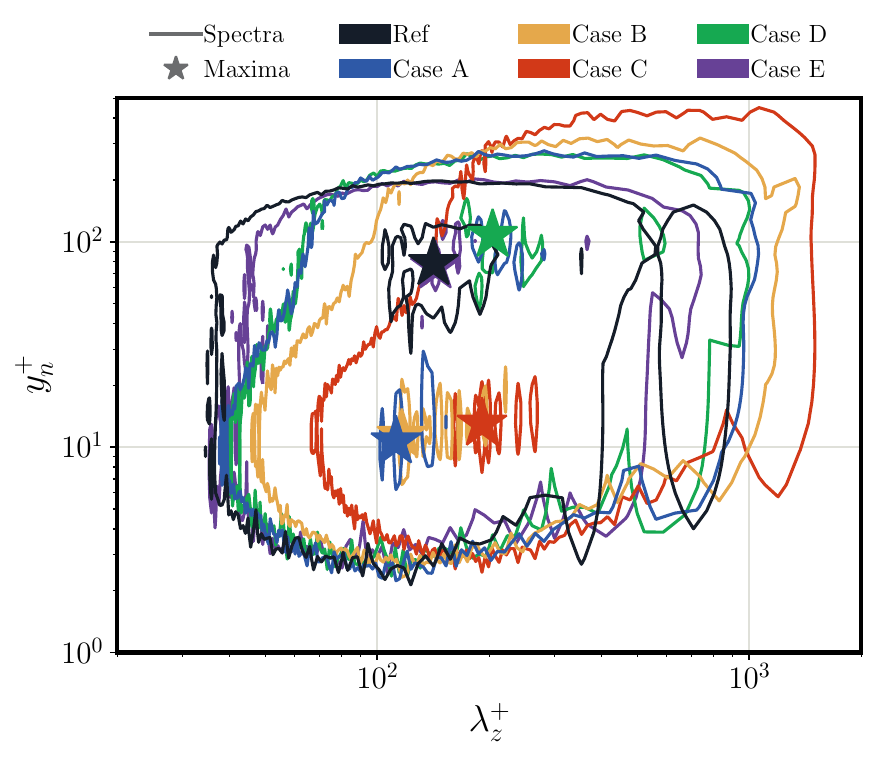}
    \caption{ 
      {
      Inner-scaled premultiplied spanwise power-spectral density (PSD) of the wall-tangential velocity fluctuation, in terms of the inner-scaled spanwise wavelength (${\lambda}^+_{z}$) and wall-normal distance ($y^+_n$) {at $x/c = 0.75$ on the suction side}.
      The contours illustrate the levels corresponding to $20\%$ and $80\%$ of the maximum power density, whereas the locations of the maxima achieved are marked with stars. 
      The color code follows~\cref{tab:ctrl_configs}.
        }
    }
    \label{fig:psd-1d}
  \end{center}
\end{figure}

First, we assess the one-dimensional power-spectral densities (PSDs) computed using the fast Fourier transform (FFT). 
\Cref{fig:psd-1d} depicts the contours of the inner-scaled premultiplied PSD of wall-tangential velocity fluctuations ($k_z\phi^+_{u_tu_t}$) at $x/c = 0.75$ on the suction side, expressed as a function of inner-scaled spanwise wavelength $\lambda^+_z$ and wall-normal distance $y^+_n$.
Under strong APG conditions near separation, the wall-normal profile of $\overline{u^2_t}$ (see~\cref{fig:uiuj-vel}) reveals that the outer peak in the overlap region dominates over the inner peak in the buffer layer. 
Spectral analysis confirms the absence of the inner peak (typically at $\lambda^+ \approx 100$), which is typically present in canonical wall-bounded flows and lower angles of attack ($AoA = 5^{\circ}$)~\citep{atzori_uniform_2021,
wang_opposition_2024,mallor_IJHFF_2024}.
Instead, the outer peak of small-scale structures dominates the energy distribution, highlighting the strong influence 
of increased wall-normal convection due to APGs.

Uniform suction over the suction side significantly modifies the energy distribution under near-separation conditions. 
By introducing high-momentum fluid near the wall, the energy peak shifts toward the buffer layer, and small-scale structures in the overlap region are attenuated, restoring characteristics of the canonical near-wall cycle. 
Cases A, B, and C shift the energy peak to $y^+_n = 10.6$, $11.5$, and $13.0$, respectively, compared to $y^+_n = 78.7$ in the uncontrolled case. 
Additionally, Case A and B yield smaller wavelengths ($\lambda^+_z = 112$ and 
$116$) compared to the reference ($\lambda^+_z = 141$), while Case C shifts 
toward a larger wavelength ($\lambda^+_z = 191$). 
Uniform suction also enhances large-scale structures in the outer region, with 
effects increasing with $\psi$.

The spectra of Cases D and E deviate from their combined counterpart (Case A). 
Case D shifts the outer peak further from the reference instead of producing an inner peak, while Case E introduces minimal modification. 
However, Case E attenuates small-scale structures in the overlap region and enhances large-scale energy in the outer region.
The results indicate that those modifications are due to the variation of $u_\tau$.

\begin{figure}[h!]
  \centering
  \begin{subfigure}{0.49\linewidth}
    \centering
    \includegraphics[width=\textwidth]{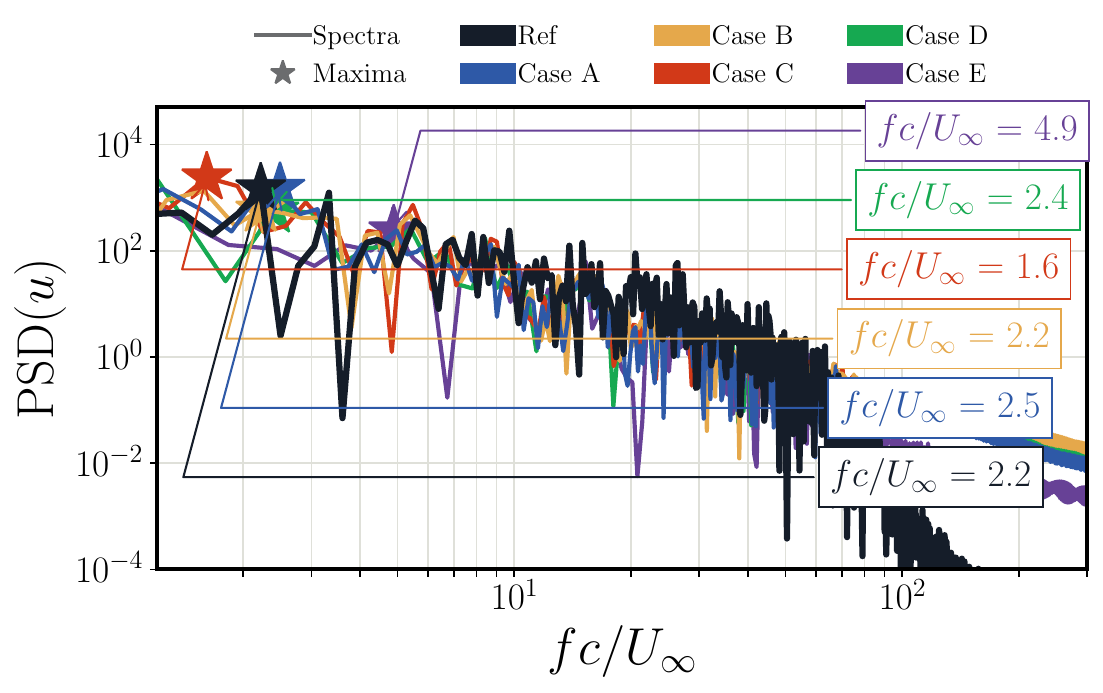}
    \subcaption{{PSD of $u$ at $x/c = 1.5$.}}
    \label{fig:FFT-u-15}
  \end{subfigure}
  \hfill
  \begin{subfigure}{0.49\linewidth}
    \centering
    \includegraphics[width=\textwidth]{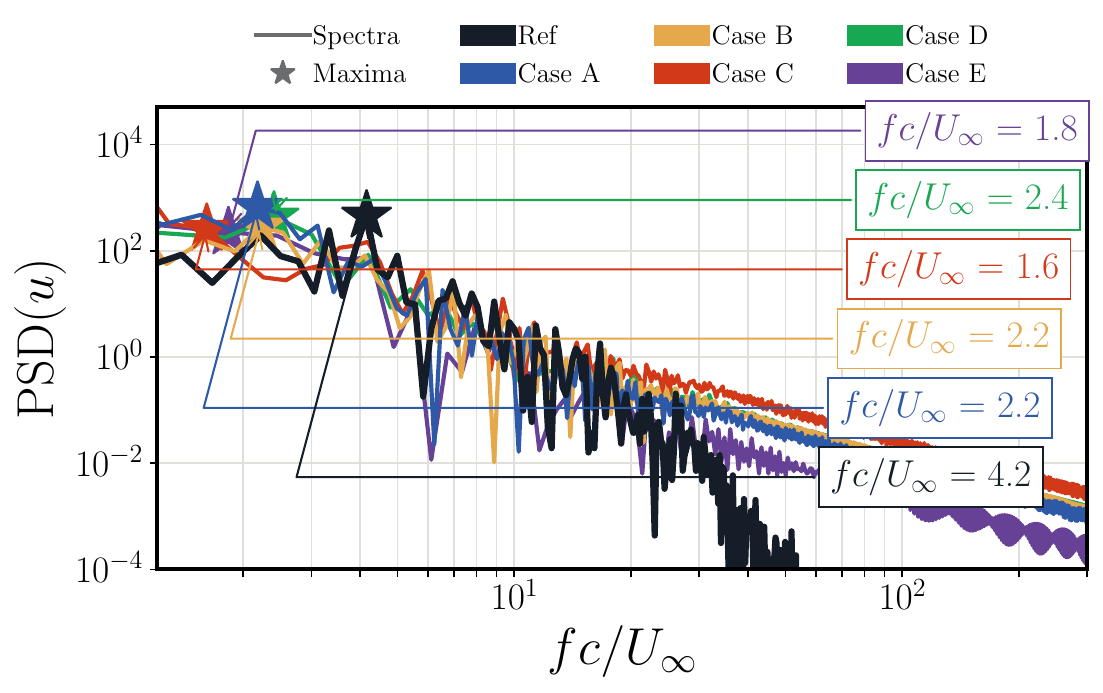}
    \subcaption{
    {PSD of $u$ at $x/c = 1.74$.}}
    \label{fig:FFT-u-17}
  \end{subfigure}
  \quad
  \vspace{0.5em}
  \begin{subfigure}{0.49\linewidth}
    \centering
    \includegraphics[width=\textwidth]{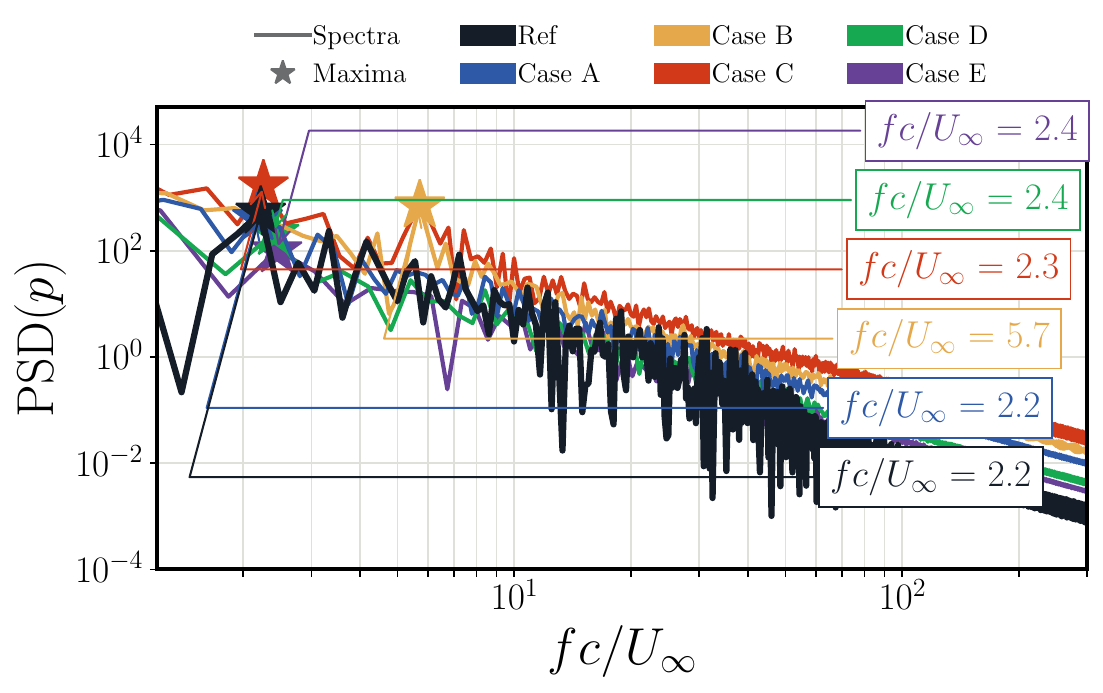}
    \subcaption{{PSD of $p$ at $x/c = 1.5$.}}
    \label{fig:FFT-p-15}
  \end{subfigure}
  \hfill
  \begin{subfigure}{0.49\linewidth}
    \centering
    \includegraphics[width=\textwidth]{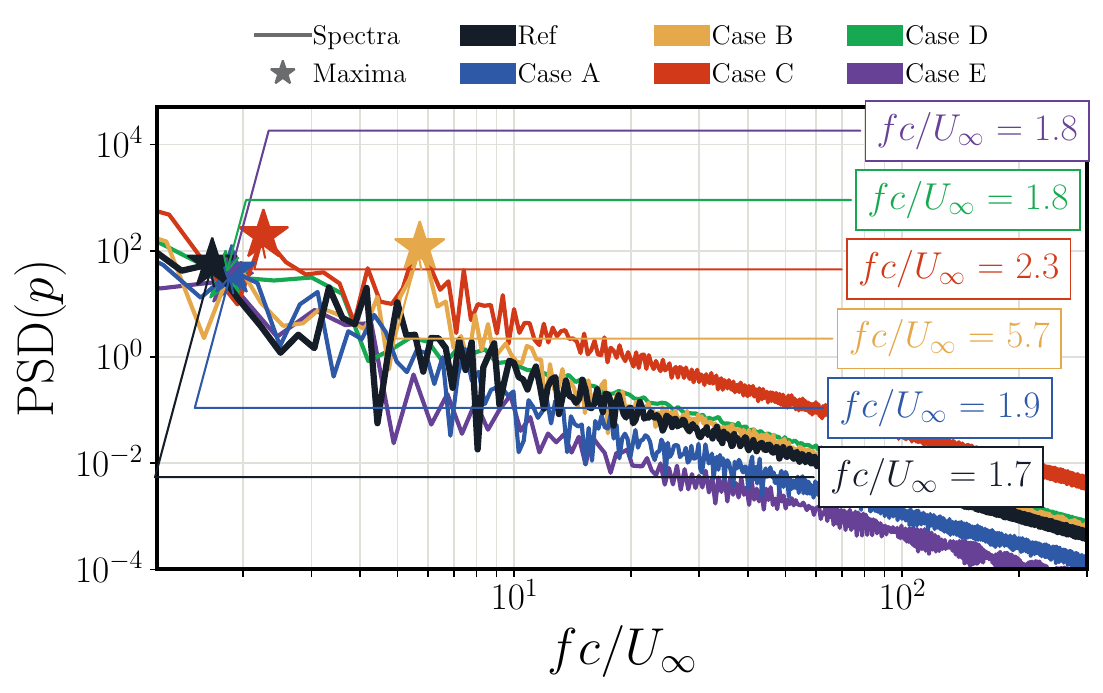}
    \subcaption{{PSD of $p$ at $x/c = 1.74$.}}
    \label{fig:FFT-p-17}
  \end{subfigure}
    \caption{ 
      {
            Power-spectral density (PSD) of ((a), (b)) fluctuating streamwise velocity and ((c), (d)) pressure in terms of the dimensionless frequency ($fc/U_{\infty}$) measured at streamwise location 
            $x/c$ of ((a), (c)) $1.5$ and ((b), (d)) $1.74$, respectively, where the locations and value of the dominant frequencies are marked with stars and denoted by panels, respectively. 
            The color code follows~\cref{tab:ctrl_configs}.
            }
    }
    \label{fig:FFT}
\end{figure}

{Next, we examine the PSDs of the fluctuating streamwise velocity ($u$) and pressure ($p$) in the wake to analyze control effects on vortex-shedding frequency. Time-series data were collected at $x/c = 1.5$ and $1.74$, at a vertical position of $y = 0$. Note that this region corresponds to the vortex-formation zone, where the downstream energy evolution of vortex shedding can be effectively captured~\citep{yarusevych_vortexShedding_2009}. Moreover, previous studies on airfoil wakes~\citep{gerontakos_nearwake_2005,yarusevych_vortexShedding_2009,liu_shedding_2017} have employed similar measurement positions within comparable Reynolds number regimes. Therefore, the current results may serve as a useful reference for future investigations.}

{\Cref{fig:FFT-u-15,fig:FFT-u-17} depict the PSDs of $u$ as a function of the dimensionless frequency $fc/U_{\infty}$ at $x/c = 1.5$ and $1.74$. }
In the uncontrolled case, the vortex-shedding frequency increases downstream, 
rising from $fc/U_{\infty} = 2.2$ at $x/c = 1.5$ to $4.2$ at $x/c = 1.74$, which is consistent with~\citet{yarusevych_vortexShedding_2009}.
Control using combined uniform suction and blowing results in minor modifications to the vortex-shedding frequency. 
Cases A, B, and C yield $fc/U_{\infty}$ values of $2.5$, $2.2$, and $1.6$ at both locations, closely matching the uncontrolled case. 
However, the effects of applying suction or blowing individually vary with the streamwise location. 
Case D exhibits an increased dominant frequency ($fc/U_{\infty} = 2.4$), similar to Case A at both locations, 
whereas Case E results in $fc/U_{\infty} = 4.9$ and $1.8$ at $x/c = 1.5$ and $1.75$, respectively. 
Additionally, all configurations amplify the PSD values for high-frequency components, which is particularly evident in \cref{fig:FFT-u-17}. 
For the pressure spectra (\Cref{fig:FFT-p-15,fig:FFT-p-17}), the dominant frequency slightly decreases downstream, being $fc/U_{\infty} = 2.2$ and $1.7$ at $x/c = 1.5$ and $1.74$, respectively. 
Similarly, the modified $fc/U_{\infty}$ exhibits a downstream reduction, 
while its deviation from the reference is minimal except in Case B, where the dominant frequency drastically increases to $fc/U_{\infty} = 5.7$. 
Interestingly, the dominant frequency of Case B resonates with that of the secondary peak of Case C, which is not observed in other configurations, 
suggesting a significant modification of the wake due to high control intensity. 
However, the modifications to the high-frequency spectra of the PSDs vary with the streamwise location. 
While all configurations energize the high-frequency motions at $x/c=1.5$, only Case C {exhibits a siginificantly energized spectrum} at $x/c=1.74$. 
These results highlight the crucial role of control intensity in modifying the vortex-shedding frequency.

\section{Summary and conclusions}\label{sec:conclusion}

In the present study, we investigated the effects of various separation-control approaches on a NACA4412 wing section at an angle of attack ($AoA$) of $11^{\circ}$ and a Reynolds number of $Re_c = 200{,}000$. 
High-resolution large-eddy simulations (LESs) were carried out using the spectral-element-method solver Nek5000~\citep{nek5000}, incorporating adaptive mesh refinement (AMR) for non-conformal meshes. 
This approach enables the use of a wide spanwise width ($L_z = 0.6$) while maintaining high accuracy, ensuring the capture of large-scale structures and accurate resolution of flow separation.

A range of control configurations were considered, categorized as follows: (1) uniform suction on the suction side, (2) uniform blowing on the pressure side, (3) combined uniform suction and blowing, and (4) periodic control on the suction side. 
The control impact on aerodynamic characteristics, including separation length ($\ell_{\rm sep}$), lift, drag components, and aerodynamic efficiency, was evaluated. 
Note that control intensities ($\psi$) and control areas were chosen based on previous studies~\citep{atzori_aerodynamic_2020,mallor_bayesian_2024}.

Overall, {the tested uniform blowing and/or suction strategies} facilitated flow reattachment but had varying effects on aerodynamic efficiency. 
Specifically, uniform suction alone produced the same reduction in $\ell_{\rm sep}$ as combined suction and blowing, 
but resulted in a smaller lift increase while causing similar drag increases, leading to a larger decrease in aerodynamic efficiency. 
Notably, uniform blowing on the pressure side was the only configuration that improved aerodynamic efficiency, albeit with a smaller reduction in $\ell_{\rm sep}$. 
The control effects on aerodynamic characteristics differed significantly from previous studies at lower $AoA$~\citep{atzori_aerodynamic_2020} and higher Reynolds numbers~\citep{mallor_bayesian_2024}. 
Furthermore, adjustments to control configurations did not yield additional benefits, and periodic control 
(\S~\ref{sec:appendix-clcd}) neither further delayed separation nor improved aerodynamic efficiency.

These findings motivated an in-depth investigation of the interaction between 
control and turbulent boundary layers (TBLs) over the wing section.
Control significantly modified the skin-friction coefficient ($c_f$) and 
pressure coefficient ($c_p$). 
Uniform suction increased both $c_f$ and $c_p$, while uniform blowing reduced $c_f$ but increased $c_p$, consistent with findings at $AoA = 5^{\circ}$~\citep{atzori_uniform_2021,fahland_drag_2023}. 
The effects on $c_p$ were equivalent to increasing the angle of attack. 
Further analysis of the Clauser pressure-gradient parameter ($\beta$) showed that uniform suction 
attenuated boundary-layer development by reducing $\beta$ and momentum thickness 
($\theta$), whereas uniform blowing on the pressure side exhibited effects 
dependent on control intensity.

Turbulence statistics at $x/c = 0.75$ were examined on both airfoil surfaces. 
The control effects on mean wall-tangential and wall-normal velocity profiles were consistent with previous studies at $AoA = 5^{\circ}$~\citep{atzori_aerodynamic_2020}. 
However, modifications to the streamwise velocity fluctuations ($\overline{u^2_t}$) on the suction side differed: uniform suction intensified the outer peak in the overlap region rather than attenuating it, influenced by strong adverse pressure gradients and flow history. 
On the pressure side, strong favorable pressure gradients (FPGs) produced Reynolds stresses with low magnitudes, whereas uniform blowing significantly increased wall-tangential fluctuations and Reynolds shear stress.

Spectral analysis further reveals the control effects on $\overline{u^2_t}$. 
One-dimensional power-spectral density (PSD) analysis at $x/c = 0.75$ revealed that uniform suction energized large-scale structures in the outer layer and small-scale structures in the buffer layer. 
Notably, an inner peak emerged only when suction was combined with uniform blowing. 
{Additionally, PSD analysis of fluctuating streamwise velocity and pressure in the wake showed that 
control significantly reduced the vortex-shedding frequency.}

An important avenue for future research is exploring control performance across a broader range of $AoA$ to provide a more general understanding. 
Comparing results at $AoA = 5^{\circ}$ and $11^{\circ}$ highlighted difference in control effects, suggesting that the need for further investigation. 
However, the present study focused on a specific $AoA$ to validate and compare against the configuration proposed by~\citet{mallor_bayesian_2024}, making broader investigations beyond the current scope.

A key challenge is optimizing control schemes to simultaneously improve aerodynamic efficiency and delay flow separation. 
Our results show that separation is delayed, but aerodynamic efficiency is reduced. 
Neither modifying configurations nor employing periodic control improved $L/D$ (\S~\ref{sec:appendix-clcd}). 
One possible explanation is that flow separation in this case is less pronounced compared to near-stall conditions 
at higher $AoA$ (e.g., $\ell_{\rm sep} = 0.51$ at $AoA = 14^{\circ}$ for higher $Re_c$~\citep{mallor_IJHFF_2024}), where lift improvement outweighs drag penalties. 
However, this reasoning does not apply to the current case, where most of the flow remains attached. 
Achieving a balance between separation delay and aerodynamic efficiency is challenging, as separation 
control inherently involves increasing near-wall momentum~\citep{greenblatt_control_2000}, which inevitably leads to higher drag.

Deep reinforcement learning (DRL) could offer a potential solution to the challenges faced by traditional control methods. 
As discussed in \S~\ref{sec:intro}, DRL can identify complex flow conditions and generate optimal responses to maximize objectives (e.g., delaying separation and improving $L/D$) by leveraging neural networks to approximate nonlinear 
functions~\citep{vinuesa_flow_2022}. 
In the context of separation control, \citet{suarez_highRe_2024} demonstrated that DRL can enhance aerodynamic 
efficiency while reducing energy consumption.

In summary, this study provides a comprehensive analysis of separation-control interactions with turbulent boundary layers over a wing section. 
To the authors' best knowledge, this is the first numerical study utilizing high-resolution LESs to provide comprehensive assessments on separation control.

\section*{Data availability}
The data that support the findings of this study will be openly available on GitHub-\href{https://github.com/KTH-FlowAI}{KTH-FlowAI} after publication. 

\section*{Competing interesting}
The authors declare that they have no competing interests. 

\section*{Acknowledgements}
Numerical studies were enabled by resources provided by the National Academic Infrastructure for Supercomputing in Sweden (NAISS) at PDC center for High Performance Computing in KTH (stockholm). RV acknowledges the financial support from ERC grant no. `2021-CoG-101043998, DEEPCONTROL'. {Views and opinions expressed are those of the authors only and do not necessarily reflect those of the European Union or the European Research Council. Neither the European Union nor the granting authority can be held responsible for them.}

\section*{CRediT authorship contribution statement}
YW: Methodology, Software, Validation, Investigation, Data curation, Data generation, Writing - Original Draft, Writing - Review \& Editing, Visualization.
FM: Methodology, Software, Validation, Investigation, Data curation, Data generation, Writing - Original Draft, Writing - Review \& Editing.
CGG: Ideation, Methodology, Validation, Investigation, Data generation, Writing - Original Draft, Writing - Review \& Editing.
RM: {Writing - Review \& Editing.}
RV: Ideation, Methodology, Validation, Data generation, Writing - Original Draft, Writing - Review \& Editing, Resources, Funding acquisition.

\section*{Declaration on artificial-intelligence (AI) assistance}
This paper has benefited from AI-assisted language refinement. 
Specifically, ChatGPT was employed for polishing the text, improving readability, and ensuring clarity. 
However, the conceptual development, data analysis, and conclusions presented in this work are the sole responsibility of the authors. 

\appendix
\section{Aerodynamic effects of periodic control and uniform suction over suction side} \label{sec:appendix-clcd}

In this section, we report the aerodynamic characteristics obtained from applying (1) periodic control and (2) uniform suction over the suction side with various configurations.
Periodic control is a well-established separation control approach that 
introduces periodic excitation to modify the spreading rate of the mixing 
layer~\citep{greenblatt_control_2000}, with successful applications in 
various studies~\citep{kim_PeriodicSeparation_2007,gul_PeriodicExperimental_2014,
tang_PeriodicUse_2014,cetin_PeriodicControl_2018,wu_ZNMFJETResponse_2022}. 
It is typically applied near the separation point within a limited control area.

We first examined the potential of periodic control in delaying separation. 
The control was implemented as a Dirichlet boundary condition at the wall, 
with the imposed control input defined as:
\begin{equation}
  V_{n, w} = \psi \sin(2\pi F^* t),
\end{equation}
\noindent where $\psi$ represents the input intensity, while $F^*$ and $t$ denote the 
dimensionless control frequency and flow-over time, respectively. 
The imposed control velocity $V_{n,w}$ is directed wall-normal but is decomposed into 
horizontal and vertical components to ensure a combined magnitude equal to 
$V_{n,w}$. 
{Note that the choice of $F^*$ depends on the vortex-shedding frequency but is not strictly bounded to it. 
In addition to $F^* = 4.2$ (i.e., the vortex-shedding frequency at $x/c=1.74$), we select various $F^*$ values such that $ F^* \sim 1 $, which is suggested as the optimal frequency for separation control~\citep{mabey_analysis_1972,roshko_controlled_1985,hsiao_Acoustic_1990,amitay_aerodynamic_2001,tuck_znmfSynJet_2008}. 
This optimal frequency is independent of curvature and the state of the separating boundary layer~\citep{greenblatt_control_2000}. 
Additionally, $F^*=10.10$ is test to explore the performance of high-frequency control.}

\begin{table}[ht]
  \begin{center}
  \resizebox*{\textwidth}{!}{
    \def~{\hphantom{0}}
    \begin{tabular}{ccccccccccc}
      Name   & $\ell_{\rm sep}$ & $C_{\mu} [\times 10^{-5}]$ & $C_l$ & $C_{d,p}$ & $C_{d,f}$ & $C_d = C_{d,p} + C_{d,f}$ & $L/D$ & $F^*$ & $\psi$ $[\%U_{\infty}]$ & Control area $[x/c]$ \\ [3pt]
      Ref    & 0.14 & 0.0 & 1.314 & 0.0450  & 0.0078 & 0.052  & 24.88 & $--$ & $--$ & $--$ \\
      Case 1 & $0.06$ (\cgreen{$-57\%$}) & $0.012$ & 1.310 (\cred{$-0.30\%$}) & 0.0470 (\cred{$+4.49\%$}) & 0.0077 (\cgreen{$-0.77\%$}) & 0.054 (\cred{$+3.71\%$}) & 23.91 (\cred{$-3.90\%$}) & $1.0$ & $0.1$ &$ 0.80 \sim 0.86 $ \\
      Case 2 & $0.05$ (\cgreen{$-64\%$}) & $0.012$ & 1.304 (\cred{$-0.76\%$}) & 0.0461 (\cred{$+2.47\%$}) & 0.0078 (\cgreen{$-0.26\%$}) & 0.054 (\cred{$+2.06\%$}) & 24.19 (\cred{$-2.77\%$}) & $4.2$ & $0.1$ &$ 0.80 \sim 0.86 $ \\
      Case 3 & $0.05$ (\cgreen{$-64\%$}) & $0.012$ & 1.308 (\cred{$-0.46\%$}) & 0.0468 (\cred{$+4.09\%$}) & 0.0078 (\cgreen{$-0.38\%$}) & 0.055 (\cred{$+3.43\%$}) & 23.95 (\cred{$-3.74\%$}) & $0.86$ & $0.1$ &$ 0.80 \sim 0.86 $  \\
      Case 4 & $0.05$ (\cgreen{$-64\%$}) & $0.012$ & 1.308 (\cred{$-0.46\%$}) & 0.0468 (\cred{$+4.09\%$}) & 0.0078 (\cgreen{$-0.51\%$}) & 0.055 (\cred{$+3.41\%$}) & 23.95 (\cred{$-3.74\%$}) & $10.10$ & $0.1$ &$ 0.80 \sim 0.86 $  \\
      Case 5 & $0.05$ (\cgreen{$-64\%$}) & $1.2$ & 1.299 (\cred{$-1.14\%$}) & 0.0451 (\cred{$+0.31\%$}) & 0.0078 (\cgreen{$-0.26\%$}) & 0.053 (\cred{$+0.23\%$}) & 24.55 (\cred{$-1.33\%$}) & $0.63$ & $1.0$ &$ 0.86 \sim 0.92 $ \\
      Case 6 & $0.05$ (\cgreen{$-64\%$}) & $1.2$ & 1.308 (\cred{$-0.46\%$}) & 0.0468 (\cred{$+4.09\%$}) & 0.0078 (\cgreen{$-0.38\%$}) & 0.055 (\cred{$+3.43\%$}) & 23.95 (\cred{$-3.74\%$}) & $0.94$ & $1.0$ &$ 0.86 \sim 0.92 $  \\
      Case 7 & $0.05$ (\cgreen{$-64\%$}) & $1.2$ & 1.301 (\cred{$-0.99\%$}) & 0.0454 (\cred{$+0.87\%$}) & 0.0078 (\cgreen{$-0.38\%$}) & 0.053 (\cred{$+0.68\%$}) & 24.46 (\cred{$-1.69\%$}) & $0.94$ & $1.0$ &$ 0.86 \sim 0.92 $  \\
      Case 8 & $0.05$ (\cgreen{$-64\%$}) & $2.8$ & 1.309 (\cred{$-0.38\%$}) & 0.0480 (\cred{$+6.71\%$}) & 0.0079 (\cgreen{$+1.02\%$}) & 0.056 (\cred{$+5.87\%$}) & 23.42 (\cred{$-5.87\%$}) & $1.16$ & $1.0$ &$ 0.86 \sim 1.00 $  \\
      Case 9 & $0.04$ (\cgreen{$-71\%$}) & $11.2$ & 1.314 (\cgreen{$0.00\%$}) & 0.0478 (\cred{$+6.25\%$}) & 0.0078 (\cgreen{$-0.51\%$}) & 0.056 (\cred{$+5.25\%$}) & 23.63 (\cred{$-5.03\%$}) & $1.16$ & $2.0$ &$ 0.86 \sim 1.00 $  \\
    \end{tabular}
    }
  \caption{Separation length ($\ell_{\rm sep}$), momentum coefficient ($C_{\mu}$), 
  lift coefficient ($C_l$), drag components ($C_{d,p}$ and $C_{d,f}$), total 
  drag ($C_d$), and aerodynamic efficiency ($L/D$) for periodic control cases on suction side of a 
  NACA4412 airfoil at $AoA = 11^{\circ}$ and $Re_c = 200,000$. 
  Percent changes from the reference case are shown in parentheses. The values in the parentheses report the relative changes obtained by control. The $F^*$ denotes the input frequency of control.}
  \label{tab:ClCd-pctrl}
  \end{center}
\end{table}
\Cref{tab:ClCd-pctrl} summarizes the aerodynamic efficiency and separation delay for the periodic control cases. Although periodic control effectively delays separation, it generally reduces lift (except for Case 9) and increases 
drag, leading to a decline in aerodynamic efficiency. 
Notably, friction drag is slightly reduced, while pressure drag increases significantly due to control application.

We also conducted additional tests to examine the effect of uniform suction 
over the suction side, specifically investigating the impact of control area 
location, which is typically adjacent to the separation point~\citep{carnarius_Suction4412Optimize_2007,
azim_Suction4412numerical_2015,oktay_Suction4412_2019}. 
A broader range of input intensities $\psi$, up to $7.5\%U_{\infty}$, was also explored.
\begin{table}[ht]
  \begin{center}
  \resizebox*{\textwidth}{!}{
    \def~{\hphantom{0}}
    \begin{tabular}{cccccccccc}
      Name   & $\ell_{\rm sep}$ & $C_{\mu}[\times 10^{-5}]$ & $C_l$ & $C_{d,p}$ & $C_{d,f}$ & $C_d = C_{d,p} + C_{d,f}$ & $L/D$ & $\psi$ $[\%U_{\infty}]$ & Control area $[x/c]$  \\ [3pt]
      Ref    & 0.14 & $--$ & 1.314 & 0.0450  & 0.0078 & 0.052  & 24.88  & $--$ & $--$ \\
      Case 1 & $0.06$ (\cgreen{$-57\%$}) & $0.6$ & 1.314 (\cgreen{$0.00\%$}) & 0.0477 (\cred{$+6.11\%$}) & 0.0080 (\cred{$+2.30\%$}) & 0.056 (\cred{$+5.55\%$}) & 23.58 (\cred{$-5.23\%$}) & $0.50$ & $0.80 \sim 0.92$ \\
      Case 2 & $0.06$ (\cgreen{$-57\%$}) & $0.012$ & 1.311 (\cred{$-0.23\%$}) & 0.0472 (\cred{$+4.91\%$}) & 0.0078 (\cgreen{$-0.77\%$}) & 0.055 (\cred{$+4.09\%$}) & 23.85 (\cred{$-4.14\%$}) & $0.10$ & $0.86 \sim 0.92$ \\
      Case 3 & $0.06$ (\cgreen{$-57\%$}) & $0.3$ & 1.311 (\cred{$-0.23\%$}) & 0.0474 (\cred{$+5.34\%$}) & 0.0078 (\cgreen{$-0.51\%$}) & 0.055 (\cred{$+4.47\%$}) & 23.77 (\cred{$-4.46\%$}) & $0.50$ & $0.86 \sim 0.92$ \\
      Case 4 & $0.05$ (\cgreen{$-64\%$}) & $1.2$ & 1.314 (\cgreen{$0.00\%$}) & 0.0479 (\cred{$+6.54\%$}) & 0.0078 (\cgreen{$-0.38\%$}) & 0.056 (\cred{$+5.51\%$}) & 23.59 (\cred{$-5.18\%$}) & $1.00$ & $0.86 \sim 0.92$ \\
      Case 5 & $0.03$ (\cgreen{$-78\%$}) & $30.0$ & 1.320 (\cgreen{$+0.46\%$}) & 0.0490 (\cred{$+9.03\%$}) & 0.0079 (\cred{$+0.26\%$}) & 0.057 (\cred{$+7.71\%$}) & 23.21 (\cred{$-6.71\%$}) & $5.00$ & $0.86 \sim 0.92$ \\
      Case 6 & $0.05$ (\cgreen{$-64\%$}) & $30.0$ & 1.334 (\cgreen{$+1.52\%$}) & 0.0520 (\cred{$+15.50\%$}) & 0.0078 (\cred{$+0.13\%$}) & 0.060 (\cred{$+13.22\%$}) & 22.31 (\cred{$-10.33\%$}) & $5.00$ & $0.86 \sim 0.92$ \\
      Case 7 & $0.05$ (\cgreen{$-64\%$}) & $11.2$ & 1.334 (\cgreen{$+1.52\%$}) & 0.0520 (\cred{$+15.50\%$}) & 0.0078 (\cred{$+0.13\%$}) & 0.060 (\cred{$+13.22\%$}) & 22.31 (\cred{$-10.33\%$}) & $2.00$ & $0.86 \sim 1.00$ \\
      Case 8 & $0.03$ (\cgreen{$-78\%$}) & $70.0$ & 1.374 (\cgreen{$+4.57\%$}) & 0.0600 (\cred{$+33.73\%$}) & 0.0080 (\cred{$+1.92\%$}) & 0.068 (\cred{$+29.00\%$}) & 20.17 (\cred{$-18.93\%$}) & $5.00$ & $0.86 \sim 1.00$ \\
      Case 9 & $0.03$ (\cgreen{$-57\%$}) & $78$ & 1.432 (\cgreen{$+8.98\%$}) & 0.0700 (\cred{$+55.31\%$}) & 0.0089 (\cred{$+13.41\%$}) & 0.079 (\cred{$+49.04\%$}) & 18.20 (\cred{$-26.85\%$}) & $7.50$ & $0.86 \sim 1.00$ \\
    \end{tabular}
  }
  \caption{Separation length ($\ell_{\rm sep}$), momentum coefficient ($C_{\mu}$), 
  lift coefficient ($C_l$), drag components ($C_{d,p}$ and $C_{d,f}$), total 
  drag ($C_d$), and aerodynamic efficiency ($L/D$) for uniform suction control cases on the suction side of a  
  NACA4412 airfoil at $AoA = 11^{\circ}$ and $Re_c = 200,000$.  
  Percent changes from the reference case are shown in parentheses. The values in the parentheses report the relative changes obtained by control. }
  \label{tab:ClCd-uctrl}
  \end{center}
\end{table}
\Cref{tab:ClCd-uctrl} summarizes the aerodynamic efficiency and separation delay 
for uniform suction cases. 
Generally, uniform suction increases lift, increases total drag, and delays separation, consistent with the results 
presented in \cref{sec:results}. 
However, applying suction adjacent to the separation point does not necessarily enhance aerodynamic efficiency or further delay separation.

\bibliographystyle{elsarticle-num-names} 
\bibliography{refs}

\end{document}